%% file: Version2.tex
\documentclass{jfm}
\usepackage{graphicx}
\usepackage{epstopdf, epsfig}
\usepackage{stmaryrd}

\usepackage{amsmath}
\usepackage{graphicx}
\usepackage{natbib}
\usepackage{pgfplots}
\usepackage{tikz}
\usetikzlibrary{patterns}
\usepackage{xcolor}
\usepgfplotslibrary{colormaps}

\usepackage{lineno,hyperref}

\pgfplotsset{compat=newest}
\usepgflibrary{decorations.markings}
\usepgflibrary{decorations.shapes}

\definecolor{green}{rgb}{0.0,.6,0.0}
\definecolor{migris}{rgb}{.85,.85,.85}%

\tikzset{dashed/.style={dash pattern=on 3pt off 1pt}}
\tikzset{dashed2/.style={dash pattern=on 2pt off 2pt}}
\tikzset{dashdot/.style={dash pattern=on .4pt off 3pt on 4pt off 3pt}}
\tikzset{,
	MyPersp1/.style={scale=1.8,x={(-0.8cm,-0.4cm)},y={(0.8cm,-0.4cm)},z={(0cm,1cm)}},
	MyPoints/.style={fill=white,draw=black,thick}
		}
\pgfplotsset{
	axis on top,
	xtick align=center,
	ytick align=center,
	xlabel near ticks,
	ylabel near ticks,
         legend cell align=left,
	width={0.9\textwidth},
    	height={0.0\textwidth},
	scale only axis,
	legend style={font=\tiny},
	title style={font=\footnotesize},
	every axis/.append style={font=\footnotesize},
	ticklabel style={font=\footnotesize},
	xlabel style={font=\footnotesize},
	ylabel style={font=\footnotesize},
	xticklabel style={font=\footnotesize},
	every axis plot/.append style={line width=.75pt,mark size=3pt},
	}
\shorttitle{Inertial and capillary migrations}
\shortauthor{Rivero-Rodriguez, J. and Scheid, B.}

\title{Bubbles dynamics in microchannels: inertial and capillary migration forces}

\author{Rivero-Rodriguez, Javier\aff{1}
  \corresp{\email{jriveror@ulb.ac.be}}
 \and Scheid, Benoit.\aff{1}}

\affiliation{\aff{1}TIPs, Universit\'e Libre de Bruxelles, C.P. 165/67, Avenue F. D. Roosevelt 50, 1050 Bruxelles, Belgium
}

\newcommand{\vOm}{\boldsymbol{\Omega}}
\newcommand{\x}{\times}

\newcommand{\vn}{\boldsymbol{n}}
\newcommand{\vnS}{\boldsymbol{n}_S}
\newcommand{\ve}{\boldsymbol{e}}
\newcommand{\vv}{\boldsymbol{v}}
\newcommand{\vx}{\boldsymbol{x}}

\newcommand{\veps}{\boldsymbol{\varepsilon}}
\newcommand{\vf}{\boldsymbol{f}}
\newcommand{\vL}{\boldsymbol{\mathcal{L}}}
\newcommand{\id}{\mathcal I}
\newcommand{\idS}{\mathcal{I}_S}

\renewcommand{\div}{\nabla \cdot}
\renewcommand{\L}{\mathcal L}
\newcommand{\Tau}{\stress}

\newcommand{\Eqref}[1]{Equation~(\ref{#1})}
\newcommand{\Eqsref}[1]{Equations~(\ref{#1})}
\newcommand{\figref}[1]{fig.~\ref{#1}}
\newcommand{\figsref}[1]{figs.~\ref{#1}}
\newcommand{\tabref}[1]{table~\ref{#1}}
\newcommand{\tabsref}[1]{tables~\ref{#1}}
\newcommand{\secref}[1]{sec.~\ref{#1}}

\newcommand{\appref}[1]{appendix~\ref{#1}}

\newcommand{\hstress}{\hat{\boldsymbol{\tau}}}
\newcommand{\stress}{\boldsymbol{\tau}}
\newcommand{\dn}{\delta}

\newcommand{\grad}{\nabla}
\newcommand{\hgrad}{\boldsymbol{D}}

\newcommand{\gradS}{\nabla_{\!S}}
\newcommand{\hgradS}{\boldsymbol{D}_{\!S}}

\newcommand{\divS}{\nabla_{\!S} \cdot}
\newcommand{\hdivS}{\boldsymbol{D}_{\!S} \cdot}

\newcommand{\rotS}{\nabla_{\!S}\times}

\newcommand{\dd}{{\rm{d}}}

\newcommand{\verde}[1]{{#1}}
\newcommand{\rojo}[1]{}

\begin{document}
\maketitle

\begin{abstract}
This work focuses on the dynamics of a train of unconfined bubbles flowing in microchannels. We investigate the transverse position of a train of bubbles, its velocity and the associated pressure drop when flowing in a microchannel depending on the internal forces due to viscosity, inertia and capillarity. Despite the small scales of the system, the inertial migration force plays a crucial role in determining the transverse equilibrium position of the bubbles. Beside inertia and viscosity, other effects may also affect the transverse migration of bubbles such as the Marangoni surface stresses and the surface deformability. We look at the influence of surfactants in the limit of infinite Marangoni effect which yields rigid bubble interface. The resulting migration force may balance external body forces if present such as buoyancy, centrifugal or magnetic ones. This balance not only determines the transverse position of the bubbles but, consequently, the surrounding flow structure, which can be determinant for any mass/heat transfer process involved. Finally, we look at the influence of the bubble deformation on the equilibrium position and compare it to the inertial migration force at the centred position, explaining the stable or unstable character of this position accordingly. A systematic study of the influence of the parameters - such as the bubble size, uniform body force, Reynolds and capillary numbers - has been carried out using numerical simulations based on the Finite Element Method, solving the full steady Navier-Stokes equations and its asymptotic counterparts for the limits of small Reynolds and/or capillary numbers. 
\end{abstract}


\section{Introduction}
   
Nowadays, microfluidic devices are increasingly used for fundamental and exploratory studies in chemistry and biology. Applications span from separation to dissolution processes including reactions which are all strongly coupled to bubble dynamics \cite[]{mikaelian2015bubblyCFD,mikaelian2015bubblyDIS}. In addition, the mixing \cite[]{gunther2004transport} and mass transfer between the disperse and continuous phases \cite[]{mikaelian2015bubblyCFD} enhanced by flow recirculation patterns provide a good ambient for microreactions. Recently, bubble dissolution in microchannels has become of great interest for its application in $CO_2$ sequestration or reactants dissolution \cite[]{shim2014dissolution}. In addition, micro-metric solid particles, drops, bubbles and more intricate objects such as capsules and cells can be easily manipulated with the help of a continuous phase. Continuous flow separation devices rely on hydrodynamics forces, known as migration forces \cite[]{pamme2007continuous} and modulated either by the geometry - such as obstacles, patterns on the channel surface or varying channel section - or by body forces of gravitational, centrifugal, electrical, magnetic or acoustical origin. 

Hydrodynamic forces have drawn the interest of the scientific community after the experiments of \citet{segre1962behaviour} on the migration of rigid spheres in Poiseuille flow of inertial origin. Subsequently, a serie of experimental data \cite[]{tachibana1973behaviour} was obtained in parallel with a theoretical background \cite[]{ho1974inertial,schonberg1989inertial} for rigid spheres. The effect of the particle rotation \cite[]{oliver1962influence}, the shear stress in linear profiles \cite[]{vasseur1976lateral,mclaughlin1991inertial} and the presence of a wall \cite[]{vasseur1977lateral} have also been addressed. \cite{vasseur1976lateral} studied the dynamics of particles sedimenting in a stagnant fluid, shear flow and parabolic flow as well as the wall effect \cite[]{vasseur1977lateral}. 
Inertial forces such as Dean ones have also been used to enhance cell ordering with applications to cell-in-droplet encapsulation \cite[]{kemna2012high_bis}. 

The effect of the particle size on its dynamics is a current field of research and, recently, the transition from small particle size to moderate size and the wall effect have been studied by \cite{hood2015inertial} showing the asymptotic behaviour for small particles. Differential behaviour with respect to particle size has been successfully exploited for particle sorting \cite[]{di2008equilibrium}. The influence of large Reynolds number flows on the transverse equilibrium positions has also been explored experimentally by \cite{matas2004inertial} showing different migration patterns depending on the Reynolds number. 

One decade after the inertial migration was discovered, the migration force induced by the deformability of drops has been address by \cite{chan1979motion}. The wall effect was shown by \cite{kennedy1994motion} to be always repulsive. Numerical simulations on the transient evolution of capillary migration of large drops have been carried out by \cite{coulliette1998motion} and \cite{mortazavi2000numerical} who revealed that the stability of the centred positions of drops strongly depends on the viscosity ratio in the limit of small Reynolds numbers. Recently, experiments and transient simulations of deformable bubbles have been carried out by \cite{chen2014inertial} who also considered the effect of inertia. In this case, they showed how equilibrium positions move to the centreline as the increasing Reynolds number causes larger deformations. They also compared to inertial migration of rigid particles which is the limit of large inner to outer viscosity ratio.

On the other limit, sufficiently small viscosity ratio represents the case of clean bubbles. In this limit, \cite{kennedy1994motion} studied the wall effect, the repulsive character of which was confirmed by  \cite{takemura2009migration} in the case of bubbles rising close to the wall. \cite{stan2011sheathless,stan2013magnitude} described several mechanisms of migration of drops and bubbles in microchannels and focused on inertial an capillary migrations. They carried out comparison between numerical and experimental results, and they concluded that analytical theories of inertial \cite[]{ho1974inertial} and capillary \cite[]{chan1979motion} migrations do not provide a satisfactory quantitative prediction of migration forces, as well as that the migration forces are very difficult to be measured experimentally. Despite these works, the dynamic of bubbles has caught few attention in the literature and a systematical study of the equilibrium state is still missing, especially in the presence of an external force. 

Migration forces have also been studied with more complex surface/bulk rheologies such as thermocapillary stresses \cite[]{Subramanian1983145}, Marangoni stresses with bulk-insoluble surfactant \cite[]{pak2014viscous}, for viscoelastic disperse media \cite[]{leshansky2007tunable} and soft capsules \cite[]{singh2014lateral}.

The general problem of the transverse migration of particles, drops and bubbles has been addressed using several approaches such as experimental, analytical and numerical ones. As analytical approaches, regular asymptotic expansion for the $Re$ numbers have been used in the inertial migration problem \cite[]{cox1968lateral}, together with matched asymptotic expansion in the bubble size, then limiting the solution to small particle/bubble sizes as well as sufficiently large separation between the bubble and the wall \cite[]{schonberg1989inertial}. Further, the transient evolution have been numerically computed using the Level Set Method \cite[]{stan2011sheathless,yang2005migration} or using surface tracking methods, such as Arbitrary Lagrangian-Eulerian Method \cite[]{yang2005migration} or the Boundary Element Method \cite[]{zhou1993flow}. If the transient behaviour is not relevant, the equilibrium solution can be obtained skipping the transient evolution by solving the steady state as performed by \cite{mikaelian2015bubblyCFD}. The latter approach allows, in terms of computational cost, systematic parametrical analysis.

In the literature, several mechanisms of the dynamics of particles, drops and bubbles have been explored with different techniques and in most of the cases, the dependence on the parameters governing the problem has been only qualitatively addressed, the main reasons being the computational cost of transient simulations, limitations of analytical techniques or experimental difficulties in the measurements of migration forces beside the relatively large number of parameters describing the problem. Even though asymptotic expansion has been addressed for the case of migration, the validity of its limits has not been reported. Also, direct coupling between inertial and capillary effects has been relatively unexplored as compared to their separate effects. 

In this paper, we study inertial and capillary migrations in steady conditions in the presence of an external force in circular microchannels, i.e. a microchannel with a circular cross section. First, we write the governing equations which consist in the Navier-Stokes equations and no-slip boundary conditions at the walls of the channel, together with periodic boundary conditions between the inlet and outlet sections of a segment of the channel\verde{, i.e. pressure gradient and velocity profile are periodic except a pressure drop,} containing one bubble and moving at the bubble velocity. We consider both rigid and stress-free interface of the bubble, as well as deformable bubbles with surface tension. We systematically explore the transverse equilibrium position depending on the Reynolds and capillary numbers, as well as the bubble size and uniform body force. We derive regular asymptotic expansions for small Reynolds and/or capillary numbers. The latter expansion involves the linearisation of boundary conditions applied at a deformed boundary and we propose a new approach for this task. We validate these expansions by comparison with the solution of the full system of equations and obtain the range of Reynolds and capillary numbers for which they are  valid. We first focus on neutral bubbles, i.e. in absence of body force, to obtain their stable and unstable positions and then explore the effect of the body force. The stability character of centred positions is also discussed. We finally studied the joint effect of inertia and capillary deformation on the equilibrium positions of neutral bubbles in the first-order expansion as well as the stability of centred bubbles in the full system of equations. 

The structure of the paper is as follows. In \secref{Modelling} we present the model we use to describe the dynamics of the train of bubbles of finite size and explain the underlying hypotheses. The boundary conditions at the bubble surface is given in \secref{ssUndef} for undeformable bubbles and in \secref{ssDef} for deformable bubbles. Additional comments on the scaling and the numerics are given in \secref{ssScaNum}. In \secref{Inertial} we study the effect of inertia, considering the limit of small Reynolds number and explore its range of validity. Analogously, in \secref{Capilar} we investigate the effect of bubble deformation in the small capillary number limit and investigate its range of validity. In both cases, we focus on the neutral bubbles as a reference and the stability of the centred position, as well as the effect of the body force. We complete our study, comparing inertial and capillary effects around the centred position solving the full system of equations and recovering the small $Re$ and $Ca$ number limits in \secref{Ohnesorge}. Conclusions are presented in \secref{Conclusions}.

\section{Modelling}  \label{Modelling}

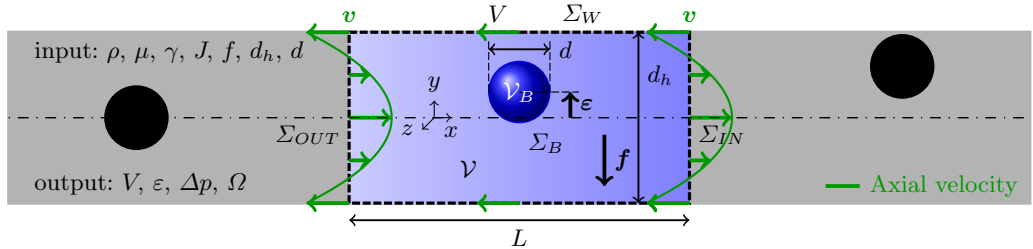
\begin{figure}
\centering
\input{./figures/Sketch.tex}
\caption{Sketch of the modelled segment of a train of bubble in a microchannel.}
\label{Sketch}
\end{figure}

We model the dynamics of a train of bubbles in a circular microchannel. Different equilibrium positions are possible depending on the bubble size and on the balance of several forces such as viscous, inertial, capillary and body forces. To model this situation, we consider a volume $\mathcal V$ containing one bubble of volume $\mathcal{V}_{B}$, or equivalent diameter  $d = \sqrt[3]{6 \mathcal{V}_B / \pi}$,  attached to it and delimited by the walls of the channel, $\Sigma_W$, two cross sections of the channel, $\Sigma_{IN}$ and $\Sigma_{OUT}$, and the bubble surface, $\Sigma_B$, as schematised in \figref{Sketch}. It is assumed that the characteristic time in which changes of volume of the bubble occurs, either due to dissolution or due to the compressibility of the gas, is small compared to the residence time \cite[]{mikaelian2015bubblyCFD,mikaelian2015bubblyDIS} and, therefore, the evolution of the bubble can be considered as quasi-steady in absence of neither vortex shedding nor turbulence, as considered in this work. We include the influence of a uniform body force $\vf \rojo{= f \ve_y}$ \verde{exerted on the liquid} (for gravity $\vf= \rho \boldsymbol{g}$) in the transverse direction. A liquid of density $\rho$, viscosity $\mu$ and surface tension $\gamma$ flows inside a channel of hydraulic diameter $d_h$ with a mean velocity $J$ producing a pressure drop due to the Poiseuille flow modified by the presence of one bubble, $\Delta p$, among a segment of length $L$ taken sufficiently large to avoid bubbles from interfering with each other. The bubble travels with a velocity $V$ and \verde{with an eccentricity} $\veps $ measured from the centre of the channel and determined by the balance of forces acting on the bubble surface in the transverse direction. The bubble might eventually rotate. \rojo{Pseudo-}Periodic boundary conditions are considered in the 'IN' and 'OUT' cross sections, velocity $V$ is imposed at the wall of the channel such that the frame of reference is moving with the bubble. The bubble velocity is determined by the balance of forces acting on the bubble surface in the streamwise direction. Two different conditions are considered at the bubble surfaces, either rigid or stress-free. In the latter case, capillary deformation of the bubble will also be investigated.

According to the previous description, the liquid flow around the bubble in the modelled segment of channel can be analysed by solving, in the reference frame attached to the bubble, the steady Navier-Stokes equations for incompressible fluids,
\begin{subequations}
\label{eq_NS}
\begin{alignat}{5}
\nabla \cdot \vv &= 0    &\qquad&&  \mbox{at $\mathcal V$} \,, \\
\rho \vv \cdot  \nabla \vv  & = \div \hstress   &\qquad&&  \mbox{at $\mathcal V$} \,,
\end{alignat}
\end{subequations}
where $\vv$ and $\hstress = - \hat{p} \id  +  \mu [\nabla \vv + (\nabla \vv)^T]$ are the velocity and the reduced stress tensor, respectively, being $\hat{p}$ the reduced pressure and $\id$ the identity tensor. The pressure then writes as $p = \hat{p} + \vf \cdot (\vx - \veps)$ with pressure reference chosen at the centre of the bubble without loss of generality. In this work, we consider that the uniform body force is transverse to the streamwise direction, i.e.
\begin{align}
\label{trans}
\vf \cdot \ve_x= 0\,.
\end{align}

In the reference frame attached to the bubble moving at the equilibrium velocity $V$, the velocity of the liquid at the wall writes
\begin{equation}
\label{wallbc}
\vv(\vx) = -V \ve_x \qquad \mbox{at $\Sigma_{W}$} \,, 
\end{equation}
whereas the equilibrium condition for the bubble, \verde{using the generalized Stokes theorem \eqref{D}}, is
\begin{equation}
\label{eqeq}
\verde{\vL =} \int_{\Sigma_B} \vn \cdot \hstress   \,\dd \Sigma  
 = \verde{ \mathcal{V}_B \vf}
\rojo{ =\int_{\Sigma_B} \vf \cdot (\vx - \veps)  \ve_{\it{x}} \cdot \vn \, \dd \Sigma }  \,, 
\end{equation}
where $\vL$, $\vn$ and $\dd \Sigma$ are the migration force, the outer normal vector and the differential surface element, respectively. Thus, we equivalently refer hereafter to the balanced body force, $\vf$, instead of the migration force, $\vL$, it is in equilibrium with.

The velocity profile and pressure gradient in Poiseuille flows are uniform, i.e. or periodic with any period, in the streamwise direction. Thus, the pressure field between two different cross-section differ by a constant pressure difference. The presence of bubbles set the period to the distance between consecutive bubbles, $L$,
\begin{subequations}
\label{pseudo}
\begin{alignat}{4}
 \hat{p}(\vx \!+\! L \ve_x) &=\hat{p}(\vx) - \Delta p + L \partial_x p_{\!\, {\rm P}}   &\qquad& \mbox{at $\Sigma_{cross}\rojo{_{OUT}}$} \,, \\
 \vv(\vx \!+\! L \ve_x) &= \vv(\vx) &\qquad& \mbox{at $\Sigma_{cross}\rojo{_{OUT}}$} \,, \\
 \partial_x \vv(\vx \!+\! L\ve_x) &= \partial_x \vv(\vx) &\qquad& \mbox{at $\Sigma_{cross}\rojo{_{OUT}}$} \,.
\end{alignat}
\end{subequations}
where $\Sigma_{cross}$ is any cross-section and the pressure drop is the Poiseuille one, $\partial_x p_{P} $,  modified by the presence of the bubble, $\Delta p$, along the period $L$. The periodic velocity profile has a mean velocity $J$ defined by 
\begin{align}
\label{flowJ}
\int_{\Sigma_{cross}\rojo{_{IN}}} (\vv \cdot \ve_x+V-J) \, \dd \Sigma = 0 \,.
\end{align}
Observe that in absence of bubbles,  $\Delta p = 0 $ and the limit of \eqref{pseudo} divided by $L$ for $L \rightarrow 0$ leads to $\partial_x \hat{p} = \partial_x p_{\!\, {\rm P}} $, $\partial_x \vv = \boldsymbol{0} $ and $\partial_{xx} \vv = \boldsymbol{0}$, which substituted in \eqref{eq_NS} and using \eqref{flowJ} yields the Poiseuille pressure drop, $\partial_x p_{P} = -32 \mu J / d_h^2 $ in circular microchannels. \Eqsref{pseudo}-\eqref{flowJ} must be imposed only at a single cross-section because of the periodicity and we choose $\Sigma_{cross}  \equiv \Sigma_{OUT}$ for convenience.

\verde{The injected energy is dissipated by the viscous forces which is modified by the presence of the bubble. It leads to an effective viscosity which can be measured as the ratio of the averaged pressure drop per unit length to the one induced by the Poiseuille flow. The averaged pressure drop per unit length, $ \Delta_x p_T$, produced in a microchannel fulfils $  \Delta_x p_T = \partial_x p_{\!\, {\rm P}} - \Delta p /L$. If one defines the bubble volumetric fraction $\alpha_G= \mathcal{V}_B / L \Sigma_{cross} $ \rojo{, the volume of the bubble $\mathcal{V}_B = \frac{\pi}{6} d^3$} and being the microchannel cross section $\Sigma_{cross} = \frac{\pi}{4} d_h^2$, the normalized averaged pressure drop writes }
\begin{align}
\label{Cubaud_eq}
\frac{\Delta_x p_T}{\partial_x p_{\!\, {\rm P}}} = 1 + \beta \alpha_G \,, 
\end{align}
where $\beta$ is the pressure correction factor $\beta=- \chi  \Delta p /  \partial_x p_{\!\, {\rm P}}$ and the geometrical factor $\chi $ is $\chi= \Sigma_{cross}/\mathcal{V}_{B} = 3 d_h^2 / 2 d^3 $ in circular microchannels. \Eqref{Cubaud_eq} has been experimentally investigated by \cite{cubaud2004transport} in the limit of small $\alpha_G$ in square microchannels with a wide dispersion around the value $\beta = 1 $ which may be due to the actual dependence of the bubble size and eccentricity. In fact, $\beta$ spans from negative to positive values, depending on $\Delta p$, as it will be shown later.

We consider different boundary conditions at the bubble surface depending on if the capillary deformation of the bubble is taken into account. 

\subsection{Undeformable bubbles}\label{ssUndef}

Firstly, we consider rigid and stress-free boundary conditions on the surface of undeformed bubbles. It is relevant for the study of the inertial migration in the cases of infinite Marangoni effect and clean surface, respectively.

On the one hand, in the limit of infinite Marangoni effect, the flow at the bubble surface corresponds to that of a rigid solid that rotates with rotational velocity 
$\vOm$ and is determined by the no-slip condition and by the equilibrium of moments, respectively, 
\begin{subequations}
\label{rigid}
\begin{alignat}{2}
            \vv(\vx) =                                              \vOm & \x (\vx-\veps)  \qquad \mbox{at $\Sigma_{B}$}  \,,   \label{rot} \\
\boldsymbol{0} = \int_{\Sigma_B}  \vn \cdot \hstress & \x (\vx-\veps)   \, \dd \Sigma \,.     \label{rotb}
\end{alignat}
\end{subequations}
\verde{These conditions are the same as those governing in the case of infinitely viscous drops or solid spheres. Thus, their interactions with the liquid are identical.} 

On the other hand, in the case of clean bubbles, \verde{the stress-free boundary condition which consists in vanishing tangential stresses and the impermeability condition,} is applied for the case of clean bubbles, i.e.
\begin{subequations}
\label{SF}
\begin{alignat}{4}
 \vn \cdot \hstress  &= - \hat{\lambda} \vn &&  \qquad \mbox{at $\Sigma_{B}$} \,,  \label{SFa}\\ 
 \vn \cdot \vv &= 0  &&  \qquad \mbox{at $\Sigma_{B}$}\,, \label{SFb}
\end{alignat}
\end{subequations}
where the surface variable $\hat{\lambda}$ is the reduced counterpart of the \verde{impermeable surface} pressure $ \lambda = \hat{\lambda} + \vf \cdot (\vx-\veps)$ that the liquid exert on the bubble surface. \verde{These conditions are the same as those governing in the case of inviscid drops.  Thus, their interactions with the liquid are identical.}


\subsection{Deformable bubbles}\label{ssDef}

Secondly, we consider the case of deformable bubbles which is relevant for the study of the capillary migration. In this case the surface position is governed by the following equilibrium of stresses and the impermeability condition
\begin{subequations}
\label{deformable}
\begin{alignat}{4}
\vn \cdot \llbracket \Tau \rrbracket &=  \hgradS \gamma  &&  \qquad \mbox{at $\Sigma_{B}$} \,,  \label{Defa}\\ 
 \vn \cdot \vv  &= 0  &&  \qquad \mbox{at $\Sigma_{B}$}\,, \label{Defb}
\end{alignat}
\end{subequations}
where $\llbracket \Tau \rrbracket =  [p_{G} - \vf \cdot (\vx-\veps) ]  \id + \hstress$ is the jump of stress tensor and \verde{the global variable} $p_G$\verde{, i.e. neither domain nor surface,} is the gas pressure governed by the Young-Laplace law which \verde{depends on the volume of the bubble, $\mathcal{V} _B$.} The exterior surface differential operator $\hgradS$ is defined by its application to a quantity $\varphi$, either scalar, vector or tensor, as \verde{
\begin{align}
\label{surfopdef}
\int_\Sigma \hgradS \varphi \,\dd \Sigma = \int_{ \Gamma} \vnS \varphi  \dd \Gamma \,,
\end{align}
where $\Sigma$ is a surface bounded by a contour $\Gamma$, the outer normal of which is denoted by $\vnS$. The vector $\vnS$ is perpendicular to the contour $\Gamma$ and to the outer normal vector of the surface $\Sigma$, $\vn$, as schematised in \figref{Sketch5}. Note that $\hgradS \gamma$ includes normal pressure proportional to the curvature and Marangoni effect (not considered in this work), $\hgradS \gamma = - \gamma  \vn \divS \vn + \gradS \gamma$, with $\gradS = (\id- \vn\vn) \cdot \grad$ the surface gradient operator.

\begin{figure}
\centering
\input{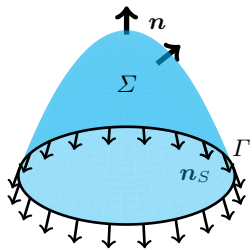}
\caption{Sketch of the contour $\Gamma$ limiting the fluid surface $\Sigma$; $\vn$ is the outer normal to the surface and $\vnS$ is the outer normal to the contour contained in $\Sigma$ such that $\vn \cdot \vnS =0$. 
}
\label{Sketch5}
\end{figure}

Since $\Sigma_B$ for a bubble is a close surface and has no contour, it can be infer from \eqref{surfopdef} that the surface tension exert no force on the bubble as a whole. Thus the integral of \eqref{Defa} along $\Sigma_B$ leads to \eqref{eqeq}, and therefore \eqref{eqeq} is redundant in the case of deformable bubbles.

Finally, since the domain is deformable, the volume and eccentricity of the deformed bubble needs to be defined,
\begin{subequations}
\label{pG}
\begin{align}
\mathcal{V}_B &= \int_{\mathcal{V}_B} \dd \mathcal{V} \,, \\
\boldsymbol{0} &= \int_{\mathcal{V}_B} (\vx - \veps) \,\dd \mathcal{V} \,, \label{pGb}
\end{align}
\end{subequations} 
where $\mathcal{V}_B$ is the domain occupied by the bubble.}


\subsection{Scaling and numerics}\label{ssScaNum}

\Eqsref{eq_NS}-\eqref{pG} can be made dimensionless with the hydraulic diameter of the channel $d_h$, the mean velocity of the flow $J$  and the viscous stress $\mu J / d_h$ as characteristic length, velocity and pressure, respectively. \Eqsref{eq_NS}-\eqref{pG} are referred hereafter as their dimensionless counterpart obtained by the substitution in these equations of
\begin{align}
\label{makedimless}
d_h \rightarrow 1  \,,\quad
J \rightarrow 1     \,,\quad
\mu \rightarrow 1 \,,\quad
\rho \rightarrow Re  \,,\quad 
\gamma \rightarrow Ca^{-1}  \,,\quad
\end{align}
where the dimensionless numbers are
\begin{align}
\label{dimless}
Re = \frac{\rho J d_h}{\mu} \,, \quad 
Ca = \frac{\mu J}{\gamma} \,. 
\end{align}
Typical values for water in a microchannel of diameter $d_h = 500 \mu$m and a flow rate of $J=0.1$ m$/$s correspond to dimensionless numbers of $Re=50$ and $Ca=1.4 \cdot 10^{-3}$. The dimensional domain variables can be recovered by
\begin{align}
\label{dimless_soldom}
\vv \leftarrow J \vv \,, \quad 
\hat{p} \leftarrow \frac{ \mu J}{ d_h} \hat{p} \,,\quad 
\end{align}
and dimensional global variables by
\begin{align}
\label{dimless_sol}
\varepsilon \leftarrow d_h \varepsilon \,,\quad 
d \leftarrow  d_h d \,, \quad
f \leftarrow \frac{\mu J}{d_h^2} f   \,, \quad
V \leftarrow J V \,, \quad 
\Delta p \leftarrow \frac{ \mu J}{ d_h} \Delta p \,,\quad 
\Omega \leftarrow \frac{J}{ d_h} \Omega \,.
\end{align}

Because of symmetry around $z=0$, $\vf$ and $\veps$ are aligned. Thus, we choose, without loss of generality, that $\ve_y$ is aligned with both. We define their magnitudes in this direction as $f=\vf \cdot \ve_y $ and $\varepsilon = \veps \cdot \ve_y $, respectively. In addition, also due to symmetry, the rotation takes places within the plane $x$-$y$ and we defined its magnitud as $\vOm = \Omega \ve_z$. 

In the numerical simulations, the relations between $\Delta p$ and $J$, $\varepsilon$ and $f$ as well as, in the case of deformable bubbles, $p_G$ and $\mathcal{V}_B$, need to be parametrised. For this purpose, we impose $J$, $\varepsilon$ and $\mathcal{V}_B$ whereas $\Delta p$, $f$ and $p_G$ are solved. This particular choice is because we make the system of equations dimensionless with $J$, the function $\varepsilon = \varepsilon(f)$ might be multi-valued whereas the inverse, $f = f(\varepsilon)$, is not, and the size of the bubble $ \mathcal{V}_B$ together with its eccentricity $\varepsilon$ fixes the domain $\mathcal V$. Although for deformable bubbles the domain deforms, it is numerically convenient to use an initial domain as similar as possible to the deformed one. Furthermore, it is the case for small $Ca$ numbers for which the bubble is spherical and the $Ca$ number can be increased using continuation methods enhancing convergence.

We consider that the bubble is undeformable for strictly $Ca = 0$, and deformable for $Ca \neq 0$. 

In the case of undeformable bubbles, the variables used in the simulations are the domain variables: $\hat{p}$, $\vv$; the surface variable: $\hat{\lambda}$ (only for stress-free interface); and the global variables: $f$, $V$, $\Delta p$ and $\Omega$ (only for rigid interface). These variables are governed by \eqref{eq_NS}, \eqref{trans}, \eqref{wallbc}, \eqref{eqeq}, \eqref{pseudo}, \eqref{flowJ} and either \eqref{rigid} or \eqref{SF}, for rigid or stress-free bubbles, respectively.

In the case of deformable bubbles, the variables are the domain variables:  $\hat{p}$, $\vv$; and the global variables: $p_G$, $f$, $V$ and $\Delta p$. These variables are governed by \eqref{eq_NS}, \eqref{trans}, \eqref{wallbc}, \eqref{pseudo}, \eqref{flowJ} and \eqref{deformable}. The deformation of the bubble is handled using the Arbitrary Lagrangian-Eulerian (ALE) method \cite[]{DONEA1982689} together with the volume and eccentricity definitions \eqref{pG}.

The aforementioned system of equations are solved using the Finite Element Method (FEM). The equations are implemented in weak form using the software COMSOL and the \emph{Moving Mesh} module that implements the ALE method. Details of the definition and discretisation using FEM of the surface operators such as $\hgradS$ and $\gradS$ are given in \appref{AppS}.

We have validated the implementation of the previous equations by comparing the equilibrium position of a particle with diameter $d=0.305$ for $Re=0.196$, in absence of body force, obtained by other authors both numerically and experimentally, see figure 2 in \cite{yang2005migration}. We benefit of the symmetry with respect to the $z=0$ plane to reduce by a factor $2$ the mesh size although computational time -which remains within the range of a few minutes (1-3 minutes) for one solution- is not limiting in the computations reported here. We have also carefully checked the convergence of a tetrahedral mesh with hexahedral elements on the bubble surface. Symmetric meshes with respect to the $x=0$ plane has been found to increase the accuracy on the solution. 


\section{Migration forces}

\begin{figure}
\centering
\input{./figures/Regimes.tex}
\caption{Considered flow regimes involving inertial and capillary migrations. Bricks and dots represent linear and nonlinear problems, respectively.}
\label{fig:casos}
\end{figure}
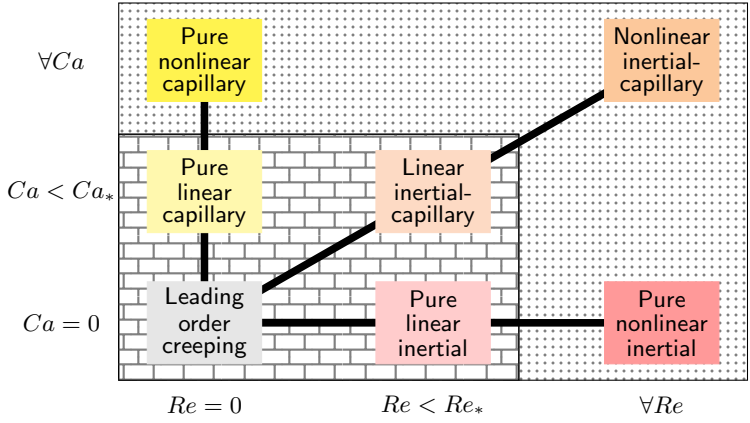

In this section, we investigate both inertial and capillary migration forces. First, we quantitatively study the influence of the bubble size and the eccentricity (equivalently the body force) in the limit of small $Re$ or $Ca$ numbers, separately. Asymptotic regular expansions are derived for these limits in dimensionless form which lead to the creeping flow around undeformed spheres for zeroth order and the inertial or capillary perturbations for first order. Their ranges of validity are studied by solving the nonlinear full system of equations \eqref{eq_NS}-\eqref{pG} and a criteria is obtained depending on the bubble diameter, i.e. $Re<Re_*(d)$ or $Ca<Ca_*(d)$, as illustrated in \figref{fig:casos} and determined later on. The stability of the centred position, namely $\varepsilon=0$, is finally studied using both limits and again comparing with the solution of the full system of equations. 

\subsection{Inertial migration.}\label{Inertial}

Straight channels are a particular geometry in which a constant flow does not depend on the Reynolds number provided no turbulence develops and it corresponds to the purely creeping flow which is reversible. However, the presence of the bubble modifies the flow structure in such a manner that inertia forces come into play and, thus, break the reversibility of the flow. In this case, the bubble experience a transverse force of inertial origin even for small but finite inertia \cite[]{segre1962behaviour}. It motivates an expansion of the system of equations \eqref{eq_NS}, \eqref{trans}, \eqref{wallbc}, \eqref{eqeq}, \eqref{pseudo}, \eqref{flowJ} and either \eqref{rigid} or \eqref{SF}, for small Reynolds number as $\psi = \sum_{j=0}^{\infty} Re^j \psi_j $, where $\psi$ represents any of the dependent variables $\hat{p}$, $\vv$, $f$, $V$, $\Delta p$ or $\hat{\lambda}$ and $\Omega$ (if applicable), that yields \cite[]{cox1968lateral}, up to first order, to
\begin{subequations}
\label{eq_NS01}
\begin{alignat}{9}
\nabla \cdot \vv_0 &= 0 \,, &\qquad &   & \nabla \cdot \vv_1 &= 0 &\qquad&&  \mbox{at $\mathcal V$}  \,, \\
\boldsymbol{0} &= \nabla \cdot  \hstress_{0} \,, &\qquad &   & \vv_0 \cdot \nabla \vv_0 &= \nabla \cdot \hstress_{1} &\qquad&&  \mbox{at $\mathcal V$}   \,, 
\end{alignat}
\end{subequations}
with vanishing body force in the axial direction \eqref{trans}
\begin{equation}
\label{trans01}
 \vf_0 \cdot \ve_x  = 0 \,,\qquad \vf_1 \cdot \ve_x  = 0  \,,  
\end{equation}
together with the linearization of the velocity at the walls \eqref{wallbc}
\begin{equation}
\label{wallbc01}
 \vv_0(\vx)  = -V_0 \ve_x \,,\qquad \vv_1(\vx)  = -V_1 \ve_x \qquad \mbox{at $\Sigma_{W}$} \,,  
\end{equation}
of the equilibrium equations for the bubble \eqref{eqeq}
\begin{equation}
 \int_{\Sigma_B} \vn \cdot \hstress_{0}  \, \dd \Sigma  = \mathcal{V}_B \vf_0   \,,\qquad \int_{\Sigma_B} \vn \cdot \stress_{1} \, \dd \Sigma = \mathcal{V}_B \vf_1 \,, \label{drag01} 
\end{equation}
of the periodic conditions \eqref{pseudo}
\begin{subequations}
\label{pseudo01}
\begin{alignat}{5}
 \hat{p}_0(\vx\!+\! L \ve_x) &\!=\!\hat{p}_0(\vx) \!-\! \Delta p_0 \!+\! L \partial_x p_{P} \,,&\,& &
 \hat{p}_1(\vx\!+\! L \ve_x) &\!=\!\hat{p}_1(\vx) \!-\! \Delta p_1    &\quad& &
  \mbox{at $\Sigma_{cross}\rojo{_{OUT}}$} \,, \\
 \vv_0(\vx\!+\! L \ve_x) &\!=\! \vv_0(\vx) \,,&\,& &
 \vv_1(\vx\!+\! L \ve_x) &\!=\! \vv_1(\vx) &\quad&  &
 \mbox{at $\Sigma_{cross}\rojo{_{OUT}}$} \,, \\
 \partial_x \vv_0(\vx\!+\!L\ve_x) &\!=\! \partial_x \vv_0(\vx) \,,&\,& &
 \partial_x \vv_1(\vx\!+\!L\ve_x) &\!=\! \partial_x \vv_1(\vx) &\quad& &
  \mbox{at $\Sigma_{cross}\rojo{_{OUT}}$} \,, 
\end{alignat}
\end{subequations}
and of the mean flow \eqref{flowJ}
\begin{align}
\label{flowJ01}
\int_{\Sigma_{cross}\rojo{_{IN}}} (\vv_0 \cdot \ve_x+V_0-1) \,\dd \Sigma = 0 \,,\qquad \int_{\Sigma_{cross}\rojo{_{IN}}} (\vv_1 \cdot \ve_x+V_1) \,\dd \Sigma = 0 \,.
\end{align}
The linearisation of the boundary conditions on the bubble surface are
\begin{subequations}
\label{rigid01}
\begin{alignat}{7}
\vv_0(\vx)         = \vOm_0 &\x (\vx-\veps)& \,&, & &\quad& \vv_1(\vx) =  \vOm_1 &\x (\vx-\veps)& \qquad& \mbox{at $\Sigma_{B}$} \,,  \\
\boldsymbol{0} = \!\int_{\Sigma_B} \!\! \vn \cdot \hstress_{0}  &\x (\vx-\veps)&  \,& \dd \Sigma \,,  &&\quad& \boldsymbol{0} = \!\int_{\Sigma_B}\!\! \vn \cdot \hstress_{1}  &\x (\vx-\veps)& \, \dd \Sigma \,,  &  
\end{alignat}
\end{subequations}
for rigid interface, or
\begin{subequations}
\label{SF01}
\begin{alignat}{13}
\vn \cdot \hstress_{0}  &=  -\hat{\lambda} _0 \vn \,, &\qquad&& \vn \cdot \hstress_{1}  &=  -\hat{\lambda} _1 \vn &&  \qquad \mbox{at $\Sigma_{B}$} \,,  \\ 
\vn \cdot \vv_0 &= {0} \,, &\qquad&& \vn \cdot \vv_1 &= {0} &&  \qquad \mbox{at $\Sigma_{B}$}\,,  
\end{alignat}
\end{subequations}
for stress-free interface. Note that the migration force can be alternatively computed using the reciprocal theorem as performed by \cite{ho1974inertial} in the limit of $Re<<1$ and $d<<1$.
  
The solution of the system \eqref{eq_NS01}-\eqref{SF01} for the inertial migration force, bubble velocity, pressure drop and rotational velocity (only applicable in the case of rigid interface) is of the form, truncated at $\mathcal{O} (Re^2)$, 
\begin{align}
\label{exp1}
f_{\rho}\equiv \ \frac{f}{Re} \approx \, f_{1}(\varepsilon,d)   \,,\quad
V \approx V_0(\varepsilon,d) \,,\quad
\beta \approx \beta_0(\varepsilon,d)  \,,\quad
\Omega \approx \Omega_0(\varepsilon,d)  \,,
\end{align}
where the removed terms have been found numerically to be vanishing, i.e. $f_0 = V_1 = \beta_1 = \Omega_1 = 0 $, as it can be inferred from the symmetries and reversibilities of the flow and shown in \appref{AppB}. Consequently, the balanced body force comes from the first-order solution whereas the velocity, pressure correction factor and the rotational velocity comes from the zeroth-order solution. Results on the inertial migration force $f$ have been recurrently reported, \cite{yang2005migration,di2009particle,stan2011sheathless}. We contribute with the systematic study of the effect of the bubble size $d$ on $f$ in addition to the functions $V(\varepsilon,d)$, $\beta(\varepsilon,d)$ and $\Omega(\varepsilon,d)$ (only for rigid bubbles).

\begin{figure}
\centering
\input{./figures/L_vs_CRd04.tex}
\input{./figures/V_c_CRd04.tex}
\input{./figures/beta_c_CRd04.tex}
\input{./figures/Om_c_CRd04.tex}
\input{./figures/Legend1.tex}
\caption{Effect of the eccentricity $\varepsilon$ of bubbles with rigid interface and size $d=0.4$ in the pure linear inertial regime on the (a) balanced body force, (b) bubble velocity, (c) pressure correction factor and (d) rotational velocity.}
\label{fig_eq_Re01}
\end{figure}
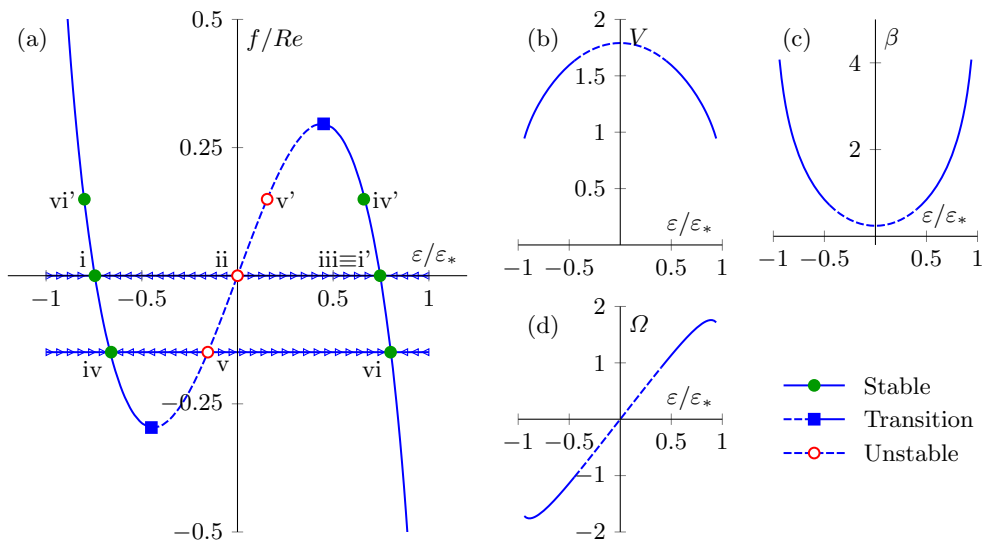

In \figref{fig_eq_Re01}, we depict the solution \eqref{exp1} for a representative bubble of size $d=0.4$ in the small Reynolds number limit for values of eccentricity within the geometrically feasible range $-\varepsilon_*<\varepsilon<\varepsilon_*$ where $\varepsilon_* = \frac12(1-d)$ corresponds to the position of bubbles touching the wall. In \figref{fig_eq_Re01}a, we can observe the existence of multiple equilibrium positions of the bubble for a range of values of the balanced body force $f$. This multiplicity of solutions is lost for sufficiently large values of the balanced body force and for which the bubbles are closer to the wall. In particular, for the case of neutral bubbles, i.e. no body force, $f=0$, there exists typically three equilibrium solutions, whose stability is determined by the slope $\partial_\varepsilon f $, two of them being stable, denoted by (i)  and (iii), whose attraction ranges are separated by the unstable one (ii). It is represented by the line decorated with arrows indicating the transverse motion of a bubble when located out of the equilibrium position. In addition, this can be observed by turning the force on towards negative values but moderate so that multiple solutions still exist, as exemplified for $f/Re = -0.15$ in \figref{fig_eq_Re01}a. In effect it results in a positive force exerted on the bubbles and, therefore, bubbles at stable positions, namely in the position (i) and (iii), are shifted upward toward positions (iv) and (vi), i.e. in the same direction than the force exerted on the bubble, namely to the right in \figref{fig_eq_Re01}a. Unstable bubbles, namely at position (ii), are shifted in the opposite direction towards (v), revealing the unstable character of this position. Analogous reasons show that the positions (iv)-(vi) show the same character than (i)-(iii). Once the equilibrium position for a given body force $f$ is known, one can obtain from \figref{fig_eq_Re01}b-d the values of $V$, $\Delta p$ and $\Omega$. In effect, at the centre of the channel, i.e. $\varepsilon = 0$, the bubble velocity is maximal, the pressure drop is minimal and the rotational velocity reverses orientation. Given the symmetries of the functions in \eqref{exp1} with respect to the position of the bubble, it is sufficient to explain them only for positive eccentricities, $\varepsilon \ge 0$. Note the odd symmetry of the balance body force for which any condition has an equivalent for the opposite position, as illustrated in \figref{fig_eq_Re01}a by the positions (iv'), (v'), (vi') in the case of moderate and positive force, $f/Re = +0.15$.
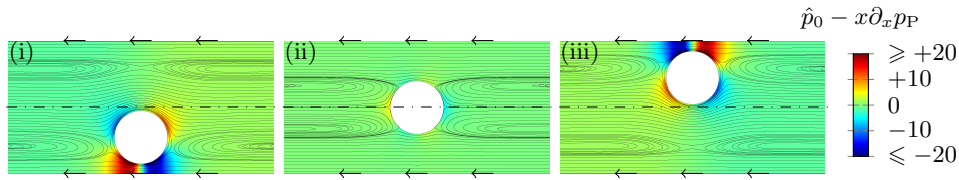
\begin{figure}
\centering
\input{./figures/pressurefield0}
\caption{Streamlines of the creeping flow, $\vv_0$, and colormap for the pressure field, $\hat{p}_0 -  x \Delta_x p_{\!\, {\rm P}}$, of the creeping flow at the symmetry plane $z=0$ for neutral bubbles $ \it{f/Re}=0$ with rigid interface and size $d=0.4$. Labels correspond to points in \figref{fig_eq_Re01}a.}
\label{pressurefield0}
\end{figure}

The flow pattern and pressure field allow to rationalise expressions \eqref{exp1}. In particular, the zeroth-order solutions, depicted in \figref{pressurefield0}, reveal the anti-symmetry and reversibility of the creeping flow. 
The positions of the bubbles in \figref{pressurefield0} correspond to the points in \figref{fig_eq_Re01} labeled with the same roman number than the picture. Observe that points (i)  and  (iii) correspond to upside down flip due to the symmetry around the plane $y=0$. 

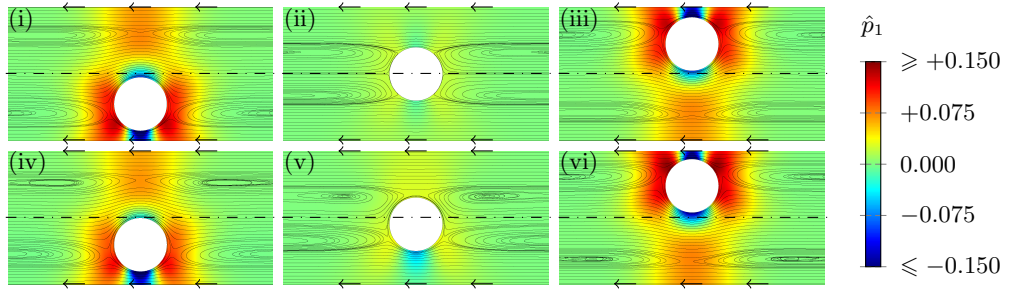
\begin{figure}
\centering
\input{./figures/pressurefield1.tex}
\caption{Streamlines of the creeping flow $\vv_0$ and colormap for the pressure field of the pure linear inertial perturbation, $\hat{p}_1$,  at the symmetry plane $z=0$ for (i-iii) neutral $f/Re=0$ and (iv-vi) non-neutral $f/Re=-0.15$ bubbles with rigid interface and size $d=0.4$. Labels correspond to point solutions marked in \figref{fig_eq_Re01}a.}
\label{fig_or_L1}
\end{figure}

The behaviour of inertial migration can be explained with the help of the zeroth-order flow pattern and the first-order correction of the pressure field shown in \figref{fig_or_L1}. In this latter case, the flow is symmetric and anti-reversible, as explained in \appref{AppB}, which enforces the first-order bubble velocity, pressure drop and rotational velocity to vanish whereas first-order balanced body force can be non-zero. In \figref{fig_eq_Re01}a, points (i)-(iii) correspond to neutral bubbles which means that underpressure and overpressure zones observed in \figref{fig_or_L1}i-iii should counterbalance each other. To understand the stability of neutral bubbles we plot in \figref{fig_or_L1}iv-vi the counterparts for a moderate negative body force, as in \figref{fig_eq_Re01}a. The displacement of bubbles from positions (i) to (iv) and (iii) to (vi), respectively, leads to a variation of the net force exerted on the bubble in the opposite direction to the displacement. It can be seen in both cases an increase of the underpressure and a decrease of the overpressure which provides the migration force balancing the body force. Contrarily, the displacement from (ii) to (iv) leads to a variation in this force in the same direction to the displacement due to an underpressure from this side making this position unstable as inferred in \figref{fig_eq_Re01}a. 

\begin{figure}
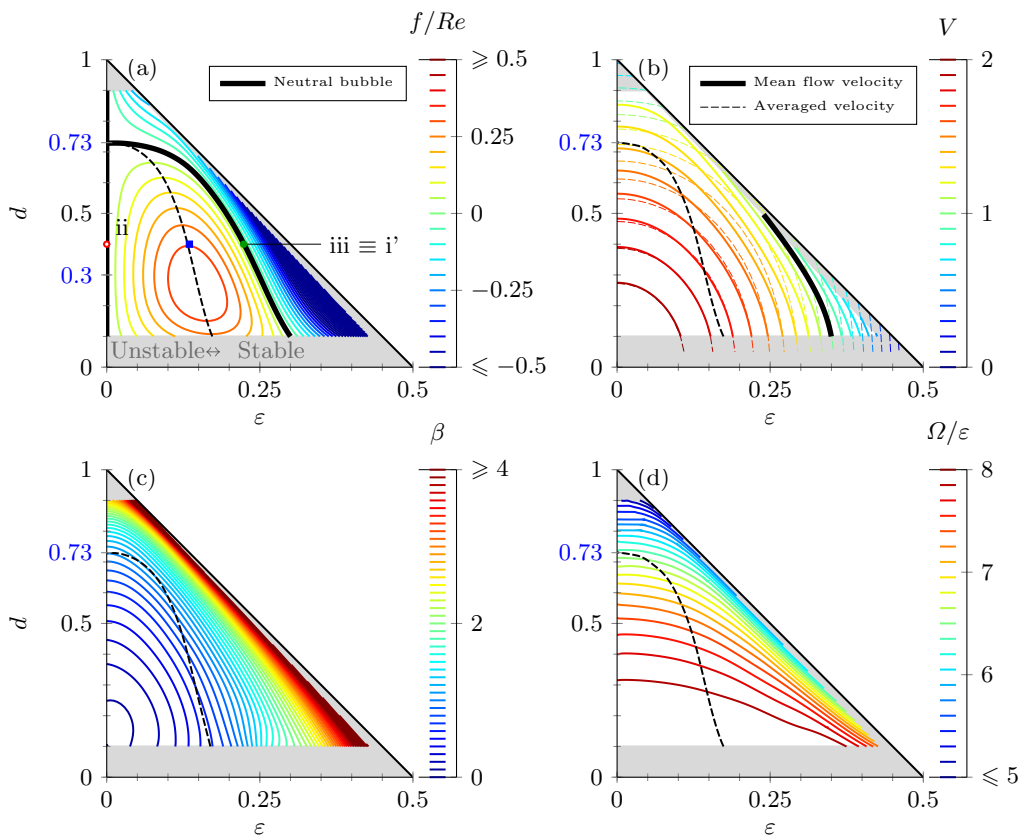

\centering
\input{./figures/CircularRigido_F_RE.tex}
\input{./figures/CircularRigido_V.tex}
\input{./figures/CircularRigido_3_64_DP_D3.tex}
\input{./figures/CircularRigido_OM_VAR.tex}
\caption{Effect of the eccentricity $\varepsilon$ and the size $d$ of bubbles with rigid interface in the pure linear inertial regime on the (a) balanced body force, (b) bubble velocity, (c) pressure correction factor and (d) rotational velocity. \ref{NotExplored} Not explored, \ref{epsast} $\varepsilon = \varepsilon_*$ and \ref{StabTrans} stability transition.}
\label{CircularRigido}
\end{figure}

In \figsref{CircularRigido}, we plot the global variables $f$, $V$, $\beta$ and $\Omega$ \eqref{exp1} describing the bubble dynamics with rigid interface \eqref{rigid01} in the pure linear inertial regime, see \figref{fig:casos}, as a function of the diameter and the position of the bubble. Because of numerical limitations, we only computed solutions within the ranges $0.1 \leq d \leq 0.9$ and $0 \leq \varepsilon \leq 0.95\varepsilon_*$. 
We observe in \figref{CircularRigido}a that neutral bubbles, $f=0$, with diameter $d  \lesssim 0.73 $ are unstable at the centred position, as revealed by the positive value of $\partial_\varepsilon f/Re>0$. We also observe that centred bubbles with diameter $d \approx 0.3$ are the most unstable and that equilibrium positions get away from the centre as its size diminishes. The transition between unstable and stable positions is depicted with black-dashed lines which correspond to the local extremum of \figref{fig_eq_Re01}a, illustrated with \ref{TransMark} for $d=0.4$. Note the imperfect Pitchfork bifurcation of the eccentricity with $d$ as the parameter and $f$ as the imperfection, which contains information of the stability of the stability of the centred branch. Some relevant conditions exhibiting these features are depicted in \figref{fig_eq_Re01}a, see points (ii) and (iii) in \figref{fig_eq_Re01}a. 
In \figref{CircularRigido}b, we can numerically observe that small bubbles follow the Poiseuille flow $V_{P}=2(1-4\varepsilon^2)$. Proximity of the bubble to the wall, either because of bubble size or due to eccentricity, turns out to make bubbles travel slower. The rotational bubble velocity corresponds to the rotational velocity of the Poiseuille flow (half of the vorticity), $ \Omega = - \frac12 \partial_\varepsilon 2[1-4 \varepsilon^2] = 8 \varepsilon$, as shown in \figref{CircularRigido}d. In fact, the bubble tends to assure continuity of the liquid at the rigid interface of the bubble with a velocity very similar to the average over the bubble surface of the Poiseuille velocity, $V \approx \Sigma^{-1}_B \int_{\Sigma_B} 2[1-4(y^2+z^2)] \, \dd \Sigma$, as plotted in \figref{CircularRigido}b in coloured thin dashed lines.  The rotational velocity, depicted in \figref{CircularRigido}d, also reduces due to the effect of the wall. Rigid bubbles behave like a plug since for large size the pressure drop increases with the bubble size as shown in \figref{CircularRigido}c. 
\begin{figure}
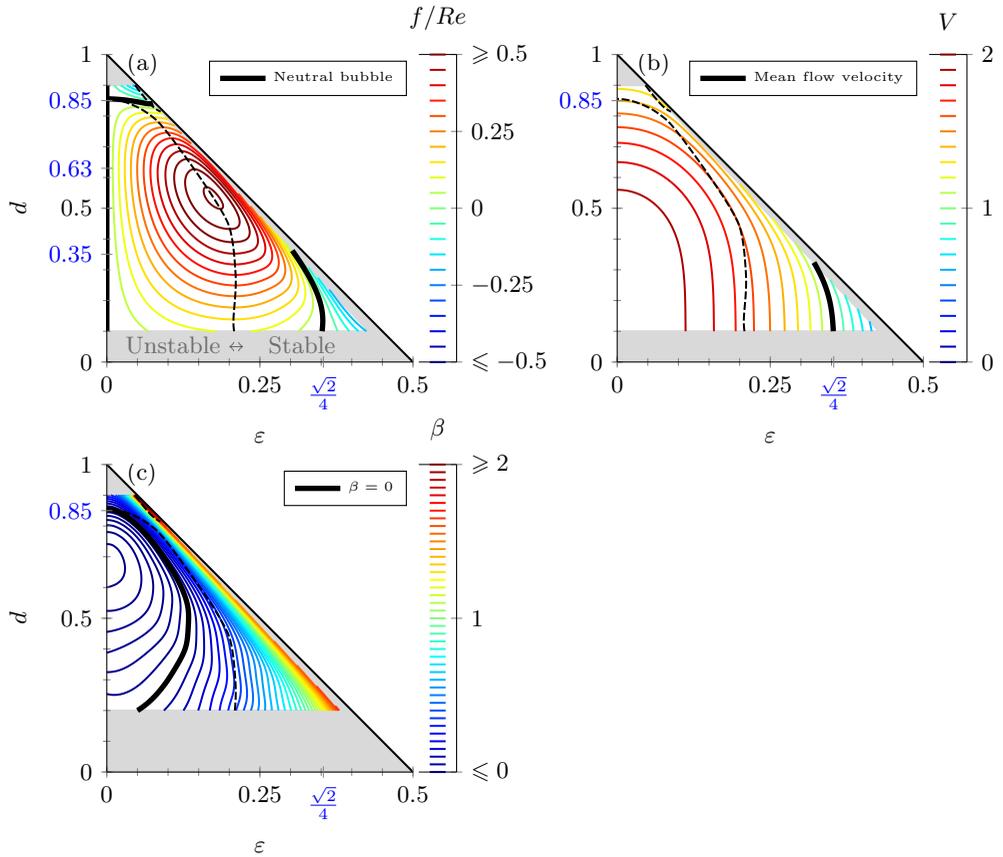

\centering
\input{./figures/CircularLibre_F_RE.tex}
\input{./figures/CircularLibre_V.tex}
\input{./figures/CircularLibre_3_64_DP.tex}
\caption{Effect of the eccentricity $\varepsilon$ and the size $d$ of bubbles with stress-free interface in the pure linear inertial regime on the (a) balanced body force, (b) bubble velocity and (c) pressure correction factor. \ref{NotExplored} Not explored, \ref{epsast} $\varepsilon = \varepsilon_*$ and \ref{StabTrans} stability transition.}
\label{CircularLibre}
\end{figure}

In \figref{CircularLibre}, we depict the influence of the bubble size on the dynamics of the bubbles with stress-free interface \eqref{SF01}. In \figref{CircularLibre}a, we plot the body force $f$, related to the migration force by \eqref{eqeq}. Analogous behaviour to the rigid interface is observed. In this case the transition of stability for centred bubbles takes place at around $d \approx 0.85$ and the most unstable centred bubble are those with diameter $d \approx 0.63$. The range of position for which the bubble is unstable is larger than for rigid bubbles. 
We observe that it exists a range of bubble size $0.35 \lesssim d \lesssim 0.83$ for which bubbles are touching the wall, i.e. in practice for $\varepsilon > 0.95 \varepsilon_{*}$. In \figref{CircularLibre}b, we can again numerically observe that small bubbles follow the Poiseuille flow $V_{P}$ and, in absence of body force, bubbles migrate to the position at which it travels at the mean flow, i.e. $V=V_P=1$ at $\varepsilon=\sqrt{2}/4$. Proximity of the bubble to the wall, either because of bubble size or due to eccentricity, turns out to make the bubbles to travel slower. In \figref{CircularLibre}c, we observe the pressure correction factor due to the presence of the bubble. We highlight the conditions for which the pressure correction factor vanishes, i.e. the presence of bubbles reduces the pressure drop. Unfortunately, these positions are all unstable and stable bubbles in the pure linear inertial regime always increase the pressure drop.

For the sake of completeness, polynomial fitting of functions \eqref{exp1} are given in \appref{AppPF} for both rigid and stress-free interfaces.

\begin{figure}
\centering
\input{./figures/CircularRigido_ReFinito_L_Re_d04.tex}
\caption{The influence of the $Re$ number on the balanced body force for bubbles of size $d=0.4$ with rigid interface in the pure nonlinear inertial regime.}
\label{fig_Re1}
\end{figure}
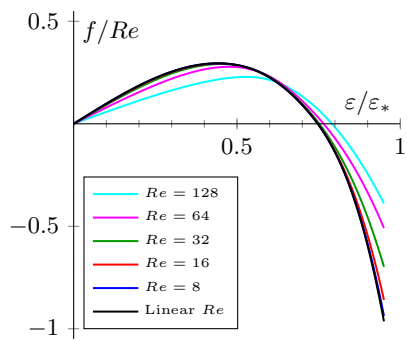

For sufficiently large $Re$ numbers, the solution \eqref{exp1} is no longer valid as we observe in \figref{fig_Re1}, where the balanced body force is depicted versus the eccentricity for various Reynolds numbers for a bubble with diameter $d = 0.4$ in the case of rigid interface bubbles and pure nonlinear inertial regime. Anyhow, the behaviour remains qualitatively similar to the pure linear inertial migration.

\begin{figure}
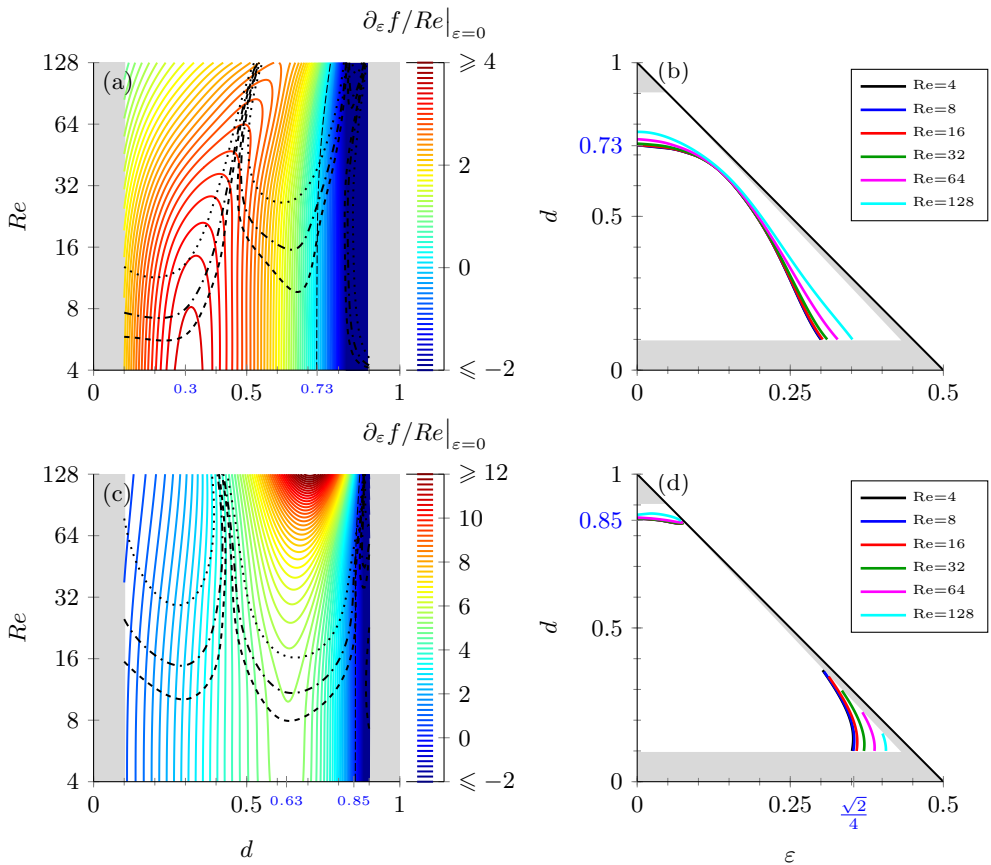

\centering
\input{./figures/CircularRigido_Stiffness_d_Re.tex}
\input{./figures/NeutralBubble_CircularRigid}
\input{./figures/CircularLibre_Stiffness_d_Re.tex}
\input{./figures/NeutralBubble_CircularFree}
\caption{Influence of the $Re$ number and the size of bubbles with (a-b) rigid or (c-d) stress-free interfaces in pure nonlinear inertial flow for bubbles on the (a,c) slope of the balanced migration force exerted on centred bubbles with respect to eccentricity (\ref{Marginal_stability} stability transition) and (b,d) on the equilibrium position of neutral bubbles. \ref{NotExplored} Not explored and validity criteria $Re<Re_*(d)$ with \ref{err5} 5\% error, \ref{err2} 2\% error and \ref{err1} 1\% error.}
\label{fig_Re2}
\end{figure}

The validity of the pure linear inertial flow, $Re<Re_*$, can be quantified as follows with two different criteria depicted in \figref{fig_Re2}. On the one hand, in \figref{fig_Re2}a,c,  we plot the influence of the Reynolds number and the bubble size on the stability of the centred position given by $\partial_\varepsilon f /Re \big\rvert_{\varepsilon = 0 }$. It can be observed that the range of validity of small $Re$ is slightly larger for stress-free bubbles than with rigid interface. \rojo{ and both limits are around, $Re<Re_* \approx 8$ as far as stable positions are concerned. Note that this limit is not valid in the vicinity of $0.3$ and $0.63$, respectively, but it only concerns unstable positions.} \verde{The variation of the slope is referred to the one of the linear regime and normalized by the value of the latter for bubbles of size $d=0.3$ and $d=0.63$, for rigid and stress-free bubbles, respectively. These values correspond to local maxima of $\partial_\varepsilon f /Re \big\rvert_{\varepsilon = 0 }$ versus $d$.} On the other hand, the equilibrium position of neutral bubbles is modified. In \figref{fig_Re2}b,d, we plot the influence of the $Re$ number on the equilibrium position of a neutral bubble of certain size, i.e. in absence of body force. \rojo{, and we observe that this position varies for $Re>8$ for small bubbles and for $Re>64$ for large bubbles, therefore, the criteria $Re_* \approx 8$ is also valid in both cases.} The nonlinear effect of the inertia turns out to be destabilising, in general, either concerning the stability of the equilibrium whose slope, $\partial_\varepsilon f$, increases, or concerning the neutral bubble position which is shifted away from the centre.   


\subsection{Capillary migration. }\label{Capilar}

\begin{figure}
\centering
\input{./figures/pressurefieldCa}
\caption{Streamlines of the pure nonlinear capillary flow, $\vv$, and colormap for the corresponding pressure field, $\hat{p} -  x \partial_x p_{\!\, {\rm P}}$, at the symmetry plane $z=0$ with bubble size $d=0.4$ centred at $\varepsilon = 0.7 \varepsilon_*$ and capillary numbers (a) $Ca=1/64$, (b) $Ca=1/16$ and (c) $Ca=1/4$.}
\label{pressurefieldCa}
\end{figure}
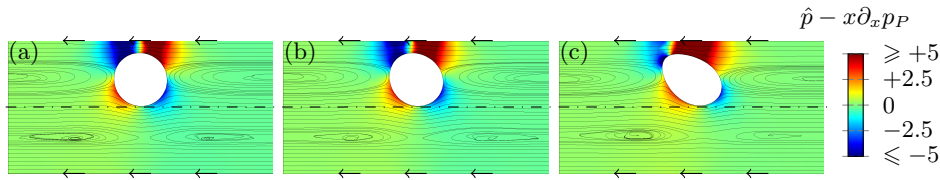

Migration forces may also arise if the bubbles are deformable. In this case, the reversibility of the creeping flow is broken due to the anti-symmetric deformation of the bubble. Indeed, the bubble deforms differently in its front and rear side, due to pressure gradients of the creeping flow along the bubble surface.  In \figref{pressurefieldCa}, the deformation of the bubble due to the viscous and pressure forces exerted on its surface in the pure nonlinear capillary regime is depicted for given values of the eccentricity and bubble size whereas the $Ca$ is varied. In particular, it can be observed that the overpressure (underpressure) for small $Ca$ numbers deforms the bubble, decreasing (increasing) locally the curvature of bubble surface curvature and the bubble deforms as shown in \figref{pressurefieldCa}. The capillary migration force exerted on the bubble surface always points towards the centre of the channel as inferred from the dominant effect of the overpressure in \figref{pressurefieldCa}b-c.

The analogous role of the $Ca$ numbers in the pure capillary regime, as compared to the $Re$ number in the pure inertial regime, motivates an expansion of the system of equations  \eqref{eq_NS}, \eqref{trans}, \eqref{wallbc}, \eqref{pseudo}, \eqref{flowJ} and \eqref{deformable} as $\psi = \sum_{j=0}^{\infty} Ca^j \psi_j $, where $\psi$ represents any of the dependent variables $\hat{p}$, $\vv$, $\vf$, \rojo{$p_G- 4 /(d \,Ca)$, $\dn /Ca$, }$V$ or $\Delta p$\rojo{ where the normal displacement of the bubble surface $\dn$ is additionally considered}. In addition, the fact that for $Ca = 0$ the dimensional pressure of the gas scales as $\gamma/d$, suggests an expansion of the pressure of the gas $p_G$ starting from minus first-order term, i.e. $p_G = \sum_{j=-1}^{\infty} Ca^j p_{Gj} $. 

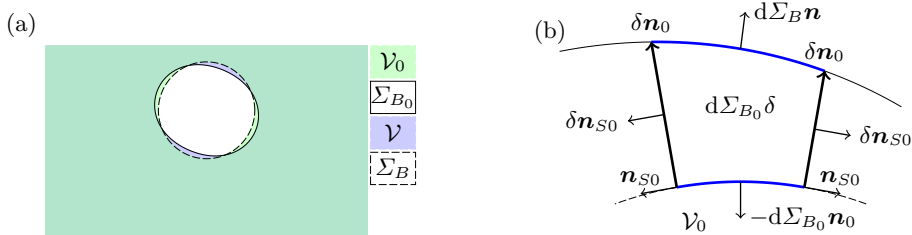
\begin{figure}
\vspace{-1.8cm}
\begin{minipage}[b][5cm][s]{.45\textwidth}
\centering
\vfill
\input{./figures/Volume2.tex}
\end{minipage}\qquad
\begin{minipage}[b][5cm][s]{.45\textwidth}
\centering
\vfill
\input{./figures/SketchSurfPert.tex}
\end{minipage}
\caption{(a) Sketch of the unperturbed volume $\mathcal{V}_0$ bounded by the unperturbed $\Sigma_0$ and their perturbed counterparts, $\mathcal{V}$ and $\Sigma$. (b) Infinitesimal portion of the perturbation of the volume $\mathcal{V}$ and its bounding boundary.}
\label{SketchSurfPert}
\end{figure}

The linearisation of the surface of the bubble $\Sigma_{B}$ writes as the displacement of the unperturbed surface $\Sigma_{B_0}$ in the normal direction an infinitesimal amount $\dn$ , i.e. $\vx = \vx_0 + \vn \dn$ where $\vx_0 \in \Sigma_{B_0}$ and $\vx \in \Sigma_{B}$. The fact that $\dn=0$ for $Ca=0$ suggests that its expansion starts from the first order, $\dn=\sum_{j=1}^{\infty} Ca^j \dn_{j}$. In \figref{SketchSurfPert}a we depict the unperturbed domain $\mathcal{V}_0$ with its bubble surface $\Sigma_{B_0}$ and their perturbed counterparts, $\mathcal{V}$ and $\Sigma_B$, respectively. In \figref{SketchSurfPert}b, we schematise an infinitesimal portion of the perturbation volume, denoted by $\Sigma_{B_0} \dn$, which is generated by the displacement $\dn \vn$ of an infinitesimal surface laying on the unperturbed one, $\dd \Sigma_{B_0}$. This volume is a truncated cone bounded by the infinitesimal surfaces lying on the unperturbed and perturbed surfaces and the generatrix with outer normal vector $\delta \vn_{S0}$. The perturbed volume $\mathcal V$ is the junction of the unperturbed and the perturbation ones, i.e. $\mathcal{V} \equiv \mathcal{V}_0 \cup \Sigma_{B_0} \dn$.

Then, the linearisation of the Navier-Stokes equations \eqref{eq_NS}, neglecting inertia $Re=0$, writes
\begin{subequations}
\label{eq_Ca_exp}
\begin{alignat}{8}
\nabla \cdot \vv_0 &= 0 \,, &\qquad &   & \boldsymbol{0}  = \nabla \cdot \hstress_{0} &\qquad&&  \mbox{at $\mathcal{V}_0$} \,,   \label{eq_Ca_exp_v} \\
\nabla \cdot \vv_1 &= 0 \,, &\qquad &   & \boldsymbol{0}  = \nabla \cdot \hstress_{1}  &\qquad&&  \mbox{at $\mathcal{V}_0$} \,,  \label{eq_Ca_exp_tau}
\end{alignat}
\end{subequations}
at the unperturbed volume $\mathcal{V}_0$ and 
\begin{subequations}
\label{Gex_NS}
\begin{alignat}{8}
 \int_{\dd \Sigma_B} \vn \cdot \vv          \,\dd\Sigma  &=&      \int_{\dd\Sigma_{B_0}} && (\vn - \hgradS \dn )  \cdot \vv         &&                \,\dd\Sigma & \,, \label{Gex_NS_v} \\
 \int_{\dd \Sigma_B} \vn \cdot \hstress  \,\dd\Sigma  &=&      \int_{\dd\Sigma_{B_0}} && (\vn - \hgradS \dn )  \cdot \hstress &&   \,\dd\Sigma  &\,, \label{Gex_NS_tau}
 \end{alignat}
 \end{subequations}
at the perturbation volume $\Sigma_{B_0} \dn$, where $\hgradS \cdot \delta \vv$ and $\hgradS \cdot \delta \hstress$ represent the terms due to the fluxes though the generatrix. For convenience, $\grad [p_G- \vf \cdot (\vx-\veps)] +\vf =0$ at the perturbation volume $\Sigma_{B_0} \dn$ writes
\begin{align}
\label{Gex_NSbis}
 \int_{\dd \Sigma_B} \vn [p_G- \vf \cdot (\vx-\veps)]  \,\dd\Sigma  =      \int_{\dd\Sigma_{B_0}} \{ (\vn - \hgradS \dn )  [p_G- \vf \cdot (\vx-\veps)]  - \dn \vf\}  \,\dd\Sigma \,. 
\end{align}

\Eqsref{trans}, \eqref{wallbc}, \eqref{pseudo} and \eqref{flowJ} are linearised as in the inertial migration case, \eqref{trans01}, \eqref{wallbc01}, \eqref{pseudo01} and \eqref{flowJ01}, because the rest of boundaries do not deform. However, the linearisation of Young-Laplace equation \eqref{Defa} are applied on the perturbed boundary $\Sigma_{B}$, and can be written on the unperturbed boundary with the help of \eqref{Gex_NS} and \eqref{Gex_NSbis} as well as with \eqref{surfopdeflin} for the surface tension term, whose details are given in \appref{AppD}, as
\begin{align}
\label{def_Ca-1}
\vn \cdot \llbracket \Tau_{-1} \rrbracket  =  \hgradS& 1  \qquad \mbox{at $\Sigma_{B_0}$} \,,  
\end{align}
for minus first-order, and
\begin{subequations}
\label{def_Ca01}
\begin{alignat}{11}
&\vn \cdot  \llbracket \Tau_0 \rrbracket   -\hgradS \cdot \big( \dn_1 \llbracket \Tau_{-1} \rrbracket & \big)   & &=  &\hgradS& \cdot \{ \vartheta_1 \id_S - [\gradS (\dn_1 \vn)]^T \} &&  \quad \mbox{at $\Sigma_{B_0}$} , \label{def_Ca01a}  \\ 
&\vn \cdot  \llbracket \Tau_1 \rrbracket   -\hgradS \cdot \big( \dn_2 \llbracket \Tau_{-1} \rrbracket & + \dn_1 \llbracket \Tau_0 \rrbracket \big) & - \dn_1 \vf_0 &= &\hgradS& \cdot \{ \vartheta_2 \id_S - [ \gradS (\dn_2 \vn)]^T \} &&  \quad \mbox{at $\Sigma_{B_0}$} ,  
\end{alignat}
\end{subequations}
for zeroth and first order, where $\hgradS 1$ is the surface mean curvature vector and $\vartheta_j = \divS (\delta_j \vn)$ is the  surface dilatation at $j$th order. Note that $ \llbracket \Tau_{-1} \rrbracket = p_{G-1} \id$. Similarly, the linearisation of the impermeability condition on the bubble surface \eqref{Defb} writes, using \eqref{Gex_NS},
\begin{subequations}
\label{deformable_Ca01_imp}
\begin{alignat}{5}
 \vn \cdot \vv_0 &                                                    &= 0  &&  \qquad \mbox{at $\Sigma_{B_0}$}\,, \\
 \vn \cdot \vv_1 &   -\hgradS \cdot ( \dn_1  \vv_0 ) &= 0  &&  \qquad \mbox{at $\Sigma_{B_0}$}\,.
\end{alignat}
\end{subequations}

The linearisation of the volume and the eccentricity definitions \eqref{pG} writes, with the help of the Reynolds transport theorem, 
\begin{subequations}
\label{pG-1}
\begin{alignat}{5}
  \int_{\mathcal{V}_{B_0}}&  \,& \dd \mathcal{V} &=& \mathcal{V}_{B}  \,, \\
  \int_{\mathcal{V}_{B_0}} &(\vx - \veps) \,& \dd \mathcal{V} &=& \boldsymbol{0}  \,, 
\end{alignat}
\end{subequations} 
where $\mathcal{V}_{B_0}$ is the domain occupied by the unperturbed bubble, and its perturbation
\begin{subequations}
\label{pG01}
\begin{alignat}{8}
  \int_{\Sigma_{B_{0}}} \dn_1 &         & \,\dd \Sigma = 0 \,, &\qquad&&   \int_{\Sigma_{B_{0}}} \dn_2 &                         & \,\dd \Sigma = 0  \,, \\
  \int_{\Sigma_{B_{0}}} \dn_1 & (\vx - \veps) & \,\dd \Sigma = \boldsymbol{0} \,, &\qquad&&   \int_{\Sigma_{B_{0}}} \dn_2 & (\vx - \veps) & \,\dd \Sigma = \boldsymbol{0}  \,.
\end{alignat}
\end{subequations} 

Since $p_{G-1}$ is a global variable, the solution of \eqref{def_Ca-1} and \eqref{pG-1}, is a sphere of volume $\mathcal{V}_B$ and at a position $\veps$,
\begin{align}
\label{deformable_Ca_1_sol}
p_{G-1}  =  \frac{4}{d}  \,, \qquad \vx _0 = \veps +\frac{d}{2} \vn  \,,
\end{align}
whose surface dilatation is $\vartheta_j = - 4 \dn_j/d$.

Solutions of the system \eqref{trans01}, \eqref{wallbc01}, \eqref{pseudo01}, \eqref{flowJ01}, \eqref{eq_Ca_exp}, \eqref{def_Ca01}, \eqref{deformable_Ca01_imp} and \eqref{pG01} for the balanced body force, bubble velocity and pressure correction factor are of the form, truncated at $\mathcal{O}(Ca^2)$,
\begin{align}
\label{exp2Ca}
f_{\gamma} \equiv f/Ca \approx   f_{1}(\varepsilon,d)    \,,\qquad 
V \approx V_{0}(\varepsilon,d)   \,, \qquad
\beta \approx   \beta_{0}(\varepsilon,d)    \,,
\end{align}
where  the removed terms have been found numerically to be vanishing, i.e. $f_0 = V_1 = \beta_1 = 0 $ as it can be inferred from the symmetries and reversibilities of the flow. Observe that the capillary migration force comes from the first-order solution whereas the velocity, pressure drop and the rotation come up from the creeping flow, analogous to the inertial migration case. Furthermore, the sum of the capillary terms of \eqref{def_Ca01a} only has normal component, i.e.
\begin{align}
\vn \x \hdivS \{ \delta_1 \llbracket \Tau_{-1} \rrbracket + \vartheta_1 \idS - [\gradS (\delta_1 \vn)]^T  \} = \boldsymbol{0} \,.
\end{align}
Thus, the zeroth-order of the capillary regime is exactly the same as the zeroth-order of the inertial regime with stress-free boundary conditions, since one can identify the capillary terms of the former as the reduced counterpart of the impermeable surface pressure $\hat{\lambda}$
\begin{align}
\hat{\lambda} = - \vn \cdot \hdivS \{ \delta_1 \llbracket \Tau_{-1} \rrbracket + \vartheta_1 \idS - [\gradS (\delta_1 \vn)]^T  \}  \,.
\end{align}
  
\begin{figure}
\centering
\input{./figures/PureNonLinearCa_d04_f.tex}
\input{./figures/PureNonLinearCa_d04_V.tex}
\input{./figures/PureNonLinearCa_d04_beta.tex}
\caption{Effect of the deformability of bubbles with size $d=0.4$ and the eccentricity in the pure nonlinear capillary regime on the (a) balanced body force, (b) bubble velocity and (c) pressure correction factor.}
\label{fig_Ca}
\end{figure}
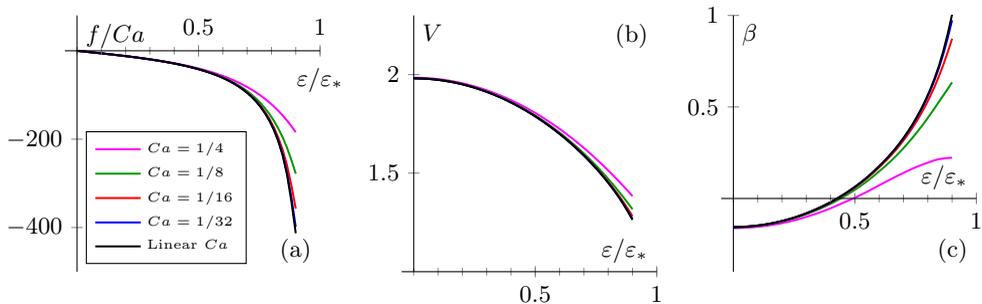

In \figref{fig_Ca}, we depict solutions of the pure nonlinear capillary flow for several $Ca$ numbers and its linear limit to which nonlinear solution converges when the $Ca$ is sufficiently small $Ca<Ca_*$, see \figref{fig:casos}. The dependence of the balanced body force, the velocity and the pressure drop on the eccentricity is shown for a given bubble size. Observe that the capillary migration is always stabilising since $\partial_\varepsilon f < 0$ as shown in \figref{fig_Ca}a. We also observe that large deformations reduce bubble migration, increase the bubble velocity and reduces the associated pressure drop. In \figref{fig_Ca}b and c, the limit of pure linear capillary regime reproduces the bubble velocity and the pressure drop for stress-free bubbles in the zeroth-order creeping flow, $V_0$ and $\Delta p_0$, the latter being negative for some conditions. Remarkably in this case, the bubble is stable for these conditions and, therefore, the presence of deformable bubbles may reduce the pressure drop.

\begin{figure}
\centering
\input{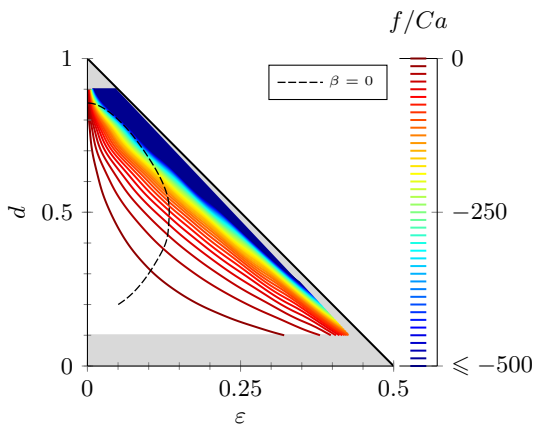}
\caption{Effect of the eccentricity $\varepsilon$ and the size $d$ of bubbles in the pure linear capillary regime on the balanced body force. \ref{NotExplored} Not explored and \ref{epsast} $\varepsilon = \varepsilon_*$.}
\label{fig_Ca_d_c}
\end{figure}

The dependence of the balanced body force on the bubble size $d$ and eccentricity $\varepsilon$ in the pure linear capillary regime is depicted as in \figref{fig_Ca_d_c}. Bubbles experience a stronger repulsion from the wall as they become closer to it either because of their size or because of their position. Velocity and pressure correction factor are the same as in the pure linear inertial regime with stress-free boundary conditions  \figref{CircularLibre}b-c. In effect, positions that reduces the pressure drop with respect to that of Poiseuille flow are stable in pure linear capillary flow.

For the sake of completeness, polynomial fitting of functions \eqref{exp2Ca} are given in \appref{AppPF}, \tabref{tab_Ca}.

\begin{figure}
\centering
\input{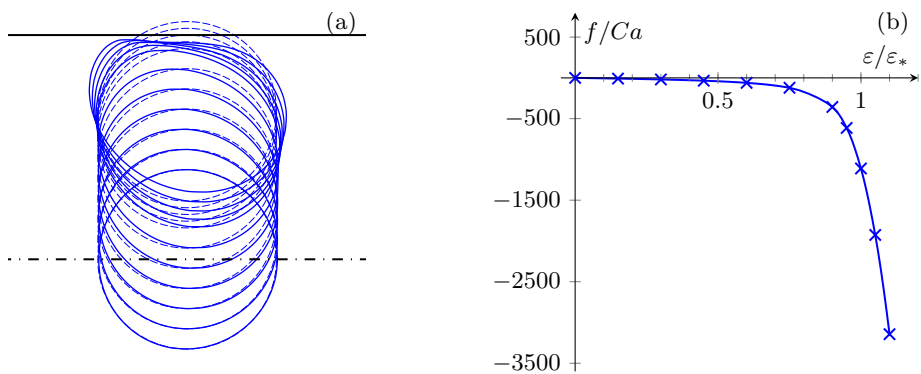} \hspace{.1\textwidth}
\input{./figures/Forma_p2_Ca_Ca_1_16_d04.tex}
\caption{Effect of the eccentricity (a) on the shape of bubbles at the symmetry plane $z=0$  and (b) on the balanced body force in the pure nonlinear capillary regime with $Ca=1/16$ for bubbles with size $d=0.4$. \ref{Undeformed} Undeformed shape, \ref{Deformed} deformed shape and \ref{discretos} discrete values of $\varepsilon/\varepsilon_*$ plotted in (a).}
\label{fig_Ca3}
\end{figure}

The validity of the pure linear capillary regime is lost either for not sufficiently small capillary numbers, see \figref{fig_Ca}a, leading to large deformation all around the bubble, or for smaller capillary numbers if the gap between the bubble and the wall is too small, leading to an increase of the local stresses, and, therefore, larger deformation in this region whereas the rest of the bubble remains mainly spherical. The latter effect is shown in \figref{fig_Ca3}, where the behaviour of a bubble with $Ca=1/16$ is depicted. In \figref{fig_Ca3}a, the bubble contour at the symmetry plane is plotted at different positions. The bubble remains almost spherical until the wall effect is noticeable for positions at which a liquid layer prevent the bubble to touch the wall by deforming it and drastically repealing it from the wall, see \figref{fig_Ca3}b. Observe, that since the bubble deforms, the centre of the bubble can get closer to the wall and hence its feasible range is $-\frac12 < \varepsilon < \frac12$ instead of the corresponding to undeformable bubbles $-\varepsilon_* < \varepsilon < \varepsilon_*$. However, in these additional regions, $-\frac12 < \varepsilon<-\varepsilon_{*}$ and $\varepsilon_{*}<\varepsilon<\frac12$, nonlinear effects are always present and the linear regime is not only nonsense but numerically impossible. 

\begin{figure}
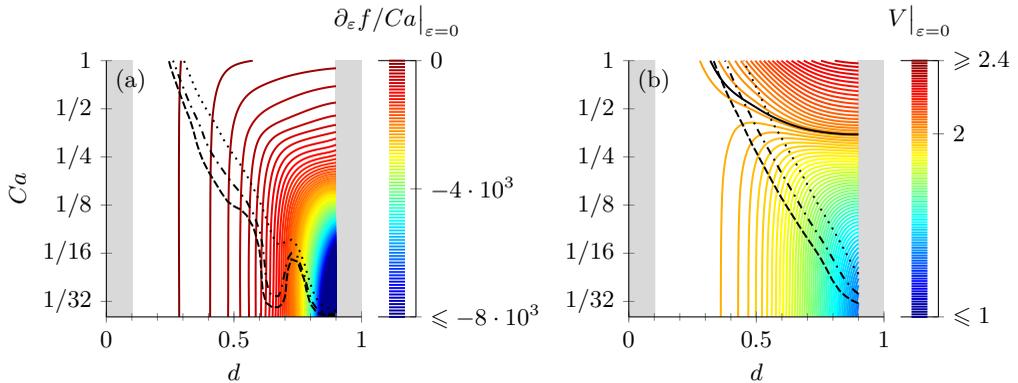

\centering
\input{./figures/CircularStiffness_Ca_d.tex}
\input{./figures/CircularV_Ca_d.tex}
\caption{Influence of the $Ca$ number and the bubble size in pure nonlinear capillary regime on the (a) slope of the balanced migration force exerted on centred bubbles ($\varepsilon = 0 $) with respect to eccentricity and (b) on the velocity of centred bubbles, \ref{Vmax} $V=2$. \ref{NotExplored} Not explored and validity criteria $Re<Re_*(d)$ with \ref{err5} 5\% error, \ref{err2} 2\% error and \ref{err1} 1\% error.}
\label{fig_Ca_Validity}
\end{figure}

Now, it remains to explore the validity of the pure linear capillary regime. On the one hand, we avoid nonlinear effect due to the proximity of the wall and study centred bubbles. We depict the stability measurement of centred positions, $\partial_\varepsilon f /Ca \big\rvert_{\varepsilon = 0 } $ and observe in \figref{fig_Ca_Validity}a that the validity range strongly depends on the bubble size. In particular, the repulsion of the wall increases the stability and produces larger deformation then decreasing the validity of the pure linear capillary flow to $\log_2 Ca_* \approx 2-9 d$ with errors smaller than 1\% for the velocity and 2\% in the slope  of the balance migration force of centred bubbles. We also observe that bubbles travel faster when they are strongly deformed, and may eventually travels faster than the maximum velocity of the liquid, i.e. twice the mean flow. 

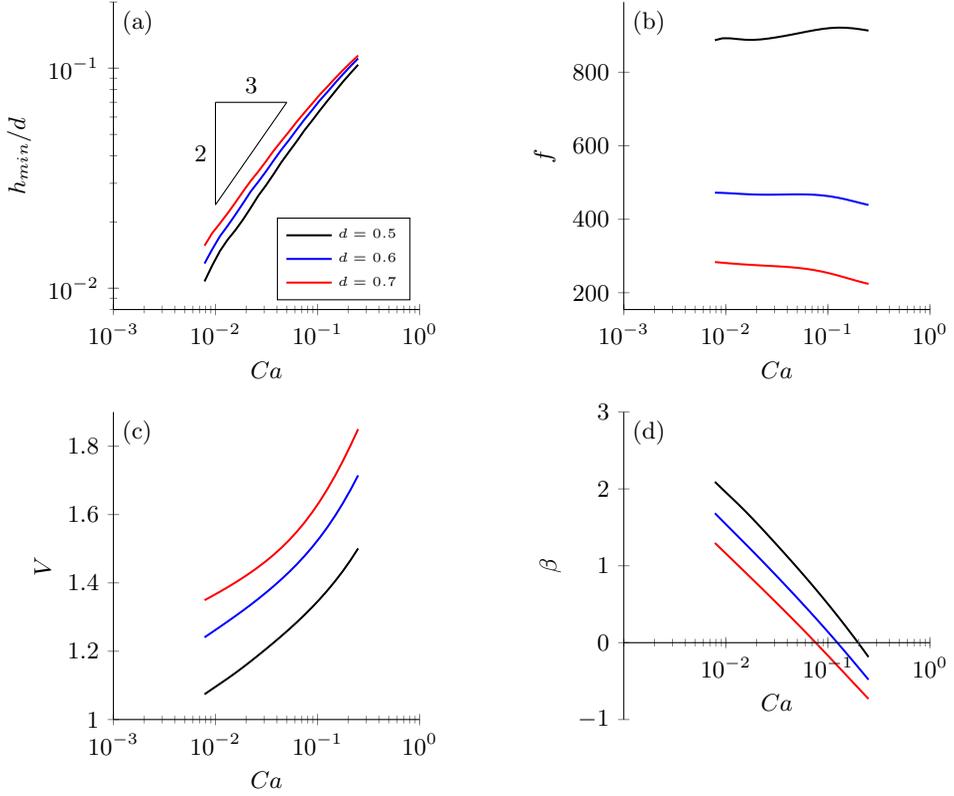
\begin{figure}
\center{
\input{./figures/LandauLevich1.tex}
\input{./figures/LandauLevich2.tex}
\input{./figures/LandauLevich3.tex}
\input{./figures/LandauLevich4a.tex}
}
\caption{Influence of the $Ca$ number in the pure nonlinear capillary regime for bubbles centred at $\varepsilon=\varepsilon_*$, where nonlinearities due to the proximity of the wall are dominant, on (a) the minimum gap between the wall and the bubble, (b) the balanced body force, (c) the bubble velocity and (d) the pressure correction factor.}
\label{fig_LandauLevich}
\end{figure}

On the other hand, we study the nonlinearities due to the proximity of the wall and the bubble is locally deformed by the presence of the wall (\figref{fig_Ca3}a). In \figref{fig_LandauLevich}, we depict the behaviour of the bubble at $\varepsilon = \varepsilon_{*}$ where the nonlinearities due to the proximity of the wall are dominant. The gap between the bubble and the wall follows the Landau-Levich $2/3$ power law at least for sufficiently small $Ca$ as shown in \figref{fig_LandauLevich}a. The balanced body force is not proportional to the $Ca$ number contrarily to the pure linear capillary regime (\figref{fig_LandauLevich}b), deformed bubbles travel faster (\figref{fig_LandauLevich}c) and the pressure drop decreases and can even take negative values (\figref{fig_LandauLevich}d).

\subsection{Inertial versus capillary migration} \label{Ohnesorge}

\begin{figure}
\centering
\input{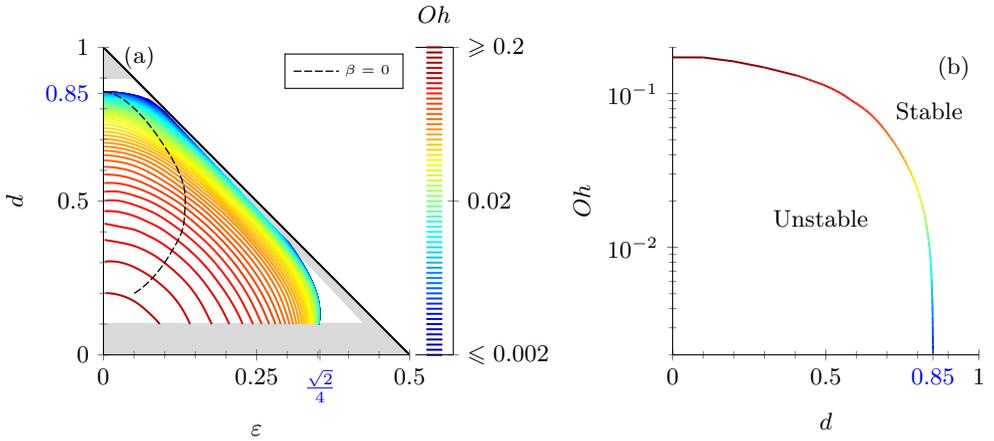}
\input{./figures/Centred_Stability_Oh.tex}
\caption{Influence of the $Oh$ number and the bubble size $d$  in the linear inertial-capillary regime. (a) Equilibrium position of neutral bubbles, $f=0$, and (b) stability of centred positions. \ref{NotExplored} Not explored and \ref{epsast} $\varepsilon = \varepsilon_*$.}
\label{fig_Oh_Neutral}
\end{figure}

We study the equilibrium position of bubbles when inertial and capillary migrations are taken into account as function of the $Re$ and $Ca$ numbers and the bubble size $d$ in the linear regime. If we focus on neutral bubbles, the balanced body force $ f \approx Re \, f_\rho + Ca\, f_\gamma$ must vanish at the equilibrium position which then depends on the Ohnesorge number, $Oh^2 = Ca/Re$, instead of the $Re$ and $Ca$ numbers separately. In \figref{fig_Oh_Neutral}a we plot the uncentred and stable equilibrium position of neutral bubbles for given $Oh$ number. We observe that as $Oh \rightarrow 0$, the equilibrium positions tend to that of the pure linear inertial regime. Contrarily, if the $Oh$ number becomes larger, the uncentred equilibrium position tends to get closer to the centre of the channel and eventually reaches the centre for sufficiently large values, as in the pure linear capillary regime. Observe that also centred positions exist, similarly to the pure inertial case, see \figref{fig_eq_Re01}a, which corresponds to the centred branch of the Pitchfork bifurcation, as illustrated in \figref{CircularLibre}a for the pure linear inertial regime. In \figref{fig_Oh_Neutral}b, we observe that there exists both threshold on the $Oh$ number and bubble size above which the centred position is always stable. This threshold also corresponds to $\partial_\varepsilon (Re \, f_\rho + Ca\, f_\gamma) = 0$.

\begin{figure}
\centering
\input{./figures/StabReCad.tex}
\caption{Stability diagram for centred bubbles in \ref{OhLinLimit} the linear and nonlinear \ref{OhNLinLimit} inertial-capillary regime. 
}
\label{Centred_Stability_Oh}
\end{figure}
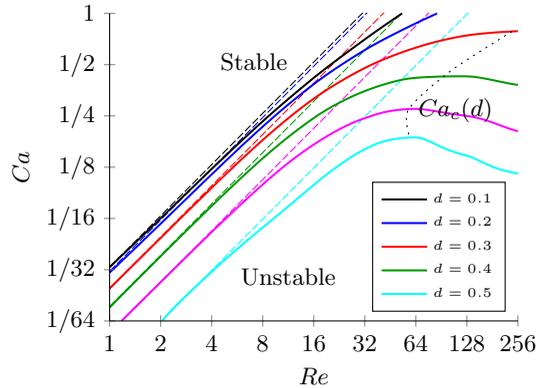

Next, we study the stability of the centred position as a function of the $Oh$ number and the bubble size $d$ in the nonlinear inertial-capillary regime. In \figref{Centred_Stability_Oh}, we depict the stability map for a few bubble sizes on the $Re$-$Ca$ plane in both linear and nonlinear inertial-capillary regime, the latter considering interactions between inertial and capillary migration as well as the effects out of the linear inertial-capillary regime. The linear inertial-capillary regime is recovered for $Re<Re_* \approx 8$, whereas the nonlinear effects are strongly stabilising above the threshold $Re>Re_* \approx 8$ due to the joint effect with capillary deformation. In particular, there exists for each bubble size, a value of the $Ca$ number depending on the bubble size $Ca_c(d)$ above which centred bubbles are always stable, regardless the value of the $Re$ number, i.e. the destabilising effect of inertia is overcome by the stabilising effect of the additional deformation of the bubble interface due to the effect of inertia. On the contrary, for values of the $Ca$ number smaller than $Ca_c$, a range of $Re$ numbers is found for which bubbles are unstable.

\section{Conclusions and discussion}\label{Conclusions}

This paper reports the dynamics of a train of bubbles regarding the flow structure and the migration forces of both an inertial and capillary deformation origin. We write the general equations governing the dynamics of a train of bubbles. We assume the change of bubble size along the channel (due to gas dissolution for instance) to be slow compared to the characteristic hydrodynamic time and that no turbulence develops. Under these assumptions, the flow can be considered quasi-stationary and time dependence can be dropped out. We quantitatively study the influence of the bubble size and of body force on the linear inertial and capillary regimes. To do this, regular asymptotic expansions of the governing equations are carried out, including the linearisation of the boundary conditions of the deformable bubble surface around the undeformed shape, for which we provide a new asymptotical approach. Then, we compare to the nonlinear regimes governed by the full system of equations, using ALE method for undeformable bubbles, in order to obtain the validity of the linear regimes for the $Re$ and $Ca$ numbers. 

In the case of pure inertial migration, we report the well-known multiple equilibrium behaviour and we illustrate the underlying hydrodynamics mechanism regarding at the first-order pressure field correction of the creeping flow. Then, we describe the influence of the bubble size and the eccentricity of the equilibrium positions on the external uniform body force that balances the migration force, the bubble velocity, the additional pressure drop (pressure correction factor) due to the presence of the bubble and the rotational velocity (only applicable in the case of rigid interface bubbles). Since the external body force is commonly given, first, the position of the bubble has to be obtained to ensure the transverse equilibrium. Once the equilibrium position is obtained, the velocity, additional pressure drop and rotation of the bubble (if applicable) can be obtained. In agreement with previous results, small bubbles follows the Poiseuille flow and neutral bubbles (in absence of body force) migrate to the position at which the velocity of the Poiseuille flow and the mean velocity coincide, i.e. at $\varepsilon = \sqrt{2}/4$ in the case of stress-free and rigid bubbles. We also observe that the surface average of the Poiseuille velocity provides a reasonably good approximation of the bubble velocity for rigid surfaces. We observe that stress-free neutral bubbles with a size smaller than $d \lesssim 0.85$ migrates out of the centre and within the range $0.35 \lesssim d \lesssim 0.83$ migrates to the wall, i.e. to $\varepsilon>0.95 \varepsilon_*$ where $\varepsilon_* = \frac12 (1-d)$. Bubbles with rigid interface and with size smaller than $d \lesssim 0.73$ migrate to intermediate positions. We also observe that the presence of bubbles may reduce the pressure drop along the channel. However, this region is unstable due to inertial migration forces but it may be stabilised by capillary migration forces. A plug effect is observed for rigid particles that dramatically increases the pressure drop with the bubble size. Small rigid particles tend to rotate with the same rotational velocity than the Poiseuille flow as the bubble size decreases. Finally, we study the validity of the linear inertial regime and obtain a criteria based on the stability of the centred position and the equilibrium position of neutral bubbles. In both rigid and stress-free cases nonlinearities break the validity of the linear regime for $Re$ numbers larger than the approximate threshold $Re_*  \approx 8$ as well as shift the equilibrium positions toward the wall. 

Additionally, we consider the behaviour of deformable bubbles. In this case, underpressure and overpressure of the creeping flow on the bubble surface make it deform leading to stabilising forces pointing towards the centre of the channel. We obtain the regular asymptotic expansion of the governing equations  for small values of the $Ca$ number. Since the boundary of the domain at the bubble surface deforms, it is necessary to linearise around the undeformed shape the boundary conditions applied at this boundary. We observe that larger bubbles balance larger outer forces at the same equilibrium position and this effect dramatically increases as the bubble approaches the wall due to either larger bubble size or eccentricity. Polynomial fitting is obtained for the limit of small bubbles $d \rightarrow 0$. The linear capillary limit is valid for centred bubbles if the $Ca$ number is smaller than the approximate threshold $Ca< Ca_*  \approx 2^{2-9d}$ which decreases with the eccentricity until it is no longer valid for undeformed bubbles touching the wall. In this case, a thin lubrication layer forms between the wall and the bubble due to the deformation of the latter. The film thickness follows the $2/3$ Landau-Levich power law  for $Ca <0.1$, the migration force very slightly depends on the surface tension (no longer proportional to the $Ca$ number) and the bubble velocity increases with its deformation while the pressure drop decreases and can even reach negative values.

Finally, we obtain the transverse equilibrium position of neutral bubbles and the stability of the centred position (for neutral bubbles) taking into account both inertia and deformability. We obtain a stability diagram depending on the $Oh= \sqrt{Ca/Re}$ number and the bubble size in the pure linear inertial-capillary regime. We observe that centred solutions are stable for either $Oh$ numbers larger than $Oh>0.2$ or for bubbles larger than $0.85 \lesssim d$, with a smooth transition between both limits. We also observe that the nonlinear effects are no longer negligible for $Re$ numbers larger than $Re_* \approx 8$ above which nonlinear inertial-capillary effects become stabilising even for given $Ca$ number, contrarily to what it is observed in the pure nonlinear inertial regime.

The dynamics of drops in microchannels is of great interest in the scientific community and this work implicitly explores the limiting cases of very high/low viscosity drops. In particular, on the one hand low viscosity drops exhibit the same behaviour of capillary deformable bubbles. On the other hand, high viscosity drops and particles exhibit the same behaviour of the rigid interface and no deformation-induced migration takes place because deformation is prevented to occur. However, the transition of the dynamics for the viscosity of the drop, i.e. for intermediate viscosities, is still to be systematically studied.

We also provide an approach for the linearisation of the boundary conditions around the equilibrium position of a deformable boundary which relies on the external surface differential operator. This approach provides a powerful and simple theoretical background for other applications relying on deformable boundaries in complex geometries, such as global stability analysis of free interfaces in the most general conditions. The validity of the proposed method has been checked by comparison with the full nonlinear equations. 

\section*{Acknowledgements}

We thank Benoit Haut, David Mikaelian and Miguel P\'erez-Saborid for useful discussions. We thank the Brussels region for the financial support of this project through the WBGreen-MicroEco project. B.S. thanks the F.R.S.-FNRS for financial support as well as the IAP-7/38 MicroMAST project for supporting this research. This work was also performed under the umbrella of COST Action MP1106. 

\appendix

\section{Polynomial fittings}\label{AppPF}

In this appendix, we provide the coefficients of the polynomial fitting, with less than 1\% error within the range of parameters for which simulations have been carried out, of \figsref{CircularRigido}, \ref{CircularLibre} and \ref{fig_Ca} in the form
\begin{align}
\varphi(d,\varepsilon) = \sum_i \sum_j \varphi_{ij}  \left( \frac{\varepsilon}{\varepsilon_*} \right) ^i d^j 
\end{align}
in \tabsref{tabCircularRigido}, \ref{tabCircularLibre} and \ref{tab_Ca}.

\begin{table}
\begin{center}
\input{./Tables/matrizLCircularLibre.tex}
\input{./Tables/matrizV_VPCircularLibre.tex}
\input{./Tables/matrizbeta100CircularLibre.tex}
 \caption{Polynomial fitting of global variables $f_1$, $V_0/V_P$ and $100 \beta$ in \eqref{exp1} for the pure linear inertial regime with stress-free boundary conditions, as plotted in \figsref{CircularRigido}}
  \label{tabCircularRigido}
  \end{center}
\end{table}

\begin{table}
\begin{center}
\input{./Tables/matrizLCircularRigido.tex}
\input{./Tables/matrizV_VPCircularRigido.tex}
\input{./Tables/matrizbeta100CircularRigido.tex}
\input{./Tables/matrizOmCircularRigido.tex}
 \caption{Polynomial fitting of global variables $f_1$, $V_0/V_P$, $100 \beta_0$ and $\Omega_0/8 \varepsilon$ in \eqref{exp1} for the pure linear inertial regime with rigid boundary conditions, as plotted in \ref{CircularLibre}.}
  \label{tabCircularLibre}
  \end{center}
\end{table}

\begin{table}
\begin{center}
\input{./Tables/matrizfdatos_Ca0.tex}
\caption{Polynomial fitting of global variables in \eqref{exp2Ca} for the pure linear capillary regime, as plotted in \ref{fig_Ca}.}
  \label{tab_Ca}
  \end{center}
\end{table}


\section{Surface operator}\label{AppS}

In this appendix, we introduce the exterior differential operator, defined by the generalized Stokes theorem, and the nabla operator, defined in terms of directional derivatives. Both are defined applied on a volume and a surface, see \tabref{tab:difop} for the nomenclature. Herein, we derive the relationships relevant in this work and the reciprocity theorem for their FEM discretization.

\begin{table}
\begin{center}
\begin{tabular}{l | cc}
      					& Volume  & Surface  \\ \hline
Directional   			& $\nabla $ & $\gradS$   \\ 
Exterior		     		& $\hgrad$ & $\hgradS$   \\ 
  \end{tabular}
 \caption{Differential operators.}
  \label{tab:difop}
  \end{center}
\end{table}

First, the exterior differential operator $\hgrad$ can be defined by the generalized Stokes theorem as
\begin{align}
\label{D}
\int_{\mathcal V} {\hgrad} \varphi \,\dd \mathcal{V} = \int_{ \Sigma} \vn \varphi \, \dd \Sigma \,,
\end{align}
where $\mathcal{V}$ is the volume bounded by the surface $\Sigma$ with outer normal vector $\vn$ and $\varphi$ can be any scalar, vectorial or tensorial variable. However, this expression is not useful as such for numerical computations. Its components in a certain basis is more convenient for these purposes. Let $\{\ve_i \}_{i=1,2,3}$ be the cartesian basis. The nabla operator $\nabla$ is then defined as 
\begin{align}
\label{nablaS}
\nabla \varphi = \ve_i \partial_{x_i} \varphi \,,
\end{align}
where $x_i$ are the cartesian coordinates and $ \partial_{x_i} \varphi = \partial_\epsilon \varphi(\vx+\epsilon \ve_i)$ is the directional derivative. Both volume differential operators are equivalent,
\begin{align}
\nabla \varphi \equiv \hgrad \varphi \,. 
\end{align}

Secondly, the exterior surface differential operator, $\hgradS$, defined on a surface sketched in \figref{Sketch5}, can be defined by
\begin{align}
\label{defhgradS}
\int_\Sigma \hgradS \varphi (\vx) \, \dd \Sigma =  \int_{ \Gamma} \vnS \varphi \, \dd \Gamma \,,
\end{align}
where $\vnS$ is the outer normal to the contour $\Gamma$ of the differential surface $\Sigma$ and contained in the subdomain, i.e. $\vn \cdot \vnS = 0 $. The surface nabla operator $\nabla_S$ is defined as
\begin{align}
\label{asdf}
\gradS \varphi (\vx) = \ve_i  \partial_{S x_i} \varphi = \id_S \cdot \ve_i \partial_{x_i} \varphi  \,,
\end{align}
 where $\partial_{S x_i} \varphi = \partial_\epsilon \varphi(\vx+\epsilon \idS \cdot \ve_i)$ is the directional derivative within the surface and $\idS$ is the surface identity $\idS = \id - \vn \vn$.  
 
Surface differential operators are related to the volume differential operators. In effect, the LHS of \eqref{defhgradS} multiplied by an infinitesimal distance $h$, writes
\begin{align}
\label{hnabSL}
h \int_\Sigma \hgradS \varphi \, \dd \Sigma = \int_{\mathcal{V}_\Sigma} \hgradS  \varphi \, \dd \mathcal{V} 
\end{align} 
where in this case $\mathcal{V}_{\Sigma}$ is the truncated cone generated by a displaced of $\Sigma$ in the direction $\vn$ and amount $h$. Similarly, the RHS of \eqref{defhgradS} multiplied by an infinitesimal distance $h$, using $\vnS = \vnS \cdot  \idS $, $\vn \cdot \idS = \boldsymbol{0}$ and \eqref{D}, and conveniently adding vanishing terms, yields
\begin{align}
\label{hnabSR}
h \int_\Gamma \vnS \cdot \idS \varphi \, \dd \Gamma \pm \int_\Sigma \vn \cdot \idS \varphi(\vx \pm \frac12 h \vn) \,\dd \Sigma 
= \int_{\mathcal{V}_\Sigma} \nabla \cdot (\idS   \varphi) \, \dd \mathcal{V} \,,
\end{align} 
where the first term of the LHS represents the flux through the generatrix and the second integral represents the flux through the top and bottom base of the cone. Note that, \eqref{hnabSR} is independent of the value of $\varphi$ outside the surface $\Sigma$, or equivalently its normal gradient $ \vn \cdot \nabla \varphi$, which might even be not defined. 

From \eqref{hnabSL}-\eqref{hnabSR}, one obtains
 \begin{align}
\label{rel}
 \hgradS \varphi = \nabla \cdot ( \idS \varphi )     \,,
\end{align}
and provided from \eqref{asdf} that
\begin{align}
\gradS \varphi =  \idS \cdot \nabla  \varphi \,,
\end{align} 
the relation between both surface operators writes
\begin{align}
\label{prop1}
 \hgradS \varphi  = \gradS \varphi + (\nabla \cdot \idS)  \varphi  \,,
\end{align}
where $\nabla \cdot \idS = \hgradS 1$ is the surface mean curvature vector. An alternative to the previous relation that avoids the definition of the surface curvature consists in the reciprocal theorem. In effect, from \eqref{rel} and \eqref{prop1}, one obtains,
\begin{align}
\label{prop2}
\psi \hgradS \varphi  = \hgradS   (\varphi \psi) -  (\gradS \psi) \varphi   \,,
\end{align}
where $\psi$ is a scalar. Integrating \eqref{prop2} among a surface $\Sigma$ with outer normal $\vnS$ to the contour $\Gamma$, leads to the reciprocal theorem in surfaces
\begin{align}
\label{prop3}
\int_\Sigma \psi \hgradS \varphi \dd \Sigma  = \int_\Gamma \vnS  (\varphi \psi)  \dd \Gamma - \int_\Sigma  (\gradS \psi)  \varphi \dd \Sigma  \,.
\end{align}

Expression \eqref{prop3} is appropriate to convert the $\hgradS$ operator into the $\gradS$ operator, in weak form, much more convenient for numerical simulations using FEM since it is defined in terms of directional derivatives.

 \section{Linearisation of surface tension}\label{AppD}

%
%

For the linearisation of the surface tension, we need to linearise the exterior differential operator $\hgradS$. For this purpose we linearised the RHS of \eqref{surfopdef},
\begin{alignat}{4}
\label{surfopdeflin00}
\int_\Gamma \vnS \varphi \,\dd \Gamma 
= \int_{\Gamma_0}   \vnS \varphi \,\dd \Gamma + \int_{\Gamma_0}  \Delta(  \vnS \varphi \,\dd \Gamma)  \,,
\end{alignat}
where $\Delta(  \vnS \varphi \,\dd \Gamma)$ is the variation of $ \vnS \varphi \,\dd \Gamma$ between $\Gamma$ and $\Gamma_0$ for which Leibniz's rule holds. The variation of $\varphi$ vanishes for variables defined only in the surface, whereas the variation of $\vnS \dd \Gamma$ is obtained as follows. Let $\dd \vx $ be an arbitrary vector on the surface with origin on $\Gamma_0$. Thus, the variation of a surface element, $ \dd \vx \cdot \vnS \dd  \Gamma$, writes \cite[]{pereira2008transport} 
 \begin{align}
\label{vnS_dG0}
\Delta ( \dd \vx \cdot \vnS \dd \Gamma) = \Delta (\dd \vx) \cdot \vnS \dd   \Gamma + \dd \vx \cdot \Delta ( \vnS \dd \Gamma) \,,
\end{align}
and therefore
 \begin{align}
\label{vnS_dG}
\vartheta \dd \vx \cdot \vnS \dd \Gamma = \dd \vx \cdot \gradS (\dn \vn) \cdot \vnS \dd   \Gamma + \dd \vx \cdot \Delta ( \vnS \dd \Gamma) \,,
\end{align}
where the surface dilatation $\vartheta = \gradS \cdot (\dn \vn)$ represents the variation of the surface elements and $\dd \vx \cdot \gradS (\dn \vn)$ represents the variation of the arbitrary vector $\dd \vx$. Thus, using \eqref{surfopdeflin00} and \eqref{vnS_dG} for arbitrary $\dd \vx$, \eqref{surfopdef} writes
\begin{alignat}{4}
\label{surfopdeflin}
\int_\Sigma \hgradS \varphi \,\dd \Sigma 
   = \int_{\Sigma_0}  \hdivS   \big{[} 
		 (1 + \vartheta ) \id 
		-  (\gradS \dn \vn)^T  
    \big{]}  \varphi \,\dd \Sigma \,.
\end{alignat}

After some algebraic manipulation, using the vector triple product and $ \vn \cdot \vnS = 0$, one obtains that the variation of the curve element is due to the dilatation of the contour and due to the rotation of the surface, respectively represented by the terms of the RHS, 
\begin{align}
\vnS \cdot [ \vartheta \id_S -  (\gradS \dn \vn)^T ]  = \vnS \cdot ( \vartheta \id_S - \dn  \gradS \vn ) - \vnS \x [ \rotS (\dn \vn) ] \,.
\end{align}

\section{Reversibilities and symmetries}\label{AppB}

Zeroth- order creeping flow, denoted with subindex $0$, is reversible. In effect, if we change the direction of the mean flow 
\begin{align}
\label{RevJ}
J \leftarrow - J \,,  
\end{align}
the reversed fields, denoted with the subindex $R$, fulfil
\begin{align}
\label{Rev}
p_0(x,y,z)= -p_R(x,y,z) \,, \qquad \vv_0(x,y,z)= - \vv_R(x,y,z) \,,
\end{align}
Furthermore, if the geometry exhibits symmetry with respect to the plane $x=0$, the flow is also anti-symmetric with respect to this plane, namely,
\begin{align}
\label{ASym}
p_0(x,y,z)=-p_0(-x,y,z) \,, \qquad \vv_0(x,y,z)= - \mathcal{S} \cdot \vv_0(-x,y,z) \,,
\end{align}
where $\mathcal{S}= - \boldsymbol{e}_x \boldsymbol{e}_x + \boldsymbol{e}_y \boldsymbol{e}_y + \boldsymbol{e}_z \boldsymbol{e}_z$ is the symmetry tensor with respect to the $x$-direction.

Evaluating expressions \eqref{ASym} at an arbitrary plane $x = h $, dividing by $h$ and evaluating the limit when $h \rightarrow 0$, leads to the symmetry conditions
\begin{align}
\label{Revlim}
p(0,y,z)=0  \,, \,\, \partial_x \vv (0,y,z) \cdot \boldsymbol{e}_x= 0 \,, \,\, \vv (0,y,z) \cdot \boldsymbol{e}_y= 0 \,, \,\, \vv (0,y,z) \cdot \boldsymbol{e}_z= 0
\end{align}

Now, pure linear inertial or capillary corrections of the creeping flow, denoted by the subindex $1$, are anti-reversible. The change of direction of the mean flow \eqref{RevJ} does not affect the linear correction since all non-homogeneous terms in the governing equations are quadratic in the creeping fields, $\hat{p}_0$, $\vv_0$ and $\dn_0$ (if applicable) and all negative signs cancel. The anti-reversibility character is described by
\begin{align}
\label{ARev}
p_1(x,y,z)= p_{AR}(x,y,z) \,, \qquad \vv_1(x,y,z)=  \vv_{AR}(x,y,z) \,
\end{align}
where subindex $AR$ refers to anti-reversible. In symmetric geometries, anti-reversible flows are also symmetric and fulfil
\begin{align}
\label{Sym}
p_1(x,y,z)=p_1(-x,y,z) \,, \qquad \vv_1(x,y,z)= \mathcal{S} \cdot \vv_1(-x,y,z) \,. 
\end{align}
The limit $h \rightarrow 0$ of the evaluation of \eqref{Sym} at the arbitrary plane $x=h$ divided by $h$ leads to the anti-symmetry conditions
\begin{align}
\label{AntiRevlim}
\partial_x p(0,y,z)=0  \,, \,\, \vv (0,y,z) \cdot \boldsymbol{e}_x= 0 \,, \,\, \partial_x \vv (0,y,z) \cdot \boldsymbol{e}_y= 0 \,, \,\, \partial_x \vv (0,y,z) \cdot \boldsymbol{e}_z= 0.
\end{align}

\begin{figure}
\centering
\input{./figures/ReversibilityExample}
\caption{Streamlines and pressure field at the $z=0$ symmetry plane (a) of the anti-symmetric and reversible flow $\hat{p}_0$ and $\vv_0$ and (b) of the symmetric and anti-reversible flow $\hat{p_1}$ and $\vv_1$ for a centred bubble of size $d=0.4$ and with rigid surface.}
\label{RevEx}
\end{figure}
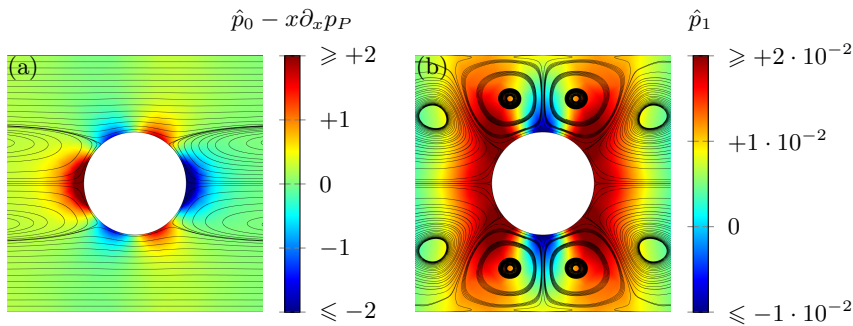

Examples of both kinds of flows are depicted in \figref{RevEx}.
In \tabref{tab:revsym}, the symmetries and reversibilities are tabulated. The zeroth-order creeping flow, which is anti-symmetric and reversible, is modified for small $Re$ and $Ca$ numbers, i.e. at first order, with a symmetric and anti-reversible flow, whereas symmetries and reversibilies are broken for large $Re$ or $Ca$ numbers. 
\begin{table}
\begin{center}
\begin{tabular}{r | ccc}
      			& $Ca= 0$ & $Ca<Ca_*$  & $\forall Ca$ \\ \hline
$Re= 0$   		& AS \& R & S \& AR & None  \\ 
$Re<Re_*$     	& S \& AR & S \& AR &  \\ 
$\forall Re$      	& None & & None      
  \end{tabular}
 \caption{Reversibilities and symmetries for the considered regimes. S: symmetric, AS: anti-symmetric, R: reversible and AR: anti-reversible.}
  \label{tab:revsym}
  \end{center}
\end{table}

\bibliographystyle{jfm}
\bibliography{jfm-instructions.bib}

\end{document}

%% file: figures/Sketch.tex
\begin{center}
\begin{tikzpicture}[x={(2.25cm,0cm)},y={(0cm,2.25cm)}]

\filldraw[color=black!30, very thick] (-3.0,0.5) rectangle (3.0,-0.5);

\shadedraw[color=black, fill=blue!20, very thick, dashed,left color=blue!20,right color=blue!50] (-1,-0.5)rectangle (1,0.5);


\draw[color=black, dashdot,line width=.15mm] (-3,0)-- (3,0) ;

\filldraw[color=blue, fill=blue,shading=ball] (0,.15) circle (.18);

\filldraw[color=black, fill=black, very thick] (-2.25,0) circle (.18);
\filldraw[color=black, fill=black, very thick] (2.25,.3) circle (.18);

\draw[color=green, line width=.5mm,->] (.0,.5)-- (-.25,0.5) node[black, midway, above] {$V$} ;
\draw[color=green, line width=.5mm,->] (.0,-.5)-- (-.25,-0.5) ;

\draw[color=black, dashed,line width=.15mm] (0,0.15)-- (0.35,0.15);
\draw[color=black, line width=.5mm,->] (0.3,0)-- (0.3,0.15) node[midway, right] {$\boldsymbol{ \varepsilon}$} ;

\draw[color=black, line width=.5mm,->] (.5,-.1)-- (0.5,-.4) node[midway, right] {$\boldsymbol{f}$} ;

\draw[color=black, line width=.25mm,<->] (-1.,-.6)-- (1.,-.6) node[midway, below] {$L$} ;
\draw[color=black, line width=.25mm,<->] (0.7,-.5)-- (0.7,.5) node[near end, right] {$d_h$} ;
\draw[color=black, line width=.25mm,<->] (-.18,.4)-- (.18,.4) node[anchor=west] {$d$} ;
\draw[color=black, dashed,line width=.15mm] (-.18,0.15)-- (-0.18,0.45);
\draw[color=black, dashed,line width=.15mm] (+.18,0.15)-- (+0.18,0.45);


\begin{scope}[xshift=-4.5cm]    

\draw[color=green, line width=.5mm,<-] (0.75,-.5)--(1,-.5); \node[below,black,anchor=east] at (1,-.1) {$\Sigma_{OUT}$};
\draw[color=green, line width=.5mm,->] (1,-.25)-- (1.125,-.25);
\draw[color=green, line width=.5mm,->] (1,0)-- (1.25,0);
\draw[color=green, line width=.5mm,->] (1,.25)-- (1.125,.25);
\draw[color=green, line width=.5mm,<-] (.75,.5)-- (1,.5) node[above, green] {$\boldsymbol{v}$};

\draw [green, thick,  domain=0:1, samples=40] 
 plot ({1.25-.5*\x^2}, {+.5*\x} );
 \draw [green, thick,  domain=0:1, samples=40] 
 plot ({1.25-.5*\x^2}, {-.5*\x} );

\end{scope}

\node[below,white,anchor=center] at (-0.0,+.15) {$\mathcal{V}_B$};

\draw[color=green, line width=.5mm,<-] (0.75,-.5)--(1,-.5); \node[below,black,anchor=west] at (-0.4,-.3) {$\mathcal V$};
\draw[color=green, line width=.5mm,<-] (0.75,-.5)--(1,-.5); \node[below,black,anchor=west] at (-0.,-.15) {$\Sigma_{B}$};
\draw[color=green, line width=.5mm,<-] (0.75,-.5)--(1,-.5); \node[below,black,anchor=south west] at (0.2,.5) {$\Sigma_{W}$};
\draw[color=green, line width=.5mm,<-] (0.75,-.5)--(1,-.5); \node[below,black,anchor=west] at (1,-.1) {$\Sigma_{IN}$};

\draw[color=green, line width=.5mm,->] (1,-.25)-- (1.125,-.25);
\draw[color=green, line width=.5mm,->] (1,0)-- (1.25,0);
\draw[color=green, line width=.5mm,->] (1,.25)-- (1.125,.25);
\draw[color=green, line width=.5mm,<-] (.75,.5)-- (1,.5) node[above, green] {$\boldsymbol{v}$};

\draw [green, thick,  domain=0:1, samples=40] 
 plot ({1.25-.5*\x^2}, {+.5*\x} );
 \draw [green, thick,  domain=0:1, samples=40] 
 plot ({1.25-.5*\x^2}, {-.5*\x} );



\draw [green, line width=.5mm] (1.8,-.4)-- (2,-.4) node[right] {Axial velocity};

\node[anchor=north west] at (-2.9,+.5) {input: $\rho, \, \mu, \, \gamma, \, J ,\, f,\, d_h,\, d$};
\node[anchor=south west] at (-2.9,-.5) {output: $V ,\, \varepsilon ,\, \Delta p  , \, \Omega $};


\draw[color=black, dashdot,line width=.15mm] (-3,0)-- (3,0) ;
\draw[color=black,->] (-.5,0) --++ (.1,.0) node[anchor=north] {\footnotesize{$x$}};
\draw[color=black,->] (-.5,0) --++ (.0,.1) node[anchor=south] {\footnotesize{$y$}};
\draw[color=black,->] (-.5,0) --++ (-.07,-.07) node[anchor=east] {\footnotesize{$z$}};

\end{tikzpicture}
\end{center}

%% file: figures/Regimes.tex
\begin{tikzpicture}[y=.4\textwidth, x=.7\textwidth,font=\sffamily]
	\def\x{.2}
	\def\y{.52}
	\def\z{.84}
	\def\f{.95}
	
	
    	
	\draw[pattern=dots, pattern color = gray] (\x-.08-.04,\x-.1-.04) rectangle (\z+.08+.04,\z+.1+.04);
	\label{patNLin}
	
	\draw[fill=white, pattern color = gray] (\x-.08-.04,\x-.1-.04) rectangle (\y+.08+.04,\y+.1+.04);
	\draw[pattern=bricks, pattern color = gray] (\x-.08-.04,\x-.1-.04) rectangle (\y+.08+.04,\y+.1+.04);
	
	\label{patLinear}

	\node[] at (\x,0) {$Re=0$};
	\node[] at (\y,0) {$Re<Re_*$};
	\node[] at (\z,0) {$\forall Re$};

	\node[] at (0,\x) {$Ca=0$};
	\node[] at (0,\y) {$Ca<Ca_*$};
	\node[] at (0,\z) {$\forall Ca$};

	\draw[|->,line width=.1cm] (\x,\x) -- (\z,\z);
	\draw[|->,line width=.1cm] (\x,\x) -- (\x,\z);
	\draw[|->,line width=.1cm] (\x,\x) -- (\z,\x);

	\draw[fill=black!10,black!10] (\x-.08,\x-.1) rectangle (\x+.08,\x+.1);
	
	\draw[fill=red!20,red!20] (\y-.08,\x-.1) rectangle (\y+.08,\x+.1);
	\draw[fill=red!40,red!40] (\z-.08,\x-.1) rectangle (\z+.08,\x+.1);

	\draw[fill=yellow!40,yellow!40] (\x-.08,\y-.1) rectangle (\x+.08,\y+.1);
	\draw[fill=yellow!80,yellow!80] (\x-.08,\z-.1) rectangle (\x+.08,\z+.1);	

	\draw[fill=yellow!40!red!40,yellow!20!red!20] (\y-.08,\y-.1) rectangle (\y+.08,\y+.1);
	\draw[fill=yellow!80!red!80,yellow!40!red!40] (\z-.08,\z-.1) rectangle (\z+.08,\z+.1);
	
	\node[] at (\x,\x+.06) {Leading};
	\node[] at (\x,\x+0) {order};
	\node[] at (\x,\x-.06) {creeping};
	
	\node[] at (\y,\x+.06) {Pure};
	\node[] at (\y,\x+0) {linear};
	\node[] at (\y,\x-.06) {inertial};

	\node[] at (\z,\x+.06) {Pure};
	\node[] at (\z,\x+0) {nonlinear};
	\node[] at (\z,\x-.06) {inertial};
	
	\node[] at (\x,\y+.06) {Pure};
	\node[] at (\x,\y+0) {linear};
	\node[] at (\x,\y-.06) {capillary};

	\node[] at (\x,\z+.06) {Pure};
	\node[] at (\x,\z+0) {nonlinear};
	\node[] at (\x,\z-.06) {capillary};

	\node[] at (\z,\z+.06) {Nonlinear};
	\node[] at (\z,\z+0) {inertial-};
	\node[] at (\z,\z-.06) {capillary};
	
	\node[] at (\y,\y+.06) {Linear};
	\node[] at (\y,\y+0) {inertial-};
	\node[] at (\y,\y-.06) {capillary};
	

\end{tikzpicture}

%% file: figures/L_vs_CRd04.tex
%
%
\begin{tikzpicture}[baseline]

\begin{axis}[%
name=one,
width=.45\textwidth,
height=.5 \textwidth,
scale only axis,
axis on top,
xlabel={$\varepsilon /\varepsilon_*$},
ylabel={$f/Re$},
every outer x axis line/.append style={black},
every x tick label/.append style={font=\color{black}},
xmin=-1.2,
xmax=1.2,
every outer y axis line/.append style={black},
every y tick label/.append style={font=\color{black}},
ymin=-.5,
ymax=.5,
axis background/.style={fill=white},
x filter/.code={\pgfmathparse{#1*1}\pgfmathresult},
y filter/.code={\pgfmathparse{#1*-29.8416}\pgfmathresult},
axis lines = center,
ytick={-.5,-.25,0,.25,.5},
axis on top,
axis line style=-,
]

\node[anchor=north west] at (rel axis cs:0,1) {(a)};


\addplot [color=blue,solid,forget plot,smooth]
  table[row sep=crcr]{%
-0.94	-0.0290471291648977\\
-0.92	-0.0235607454774573\\
-0.9	-0.019146825743348\\
-0.88	-0.015444226893984\\
-0.86	-0.0122623855155063\\
-0.84	-0.00948436445136234\\
-0.82	-0.00702936168074798\\
-0.8	-0.00484343041130743\\
-0.75	-0.000327506172111871\\
-0.7	0.0031217817230984\\
-0.65	0.00571899568609368\\
-0.6	0.00760208259542743\\
-0.55	0.00888267843786308\\
-0.5	0.00963648714895407\\
-0.45	0.00992762105375546\\
};
\addplot [color=blue,dashed,forget plot,smooth]
  table[row sep=crcr]{%
-0.45	0.00992762105375546\\
-0.4	0.00981113823823544\\
-0.35	0.00934223043161906\\
-0.3	0.00856667218132908\\
-0.25	0.00753274187009934\\
-0.2	0.00628210489051515\\
-0.15	0.00486125889219309\\
-0.1	0.00331184293195512\\
-0.05	0.00167690455830796\\
-0	-1.41792078970038e-07\\
0	1.41792078970038e-07\\
0.05	-0.00167690455830796\\
0.1	-0.00331184293195512\\
0.15	-0.00486125889219309\\
0.2	-0.00628210489051515\\
0.25	-0.00753274187009934\\
0.3	-0.00856667218132908\\
0.35	-0.00934223043161906\\
0.4	-0.00981113823823544\\
0.45	-0.00992762105375546\\
};
\addplot [color=blue,solid,forget plot,smooth]
  table[row sep=crcr]{%
0.45	-0.00992762105375546\\
0.5	-0.00963648714895407\\
0.55	-0.00888267843786308\\
0.6	-0.00760208259542743\\
0.65	-0.00571899568609368\\
0.7	-0.0031217817230984\\
0.75	0.000327506172111871\\
0.8	0.00484343041130743\\
0.82	0.00702936168074798\\
0.84	0.00948436445136234\\
0.86	0.0122623855155063\\
0.88	0.015444226893984\\
0.9	0.019146825743348\\
0.92	0.0235607454774573\\
0.94	0.0290471291648977\\
};

\draw[color=blue,decorate,decoration={triangles,segment length=4}] (axis cs:-1, -.00) -- (axis cs:-.745,-.00);
\draw[color=blue,decorate,decoration={triangles,segment length=4}] (axis cs:-.0,-.00) -- (axis cs:-.745,-.00);
\draw[color=blue,decorate,decoration={triangles,segment length=4}] (axis cs:-.0,-.00) -- (axis cs:+.745,-.00);
\draw[color=blue,decorate,decoration={triangles,segment length=4}] (axis cs:1,-.00) -- (axis cs:.745,-.00);
\addplot [color=blue,solid,forget plot]
  table[row sep=crcr]{%
-1	+0.0\\
1	+0.0\\
};

\addplot [color=green,only marks,mark=*,mark options={solid},forget plot,mark size=2pt]
  table[row sep=crcr]{%
-0.745 +0.00\\
+0.745	 +0.00\\
};
\addplot [color=red,only marks,mark=*,mark options={solid,fill=white},forget plot,mark size=2pt]
  table[row sep=crcr]{%
-0.0 0.00\\
};
\node at (-.745,-0.00) [anchor=south east] {i};
\node at (-.000,-0.00) [anchor=south east] {ii};
\node at (+.745,-0.00) [anchor=south east] {iii$ \equiv $i'};

%

%
%

\draw[color=blue,decorate,decoration={triangles,segment length=4}] (axis cs:-1, -0.1492) -- (axis cs:-.66,-0.1492);
\draw[color=blue,decorate,decoration={triangles,segment length=4}] (axis cs:-.155,-0.1492) -- (axis cs:-.66,-0.1492);
\draw[color=blue,decorate,decoration={triangles,segment length=4}] (axis cs:-.155,-0.1492) -- (axis cs:+.8,-0.1492);
\draw[color=blue,decorate,decoration={triangles,segment length=4}] (axis cs:1,-0.1492) -- (axis cs:.8,-0.1492);
\addplot [color=blue,solid,forget plot]
  table[row sep=crcr]{%
-1	+0.005\\
1	+0.005\\
};

\addplot [color=green,only marks,mark=*,mark options={solid},forget plot,mark size=2pt]
  table[row sep=crcr]{%
-0.66	 +0.005\\
+0.8	 +0.005\\
};
\addplot [color=red,only marks,mark=*,mark options={solid,fill=white},forget plot,mark size=2pt]
  table[row sep=crcr]{%
-0.155 0.005\\
};
\node at (-.66,-0.1492) [anchor=north east] {iv};
\node at (-.155,-0.1492) [anchor=north west] {v};
\node at (+.8,-0.1492) [anchor=north east] {vi};


%
%
%


\addplot [color=blue,only marks,mark=square*,mark options={solid},forget plot,mark size=2pt]
  table[row sep=crcr]{%
-0.45 	+0.00992762105375546\\
+0.45 	-0.00992762105375546\\
};
\label{TransMark};

\addplot [color=green,only marks,mark=*,mark options={solid},forget plot,mark size=2pt]
  table[row sep=crcr]{%
+0.66	 -0.005\\
-0.80 	 -0.005\\
};
\addplot [color=red,only marks,mark=*,mark options={solid,fill=white},forget plot,mark size=2pt]
  table[row sep=crcr]{%
+0.155 -0.005\\
};
\node at (+.66,+0.1492) [anchor=west] {iv'};
\node at (+.155,+0.1492) [anchor=west] {v'};
\node at (-.8,+0.1492) [anchor=east] {vi'};

\end{axis}

%% file: figures/V_c_CRd04.tex
%
%

\begin{axis}[%
name=two,
at=(one.north east), anchor=north west, xshift=.05\textwidth, 
width=.2\textwidth,
height=.22 \textwidth,
scale only axis,
axis lines = center,
axis on top,
xlabel={$\varepsilon /\varepsilon_*$},
ylabel={$V$},
every outer x axis line/.append style={black},
every x tick label/.append style={font=\color{black}},
xmin=-1,
xmax=1,
every outer y axis line/.append style={black},
every y tick label/.append style={font=\color{black}},
ymin=0,
ymax=2,
axis background/.style={fill=white},
axis line style=-,
]
\node[anchor=north west] at (rel axis cs:0,1) {(b)};
\addplot [color=blue,solid,forget plot]
  table[row sep=crcr]{%
-0.94	0.944192659216269\\
-0.92	1.00217749688946\\
-0.9	1.05256650366987\\
-0.88	1.09771529730294\\
-0.86	1.13896243995823\\
-0.84	1.17712296519986\\
-0.82	1.21276400724855\\
-0.8	1.24626515407659\\
-0.75	1.32235384174167\\
-0.7	1.38960423441092\\
-0.65	1.44961060734428\\
-0.6	1.50338549901726\\
-0.55	1.55160232796\\
-0.5	1.59471785450127\\
-0.45	1.63307492941196\\
};
\addplot [color=blue,dashed,forget plot]
 table[row sep=crcr]{%
-0.45	1.63307492941196\\
-0.4	1.66691474755691\\
-0.35	1.69643085683752\\
-0.3	1.72176943521855\\
-0.25	1.74304844074126\\
-0.2	1.76035120181506\\
-0.15	1.77374560446177\\
-0.1	1.78327851919158\\
-0.05	1.7889851363607\\
-0	1.79088586512191\\
0	1.79088586512191\\
0.05	1.7889851363607\\
0.1	1.78327851919158\\
0.15	1.77374560446177\\
0.2	1.76035120181506\\
0.25	1.74304844074126\\
0.3	1.72176943521855\\
0.35	1.69643085683752\\
0.4	1.66691474755691\\
0.45	1.63307492941196\\
};
\addplot [color=blue,solid,forget plot]
 table[row sep=crcr]{%
0.45	1.63307492941196\\
0.5	1.59471785450127\\
0.55	1.55160232796\\
0.6	1.50338549901726\\
0.65	1.44961060734428\\
0.7	1.38960423441092\\
0.75	1.32235384174167\\
0.8	1.24626515407659\\
0.82	1.21276400724855\\
0.84	1.17712296519986\\
0.86	1.13896243995823\\
0.88	1.09771529730294\\
0.9	1.05256650366987\\
0.92	1.00217749688946\\
0.94	0.944192659216269\\
};
\end{axis}

%% file: figures/beta_c_CRd04.tex
%
%

\begin{axis}[%
at=(two.north east), anchor=north west, xshift=.05\textwidth, 
width=.2\textwidth,
height=.22 \textwidth,
scale only axis,
axis lines = center,
axis on top,
xlabel={$\varepsilon /\varepsilon_*$},
ylabel={$\beta$},
every outer x axis line/.append style={black},
every x tick label/.append style={font=\color{black}},
xmin=-1,
xmax=1,
every outer y axis line/.append style={black},
every y tick label/.append style={font=\color{black}},
ymin=-.2,
ymax=5,
axis background/.style={fill=white},
x filter/.code={\pgfmathparse{#1*1}\pgfmathresult},
y filter/.code={\pgfmathparse{(#1-96)*0.732421875}\pgfmathresult},
axis line style=-,
]
\node[anchor=north west] at (rel axis cs:0,1) {(c)};
\addplot [color=blue,solid,forget plot]
  table[row sep=crcr]{%
-0.94	101.563048316017\\
-0.92	100.973238263544\\
-0.9	100.506087654162\\
-0.88	100.119807030086\\
-0.86	99.7909595192187\\
-0.84	99.5051911410012\\
-0.82	99.25282753311\\
-0.8	99.0272916840734\\
-0.75	98.5521949342708\\
-0.7	98.1686872948089\\
-0.65	97.8505365944699\\
-0.6	97.5818574244132\\
-0.55	97.3522038610317\\
-0.5	97.154711364786\\
-0.45	96.9845792330771\\
};
\addplot [color=blue,dashed,forget plot]
 table[row sep=crcr]{%
-0.45	96.9845792330771\\
-0.4	96.8384368016311\\
-0.35	96.7137364065329\\
-0.3	96.6085765933246\\
-0.25	96.5214829000708\\
-0.2	96.4514764673547\\
-0.15	96.3977349665749\\
-0.1	96.359746185164\\
-0.05	96.3371056686904\\
-0	96.3295528997321\\
0	96.3295528997321\\
0.05	96.3371056686904\\
0.1	96.359746185164\\
0.15	96.3977349665749\\
0.2	96.4514764673547\\
0.25	96.5214829000708\\
0.3	96.6085765933246\\
0.35	96.7137364065329\\
0.4	96.8384368016311\\
};
\addplot [color=blue,solid,forget plot]
 table[row sep=crcr]{%
0.4	96.8384368016311\\
0.45	96.9845792330771\\
0.5	97.154711364786\\
0.55	97.3522038610317\\
0.6	97.5818574244132\\
0.65	97.8505365944699\\
0.7	98.1686872948089\\
0.75	98.5521949342708\\
0.8	99.0272916840734\\
0.82	99.25282753311\\
0.84	99.5051911410012\\
0.86	99.7909595192187\\
0.88	100.119807030086\\
0.9	100.506087654162\\
0.92	100.973238263544\\
0.94	101.563048316017\\
};
\end{axis}

%% file: figures/Om_c_CRd04.tex
%
%

\begin{axis}[%
name=three,
at=(one.south east), anchor=south west, xshift=.05\textwidth,
width=.2\textwidth,
height=.22\textwidth,
scale only axis,
axis lines = center,
axis on top,
xlabel={$\varepsilon /\varepsilon_*$},
ylabel={$\Omega$},
every outer x axis line/.append style={black},
every x tick label/.append style={font=\color{black}},
xmin=-1,
xmax=1,
every outer y axis line/.append style={black},
every y tick label/.append style={font=\color{black}},
ymin=-2,
ymax=2,
axis background/.style={fill=white},
axis line style=-,
]
\node[anchor=north west] at (rel axis cs:0,1) {(d)};
\addplot [color=blue,solid,forget plot]
  table[row sep=crcr]{%
-0.94	-1.71815587387788\\
-0.92	-1.74651867269338\\
-0.9	-1.7580447983406\\
-0.88	-1.75829129288416\\
-0.86	-1.75022150775833\\
-0.84	-1.73592838252293\\
-0.82	-1.7166805602138\\
-0.8	-1.69347363997762\\
-0.75	-1.62262933136596\\
-0.7	-1.53855794634169\\
-0.65	-1.44571559718891\\
-0.6	-1.34658904865125\\
-0.55	-1.24300374639738\\
-0.5	-1.13623649632752\\
-0.45	-1.02697618966342\\
};
\addplot [color=blue,dashed,forget plot]
 table[row sep=crcr]{%
-0.45	-1.02697618966342\\
-0.4	-0.915963308431131\\
-0.35	-0.803624805881918\\
-0.3	-0.690281105777337\\
-0.25	-0.576194177871265\\
-0.2	-0.461530949009765\\
-0.15	-0.34647161061901\\
-0.1	-0.231124947005933\\
-0.05	-0.115603745579936\\
-0	1.27819445167773e-07\\
0	-1.27819445167773e-07\\
0.05	0.115603745579936\\
0.1	0.231124947005933\\
0.15	0.34647161061901\\
0.2	0.461530949009765\\
0.25	0.576194177871265\\
0.3	0.690281105777337\\
0.35	0.803624805881918\\
0.4	0.915963308431131\\
};
\addplot [color=blue,solid,forget plot]
 table[row sep=crcr]{%
0.4	0.915963308431131\\
0.45	1.02697618966342\\
0.5	1.13623649632752\\
0.55	1.24300374639738\\
0.6	1.34658904865125\\
0.65	1.44571559718891\\
0.7	1.53855794634169\\
0.75	1.62262933136596\\
0.8	1.69347363997762\\
0.82	1.7166805602138\\
0.84	1.73592838252293\\
0.86	1.75022150775833\\
0.88	1.75829129288416\\
0.9	1.7580447983406\\
0.92	1.74651867269338\\
0.94	1.71815587387788\\
};
\end{axis}

%% file: figures/Legend1.tex
%
%

\begin{axis}[%
at=(three.north east), anchor=north west, xshift=.05\textwidth, 
width=.2\textwidth,
height=.22\textwidth,
scale only axis,
axis lines = center,
axis on top,
xlabel={$\varepsilon /\varepsilon_*$},
ylabel={$\Omega$},
every outer x axis line/.append style={black},
every x tick label/.append style={font=\color{black}},
xmin=-1,
xmax=1,
every outer y axis line/.append style={black},
every y tick label/.append style={font=\color{black}},
ymin=-1,
ymax=1,
axis background/.style={fill=white},
hide axis,
]


\addplot [color=blue,forget plot]
  table[row sep=crcr]{%
-.9	.3\\
-.3	.3\\
};
\addplot [color=green,only marks,mark=*,mark options={solid},forget plot,mark size=2pt]
  table[row sep=crcr]{%
-0.6	.3\\
};

\addplot [color=blue,dashed,forget plot]
  table[row sep=crcr]{%
-.9	-.3\\
-.3	-.3\\
};
\addplot [color=red,only marks,mark=*,mark options={solid,fill=white},forget plot,mark size=2pt]
  table[row sep=crcr]{%
-0.6	-.3\\
};

\addplot [color=blue,forget plot]
  table[row sep=crcr]{%
-.6	.0\\
-.3	.0\\
};
\addplot [color=blue,dashed,forget plot]
  table[row sep=crcr]{%
-.9	-.0\\
-.6	-.0\\
};
\addplot [color=blue,only marks,mark=square*,mark options={solid},forget plot,mark size=2pt]
  table[row sep=crcr]{%
-0.6	-.0\\
};

\node[anchor=west] at (-.2,+.3) {Stable};
\node[anchor=west] at (-.2,-.3) {Unstable};
\node[anchor=west] at (-.2,-.0) {Transition};

%

%

\end{axis}
\end{tikzpicture}%

%% file: figures/pressurefield0.tex
\begin{tikzpicture}[scale=1.08,]
\node[inner sep=0pt,anchor=north west] (fig1) at (.00\textwidth,.12\textwidth)   {\includegraphics[width=.2592\textwidth,angle=180]{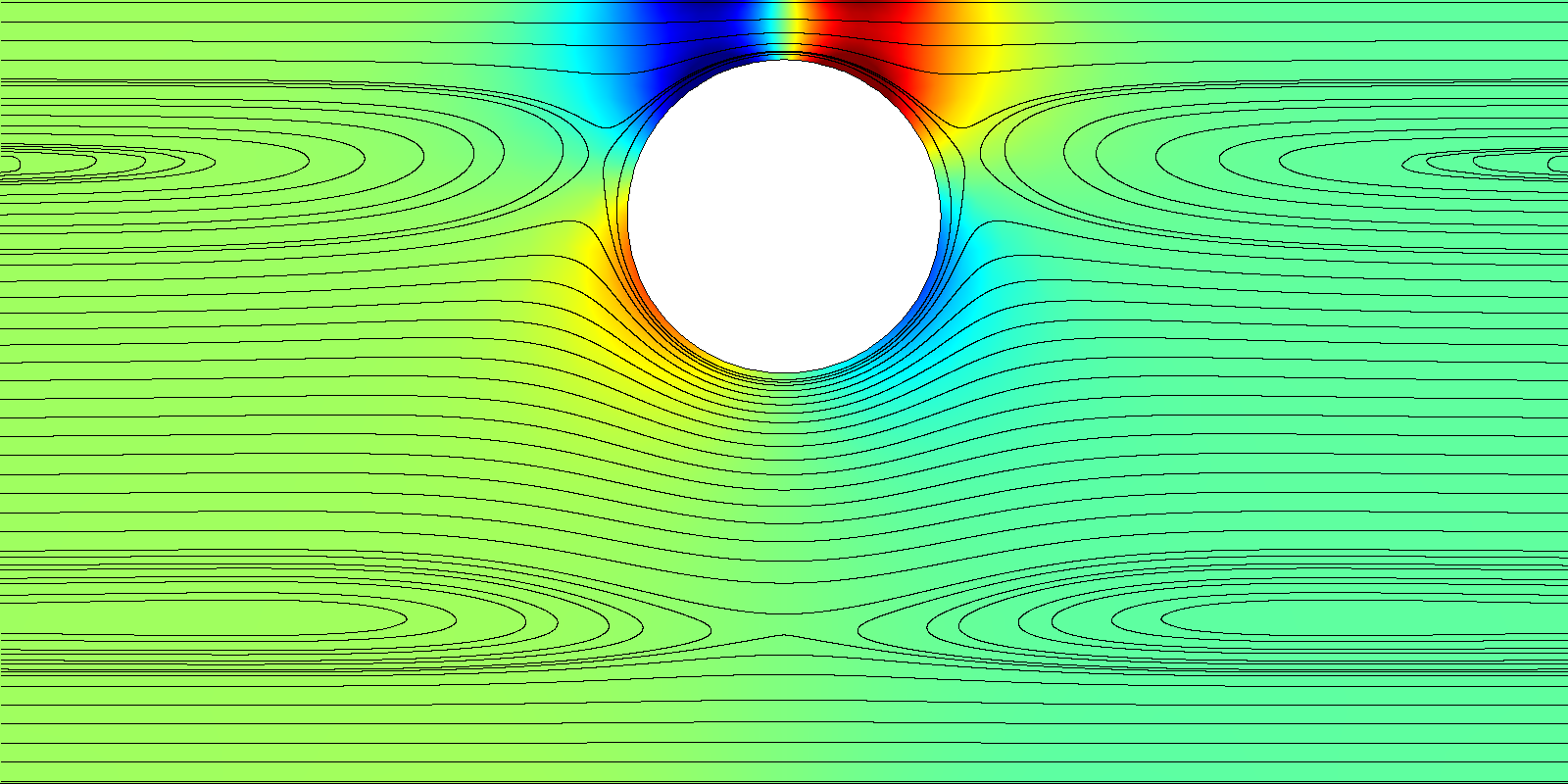}};
\node[inner sep=0pt,anchor=north west] (fig2) at (.25\textwidth,.12\textwidth)   {\includegraphics[width=.2592\textwidth]{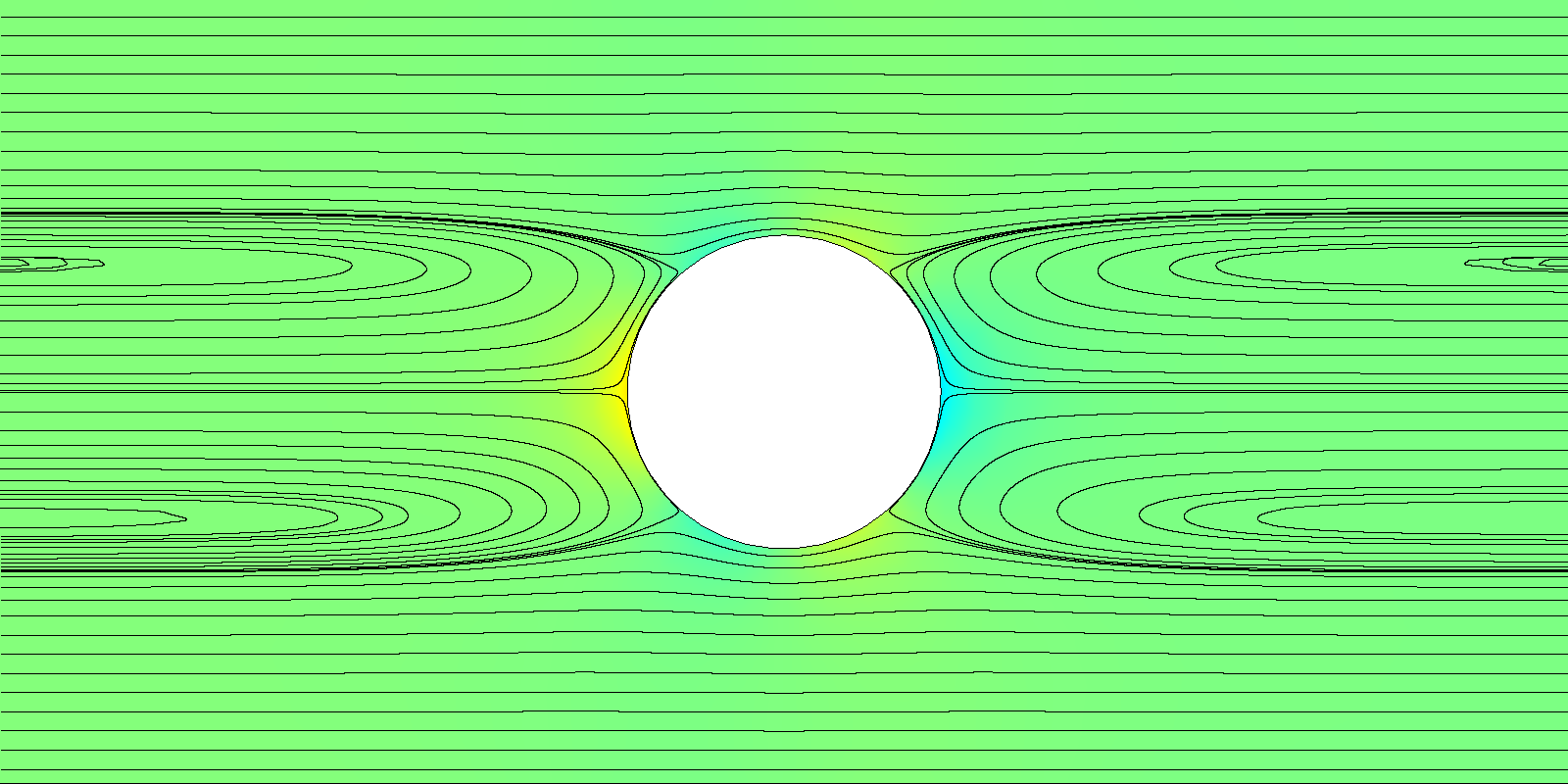}};
\node[inner sep=0pt,anchor=north west] (fig3) at (.50\textwidth,.12\textwidth)   {\includegraphics[width=.2592\textwidth]{./figures/p0a_p_20.png}};

\draw[dashdot] (-.01\textwidth,.06\textwidth) -- (.75\textwidth,.06\textwidth);
\node[inner sep=0pt,anchor=north west] (fig1) at (.00\textwidth,.12\textwidth)   {(i)}; 
\node[inner sep=0pt,anchor=north west] (fig2) at (.25\textwidth,.12\textwidth)   {(ii)}; 
\node[inner sep=0pt,anchor=north west] (fig3) at (.50\textwidth,.12\textwidth)   {(iii)}; 

\draw[<-] (.05\textwidth,.12\textwidth)  -- (.07\textwidth,.12\textwidth);
\draw[<-] (.11\textwidth,.12\textwidth)  -- (.13\textwidth,.12\textwidth);
\draw[<-] (.17\textwidth,.12\textwidth)  -- (.19\textwidth,.12\textwidth);
\draw[<-] (.05\textwidth,.00\textwidth)  -- (.07\textwidth,.00\textwidth);
\draw[<-] (.11\textwidth,.00\textwidth)  -- (.13\textwidth,.00\textwidth);
\draw[<-] (.17\textwidth,.00\textwidth)  -- (.19\textwidth,.00\textwidth);

\draw[<-] (.30\textwidth,.12\textwidth)  -- (.32\textwidth,.12\textwidth);
\draw[<-] (.36\textwidth,.12\textwidth)  -- (.38\textwidth,.12\textwidth);
\draw[<-] (.42\textwidth,.12\textwidth)  -- (.44\textwidth,.12\textwidth);
\draw[<-] (.30\textwidth,.00\textwidth)  -- (.32\textwidth,.00\textwidth);
\draw[<-] (.36\textwidth,.00\textwidth)  -- (.38\textwidth,.00\textwidth);
\draw[<-] (.42\textwidth,.00\textwidth)  -- (.44\textwidth,.00\textwidth);

\draw[<-] (.55\textwidth,.12\textwidth)  -- (.57\textwidth,.12\textwidth);
\draw[<-] (.61\textwidth,.12\textwidth)  -- (.63\textwidth,.12\textwidth);
\draw[<-] (.67\textwidth,.12\textwidth)  -- (.69\textwidth,.12\textwidth);
\draw[<-] (.55\textwidth,.00\textwidth)  -- (.57\textwidth,.00\textwidth);
\draw[<-] (.61\textwidth,.00\textwidth)  -- (.63\textwidth,.00\textwidth);
\draw[<-] (.67\textwidth,.00\textwidth)  -- (.69\textwidth,.00\textwidth);
\end{tikzpicture}
\begin{tikzpicture}[yscale=1.0]
\hspace{-.05\textwidth}
\begin{axis}[hide axis ,  scale only axis,    height=.100\textwidth,    width=.2cm,    colormap/jet,    colorbar,    point meta min=-1,    point meta max=1,
    colorbar style={  title=$\hat{p}_0- x \partial_x p_{\!\, {\rm P}}$,        width=.2cm,        height=.100\textwidth,        ytick={-1,-.5,0,.5,1},        yticklabels={$\leq-20$,$-10$,$0$,$+10$,$\geq+20$},   yticklabel style={xshift=0.5ex} }
    ]
\end{axis}
\end{tikzpicture}

%% file: figures/pressurefield1.tex
\begin{tikzpicture}[scale=1.08,]
\node[inner sep=0pt,anchor=north west] (fig1) at (.00\textwidth,.13\textwidth)   {\includegraphics[width=.2592\textwidth,angle=180]{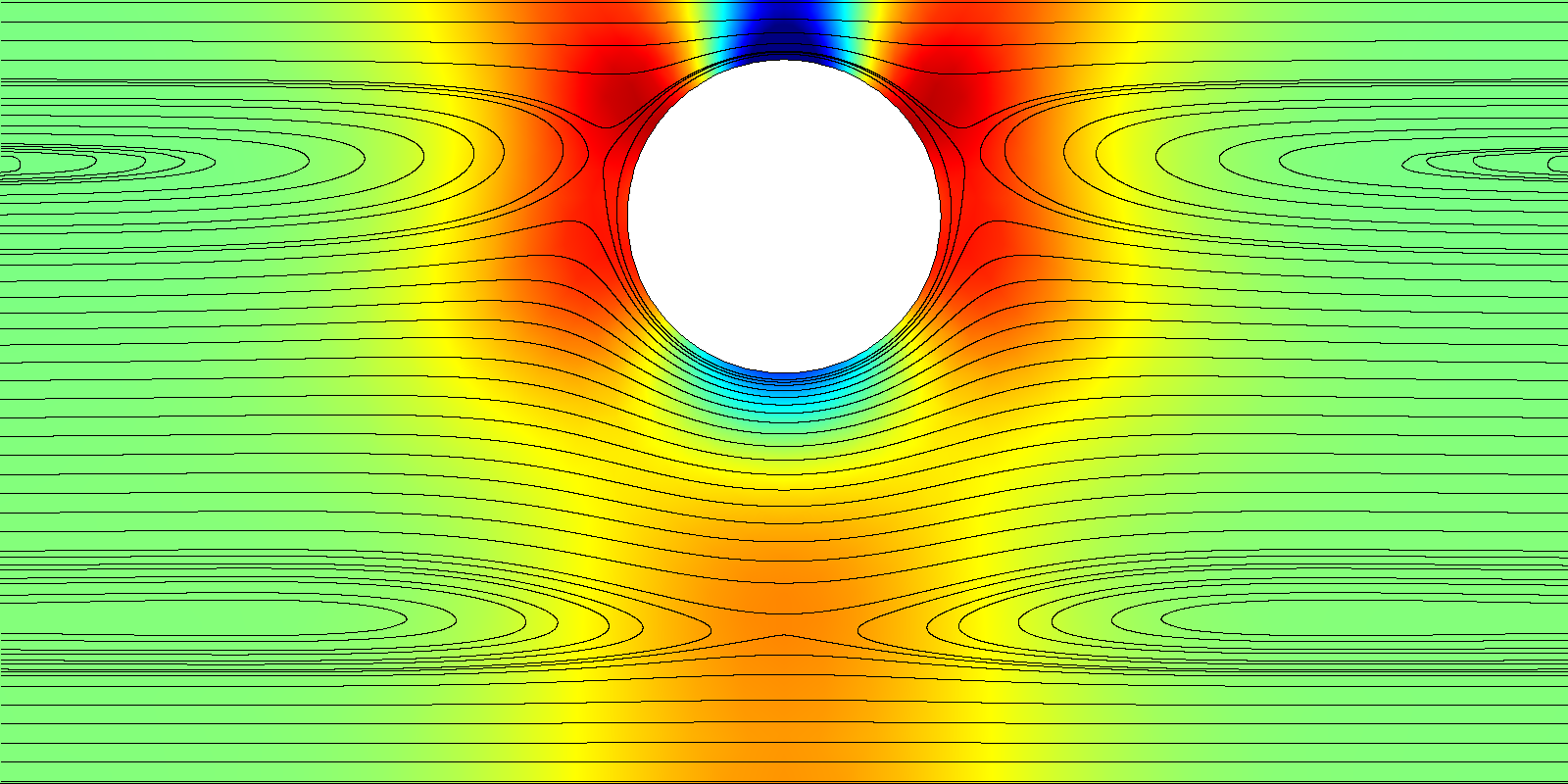}};
\node[inner sep=0pt,anchor=north west] (fig2) at (.25\textwidth,.13\textwidth)   {\includegraphics[width=.2592\textwidth]{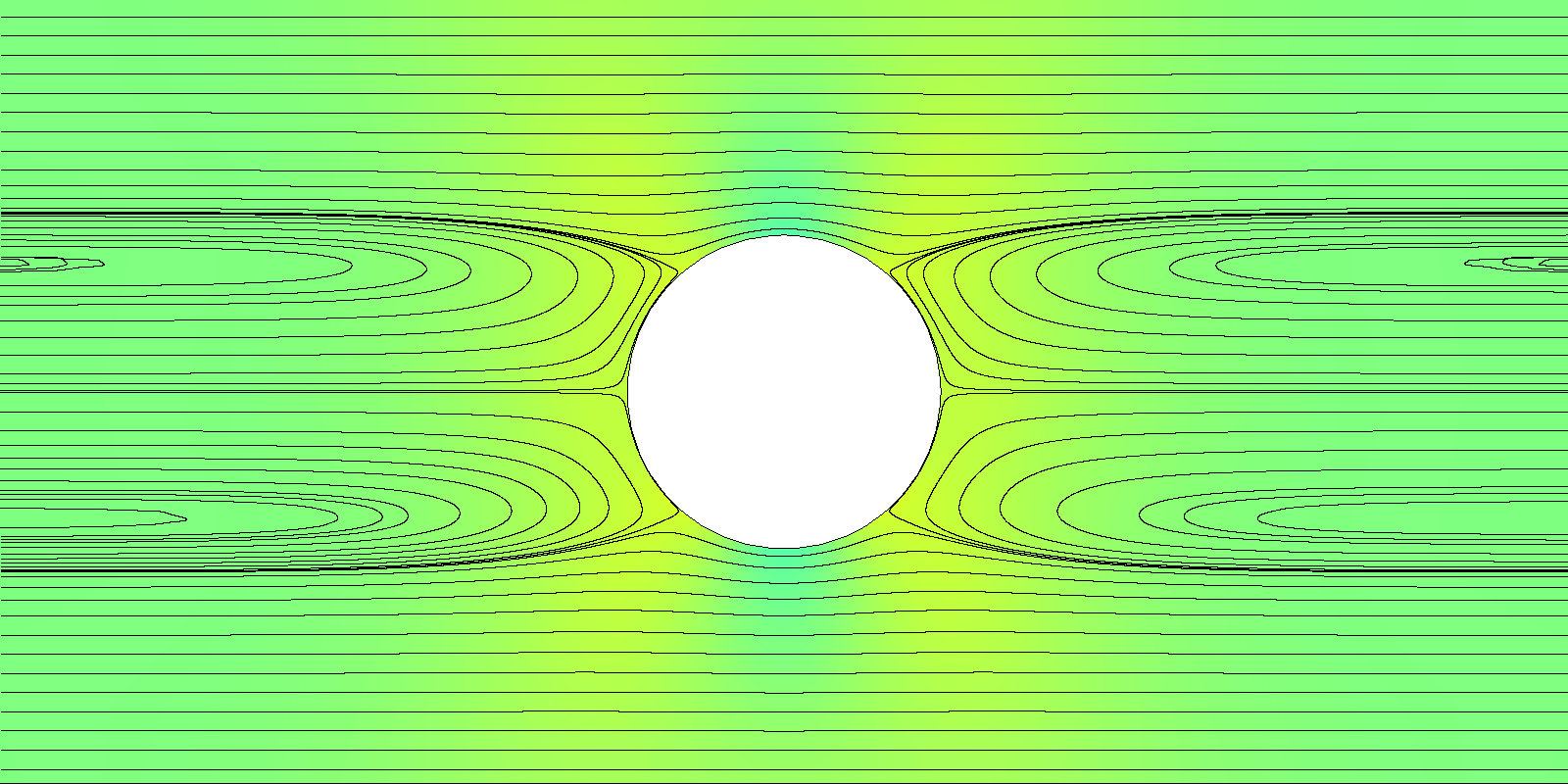}};
\node[inner sep=0pt,anchor=north west] (fig3) at (.50\textwidth,.13\textwidth)   {\includegraphics[width=.2592\textwidth]{./figures/p1a_p_015.png}};
\draw[dashdot] (-.01\textwidth,.07\textwidth) -- (.75\textwidth,.07\textwidth);
\node[inner sep=0pt,anchor=north west] (fig1) at (.00\textwidth,.13\textwidth)   {(i)};
\node[inner sep=0pt,anchor=north west] (fig2) at (.25\textwidth,.13\textwidth)   {(ii)};
\node[inner sep=0pt,anchor=north west] (fig3) at (.50\textwidth,.13\textwidth)   {(iii)};
\node[inner sep=0pt,anchor=north west] (fig1) at (.00\textwidth,.0\textwidth)   {\includegraphics[width=.2592\textwidth]{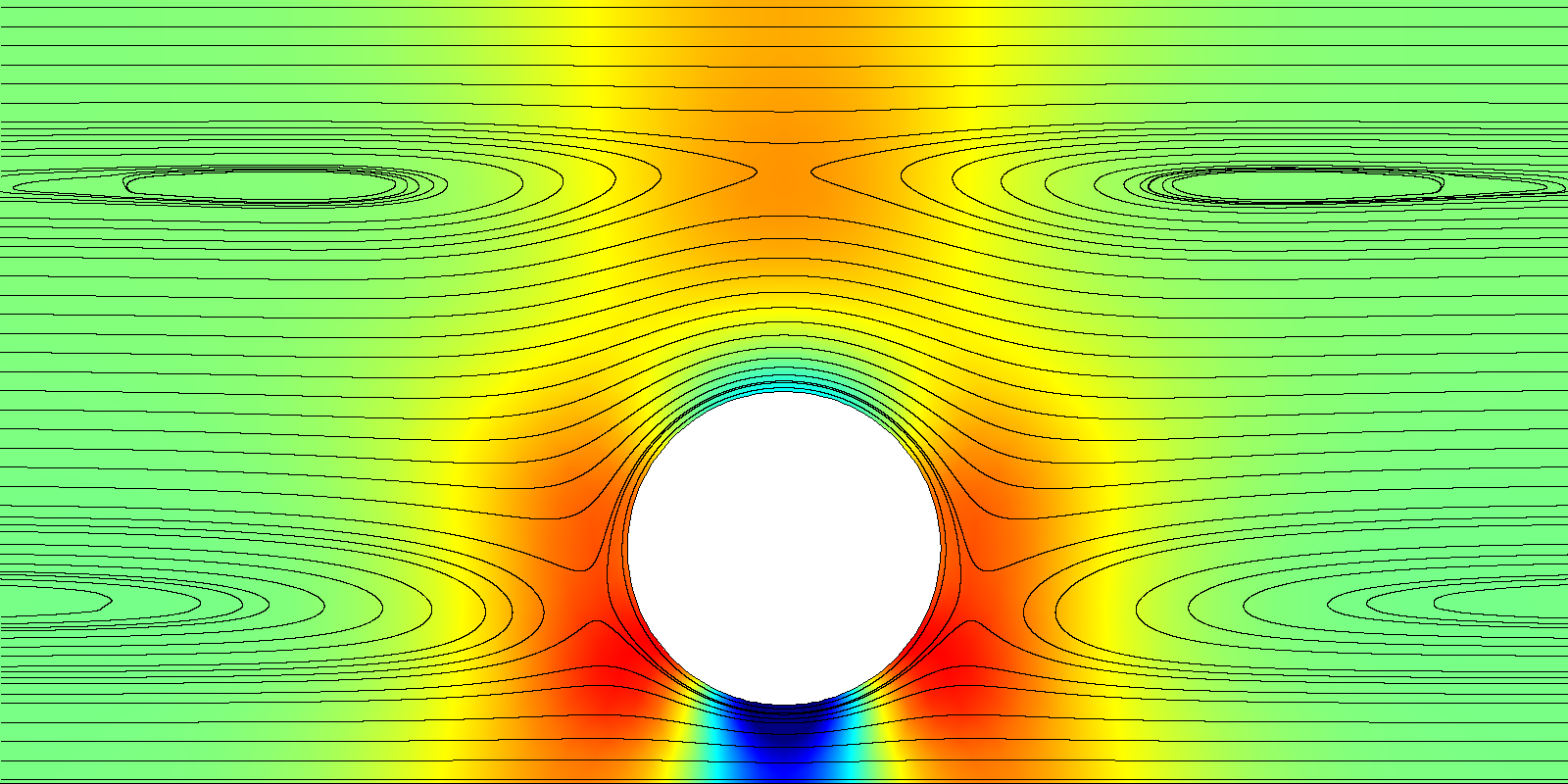}};
\node[inner sep=0pt,anchor=north west] (fig2) at (.25\textwidth,.0\textwidth)   {\includegraphics[width=.2592\textwidth]{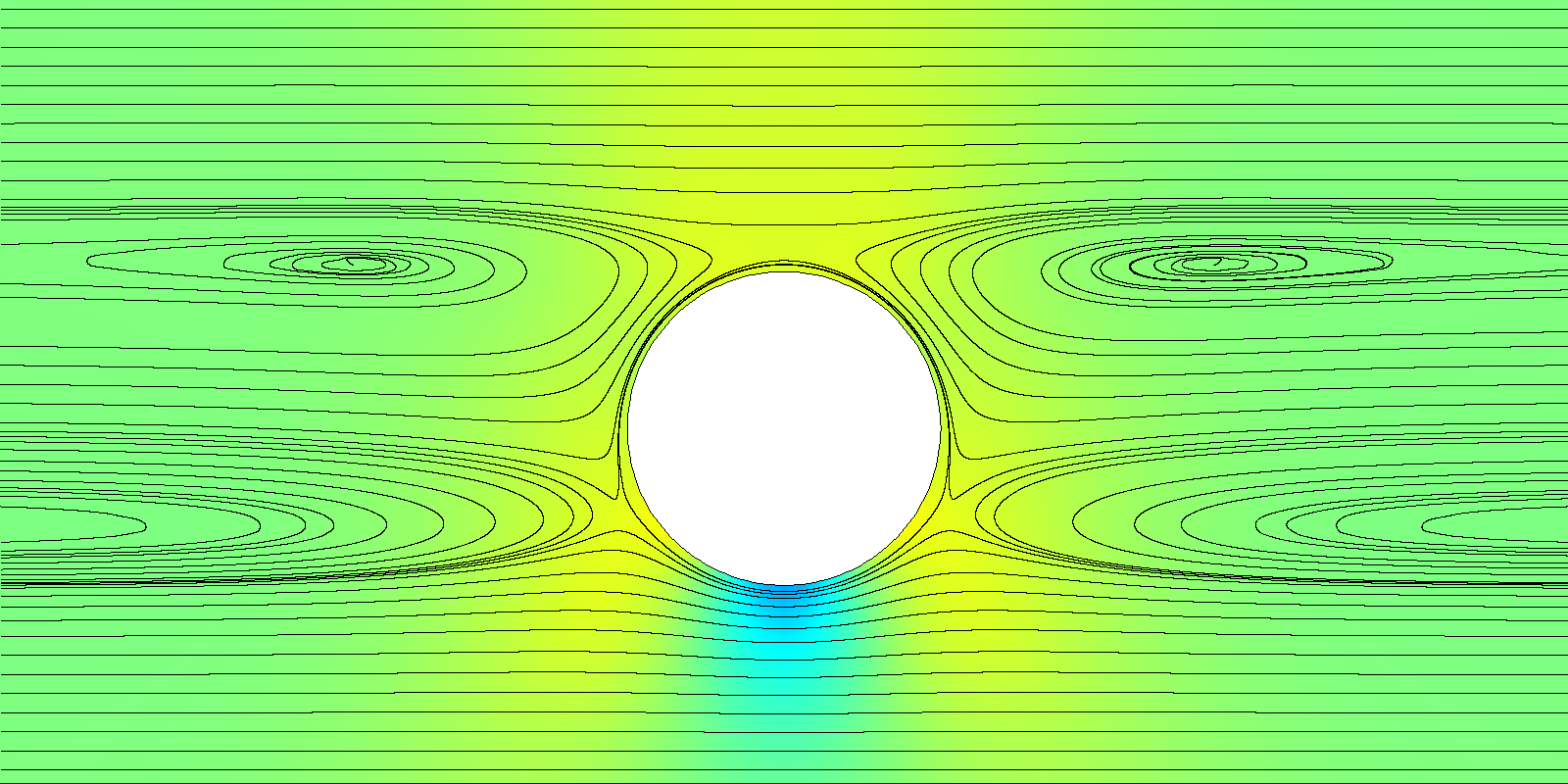}};
\node[inner sep=0pt,anchor=north west] (fig3) at (.50\textwidth,.0\textwidth)   {\includegraphics[width=.2592\textwidth]{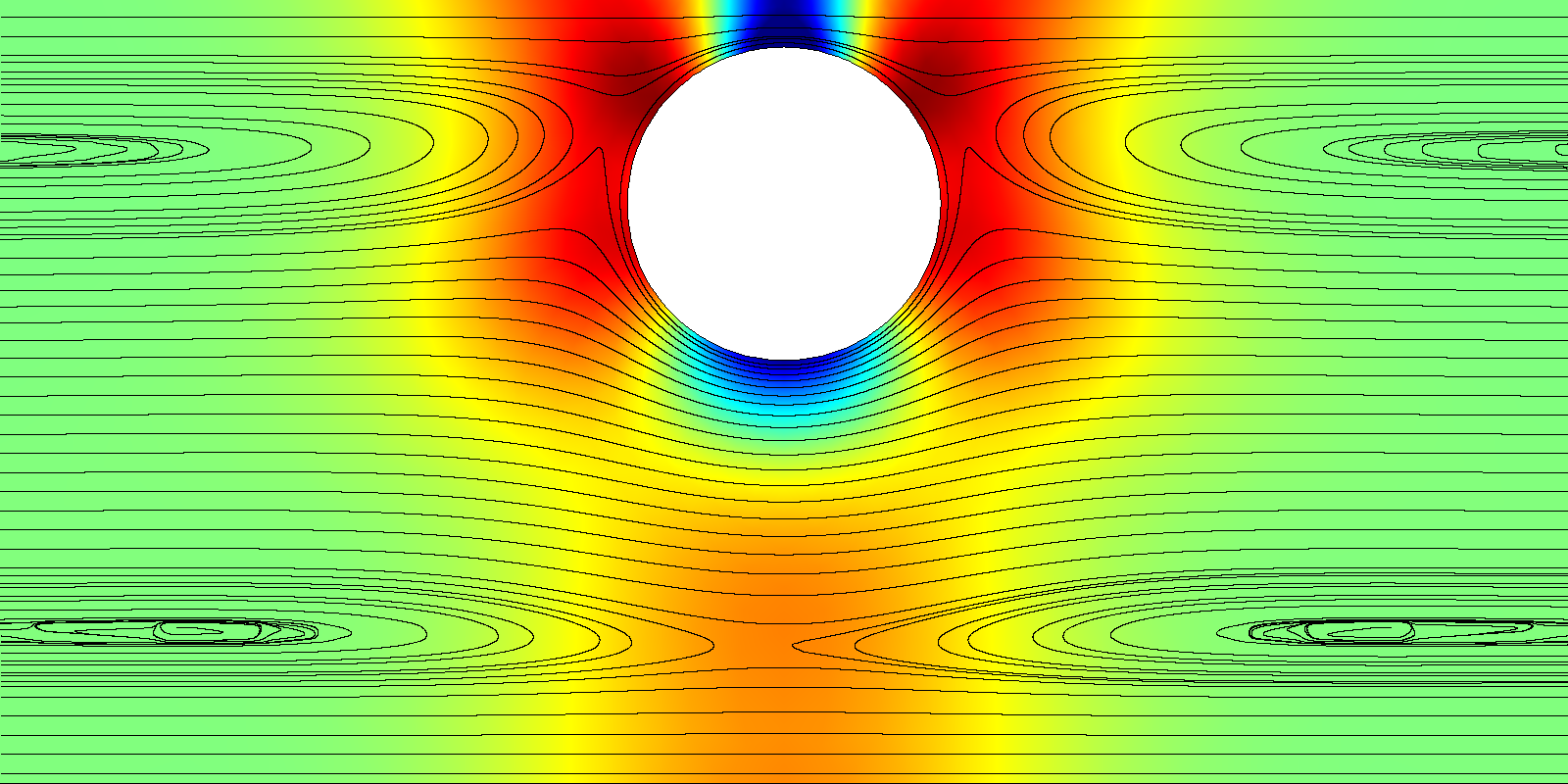}};
\draw[dashdot] (-.01\textwidth,-.06\textwidth) -- (.75\textwidth,-.06\textwidth);
\node[inner sep=0pt,anchor=north west] (fig1) at (.00\textwidth,.0\textwidth)   {(iv)};  
\node[inner sep=0pt,anchor=north west] (fig2) at (.25\textwidth,.0\textwidth)   {(v)};  
\node[inner sep=0pt,anchor=north west] (fig3) at (.50\textwidth,.0\textwidth)   {(vi)};   

\draw[<-] (.05\textwidth,.13\textwidth)  -- (.07\textwidth,.13\textwidth);
\draw[<-] (.11\textwidth,.13\textwidth)  -- (.13\textwidth,.13\textwidth);
\draw[<-] (.17\textwidth,.13\textwidth)  -- (.19\textwidth,.13\textwidth);
\draw[<-] (.05\textwidth,.01\textwidth)  -- (.07\textwidth,.01\textwidth);
\draw[<-] (.11\textwidth,.01\textwidth)  -- (.13\textwidth,.01\textwidth);
\draw[<-] (.17\textwidth,.01\textwidth)  -- (.19\textwidth,.01\textwidth);

\draw[<-] (.30\textwidth,.13\textwidth)  -- (.32\textwidth,.13\textwidth);
\draw[<-] (.36\textwidth,.13\textwidth)  -- (.38\textwidth,.13\textwidth);
\draw[<-] (.42\textwidth,.13\textwidth)  -- (.44\textwidth,.13\textwidth);
\draw[<-] (.30\textwidth,.01\textwidth)  -- (.32\textwidth,.01\textwidth);
\draw[<-] (.36\textwidth,.01\textwidth)  -- (.38\textwidth,.01\textwidth);
\draw[<-] (.42\textwidth,.01\textwidth)  -- (.44\textwidth,.01\textwidth);

\draw[<-] (.55\textwidth,.13\textwidth)  -- (.57\textwidth,.13\textwidth);
\draw[<-] (.61\textwidth,.13\textwidth)  -- (.63\textwidth,.13\textwidth);
\draw[<-] (.67\textwidth,.13\textwidth)  -- (.69\textwidth,.13\textwidth);
\draw[<-] (.55\textwidth,.01\textwidth)  -- (.57\textwidth,.01\textwidth);
\draw[<-] (.61\textwidth,.01\textwidth)  -- (.63\textwidth,.01\textwidth);
\draw[<-] (.67\textwidth,.01\textwidth)  -- (.69\textwidth,.01\textwidth);

\begin{scope}[yshift=-.13\textwidth]
\draw[<-] (.05\textwidth,.13\textwidth)  -- (.07\textwidth,.13\textwidth);
\draw[<-] (.11\textwidth,.13\textwidth)  -- (.13\textwidth,.13\textwidth);
\draw[<-] (.17\textwidth,.13\textwidth)  -- (.19\textwidth,.13\textwidth);
\draw[<-] (.05\textwidth,.01\textwidth)  -- (.07\textwidth,.01\textwidth);
\draw[<-] (.11\textwidth,.01\textwidth)  -- (.13\textwidth,.01\textwidth);
\draw[<-] (.17\textwidth,.01\textwidth)  -- (.19\textwidth,.01\textwidth);

\draw[<-] (.30\textwidth,.13\textwidth)  -- (.32\textwidth,.13\textwidth);
\draw[<-] (.36\textwidth,.13\textwidth)  -- (.38\textwidth,.13\textwidth);
\draw[<-] (.42\textwidth,.13\textwidth)  -- (.44\textwidth,.13\textwidth);
\draw[<-] (.30\textwidth,.01\textwidth)  -- (.32\textwidth,.01\textwidth);
\draw[<-] (.36\textwidth,.01\textwidth)  -- (.38\textwidth,.01\textwidth);
\draw[<-] (.42\textwidth,.01\textwidth)  -- (.44\textwidth,.01\textwidth);

\draw[<-] (.55\textwidth,.13\textwidth)  -- (.57\textwidth,.13\textwidth);
\draw[<-] (.61\textwidth,.13\textwidth)  -- (.63\textwidth,.13\textwidth);
\draw[<-] (.67\textwidth,.13\textwidth)  -- (.69\textwidth,.13\textwidth);
\draw[<-] (.55\textwidth,.01\textwidth)  -- (.57\textwidth,.01\textwidth);
\draw[<-] (.61\textwidth,.01\textwidth)  -- (.63\textwidth,.01\textwidth);
\draw[<-] (.67\textwidth,.01\textwidth)  -- (.69\textwidth,.01\textwidth);
\end{scope}

\end{tikzpicture}
\begin{tikzpicture}
\begin{axis}[hide axis ,  scale only axis,    height=.080\textwidth,    width=.2cm,    colormap/jet,    colorbar,    point meta min=-1,    point meta max=1,
    colorbar style={  title=$\hat{p}_1$,        width=.2cm,        height=.200\textwidth,        ytick={-1,-.5,0,.5,1},        yticklabels={$\leq-0.150$,$-0.075$,$0.000$,$+0.075$,$\geq+0.150$},   yticklabel style={xshift=0.5ex} }
    ]
\end{axis}
\end{tikzpicture}

%% file: figures/CircularRigido_ReFinito_L_Re_d04.tex
%
%
\definecolor{mycolor1}{rgb}{0.00000,1.00000,1.00000}%
\definecolor{mycolor2}{rgb}{1.00000,0.00000,1.00000}%
\begin{tikzpicture}

\begin{axis}[%
width=0.32\textwidth,
height=0.32\textwidth,
at={(0\textwidth,0\textwidth)},
scale only axis,
every outer x axis line/.append style={black},
every x tick label/.append style={font=\color{black}},
xmin=0,
xmax=1,
xlabel={$\varepsilon/\varepsilon_*$},
every outer y axis line/.append style={black},
every y tick label/.append style={font=\color{black}},
ymax=+.55,
ymin=-1.05,
ylabel={$f/Re$},
axis background/.style={fill=white},
axis x line*=bottom,
axis y line*=left,
legend style={legend cell align=left,align=left,fill=white},
legend pos=south west,
axis lines = center,
x filter/.code={\pgfmathparse{#1/0.3000}\pgfmathresult},
axis on top,
minor xtick={ 0, 0.1, 0.2, 0.3, 0.4, 0.50, 0.6, 0.70, 0.8, .90,1},
xtick={ 0, 0.5, 1},
xticklabels={ $0$, $0.5$, $1$},
ytick={-1, -.5, 0, .5},
yticklabels={ $-1$, $-0.5$, $0$, $0.5$},
axis line style=-,
]

\addplot[solid,draw=mycolor1]
table[row sep=crcr] {%
x	y\\
2.85e-05	5.55487167013952e-05\\
0.001809571875	0.00352649535841561\\
0.00359064375	0.00699610533457076\\
0.005371715625	0.0104638570674693\\
0.0071527875	0.0139292289794136\\
0.00893385937500001	0.0173916994927062\\
0.01071493125	0.0208507470296494\\
0.012496003125	0.0243058500125457\\
0.014277075	0.0277564868636976\\
0.016058146875	0.0312021360054074\\
0.01783921875	0.0346422758599776\\
0.019620290625	0.0380763848497107\\
0.0214013625	0.0415039413969089\\
0.023182434375	0.0449244239238749\\
0.02496350625	0.048337310852911\\
0.026744578125	0.0517420806063197\\
0.02852565	0.0551382116064033\\
0.030306721875	0.0585251822754643\\
0.03208779375	0.0619024710358052\\
0.033868865625	0.0652695563097284\\
0.0356499375	0.0686259165195363\\
0.037431009375	0.0719710300875313\\
0.03921208125	0.0753043754360159\\
0.040993153125	0.0786254309872925\\
0.042774225	0.0819336751636635\\
0.044555296875	0.0852285863874314\\
0.04633636875	0.0885096430808986\\
0.048117440625	0.0917763236663676\\
0.0498985125	0.0950281065661406\\
0.051679584375	0.0982644702025203\\
0.05346065625	0.101484892997809\\
0.055241728125	0.104688853374309\\
0.0570228	0.107875829754323\\
0.058803871875	0.111045300560154\\
0.06058494375	0.114196744214103\\
0.062366015625	0.117329639138473\\
0.0641470875	0.120443463755567\\
0.065928159375	0.123537696487687\\
0.06770923125	0.126611815757135\\
0.069490303125	0.129665299986215\\
0.071271375	0.132697627597064\\
0.073052446875	0.135708179102399\\
0.07483351875	0.138695957224437\\
0.076614590625	0.141659873578015\\
0.0783956625	0.144598839777973\\
0.080176734375	0.147511767439146\\
0.08195780625	0.150397568176373\\
0.083738878125	0.15325515360449\\
0.08551995	0.156083435338336\\
0.087301021875	0.158881324992748\\
0.08908209375	0.161647734182563\\
0.090863165625	0.164381574522619\\
0.0926442375	0.167081757627754\\
0.094425309375	0.169747195112804\\
0.09620638125	0.172376798592608\\
0.097987453125	0.174969479682002\\
0.099768525	0.177524149995825\\
0.101549596875	0.180039721148914\\
0.10333066875	0.182515104756106\\
0.105111740625	0.184949212432239\\
0.1068928125	0.187340955791996\\
0.108673884375	0.189689087938239\\
0.11045495625	0.191991746665078\\
0.112236028125	0.194246920486428\\
0.1140171	0.196452597916201\\
0.115798171875	0.198606767468312\\
0.11757924375	0.200707417656673\\
0.119360315625	0.202752536995199\\
0.1211413875	0.204740113997802\\
0.122922459375	0.206668137178397\\
0.12470353125	0.208534595050896\\
0.126484603125	0.210337476129213\\
0.128265675	0.212074768927262\\
0.130046746875	0.213744461958956\\
0.13182781875	0.215344543738208\\
0.133608890625	0.216873002778933\\
0.1353899625	0.218327827595043\\
0.137171034375	0.219707006700452\\
0.13895210625	0.221008528609073\\
0.140733178125	0.22223038183482\\
0.14251425	0.223370554891019\\
0.144295321875	0.224425861385635\\
0.14607639375	0.225388526755806\\
0.147857465625	0.226249656600958\\
0.1496385375	0.227000356520515\\
0.151419609375	0.227631732113902\\
0.15320068125	0.228134888980544\\
0.154981753125	0.228500932719867\\
0.156762825	0.228720968931296\\
0.158543896875	0.228786103214256\\
0.16032496875	0.228687441168171\\
0.162106040625	0.228416088392467\\
0.1638871125	0.227963150486569\\
0.165668184375	0.227319733049902\\
0.16744925625	0.226476941681891\\
0.169230328125	0.225425881981962\\
0.1710114	0.224157659549539\\
0.172792471875	0.222663379984047\\
0.17457354375	0.220934148884911\\
0.176354615625	0.218961071851557\\
0.1781356875	0.21673525448377\\
0.179916759375	0.214249499049345\\
0.18169783125	0.211503273414194\\
0.183478903125	0.208497682113346\\
0.185259975	0.20523382968183\\
0.187041046875	0.201712820654675\\
0.18882211875	0.197935759566908\\
0.190603190625	0.19390375095356\\
0.1923842625	0.189617899349659\\
0.194165334375	0.185079309290233\\
0.19594640625	0.180289085310312\\
0.197727478125	0.175248331944925\\
0.19950855	0.169958153729099\\
0.201289621875	0.164419655197865\\
0.20307069375	0.15863394088625\\
0.204851765625	0.152602115329284\\
0.2066328375	0.146325283061996\\
0.208413909375	0.139804548619413\\
0.21019498125	0.133041016536566\\
0.211976053125	0.126035791348482\\
0.213757125	0.118789977590159\\
0.215538196875	0.111304155492529\\
0.21731926875	0.103576833088478\\
0.219100340625	0.0956060065414944\\
0.2208814125	0.0873896720150643\\
0.222662484375	0.078925825672675\\
0.22444355625	0.0702124636778139\\
0.226224628125	0.0612475821939687\\
0.2280057	0.0520291773846265\\
0.229786771875	0.0425552454132744\\
0.23156784375	0.0328237824433999\\
0.233348915625	0.0228327846384902\\
0.2351299875	0.0125802481620326\\
0.236911059375	0.00206416917751434\\
0.23869213125	-0.00871745615157703\\
0.240473203125	-0.0197666316617543\\
0.242254275	-0.0310853611895301\\
0.244035346875	-0.0426756485714174\\
0.24581641875	-0.0545394976439289\\
0.247597490625	-0.066678912243577\\
0.2493785625	-0.0790958962068746\\
0.251159634375	-0.0917924533703346\\
0.25294070625	-0.104770587570469\\
0.254721778125	-0.118032302643791\\
0.25650285	-0.131579602426814\\
0.258283921875	-0.145414490756049\\
0.26006499375	-0.15953897146801\\
0.261846065625	-0.173955048399209\\
0.2636271375	-0.18866472538616\\
0.265408209375	-0.203670006265374\\
0.26718928125	-0.218972894873365\\
0.268970353125	-0.234575395046645\\
0.270751425	-0.250479510621726\\
0.272532496875	-0.266687245435123\\
0.27431356875	-0.283200603323346\\
0.276094640625	-0.30002158812291\\
0.2778757125	-0.317152203670326\\
0.279656784375	-0.334594453802107\\
0.28143785625	-0.352350342354767\\
0.283218928125	-0.370421873164818\\
0.285	-0.388811050068771\\
nan	nan\\
};
\addlegendentry{$Re=128$};

\addplot[solid,draw=mycolor2]
table[row sep=crcr] {%
x	y\\
2.85e-05	7.556838311297e-05\\
0.001809571875	0.00480028770579546\\
0.00359064375	0.00952850513764449\\
0.005371715625	0.0142590763763785\\
0.0071527875	0.018990857119716\\
0.00893385937500001	0.0237227030653754\\
0.01071493125	0.0284534699110751\\
0.012496003125	0.0331820133545337\\
0.014277075	0.0379071890934695\\
0.016058146875	0.0426278528256011\\
0.01783921875	0.0473428602486469\\
0.019620290625	0.0520510670603254\\
0.0214013625	0.0567513289583549\\
0.023182434375	0.0614425016404541\\
0.02496350625	0.0661234408043413\\
0.026744578125	0.070793002147735\\
0.02852565	0.0754500413683536\\
0.030306721875	0.0800934141639156\\
0.03208779375	0.0847219762321396\\
0.033868865625	0.0893345832707438\\
0.0356499375	0.0939300909774468\\
0.037431009375	0.0985073550499671\\
0.03921208125	0.103065231186023\\
0.040993153125	0.107602575083333\\
0.042774225	0.112118242439616\\
0.044555296875	0.11661108895259\\
0.04633636875	0.121079970319973\\
0.048117440625	0.125523742239485\\
0.0498985125	0.129941260408842\\
0.051679584375	0.134331380525765\\
0.05346065625	0.138692958287971\\
0.055241728125	0.143024849393179\\
0.0570228	0.147325909539107\\
0.058803871875	0.151594994423474\\
0.06058494375	0.155830959743999\\
0.062366015625	0.160032661198398\\
0.0641470875	0.164198954484392\\
0.065928159375	0.168328695299699\\
0.06770923125	0.172420739342036\\
0.069490303125	0.176473942309124\\
0.071271375	0.180487159898255\\
0.073052446875	0.184458993801001\\
0.07483351875	0.188387065609726\\
0.076614590625	0.192268760557561\\
0.0783956625	0.196101463877641\\
0.080176734375	0.199882560803095\\
0.08195780625	0.203609436567057\\
0.083738878125	0.20727947640266\\
0.08551995	0.210890065543034\\
0.087301021875	0.214438589221312\\
0.08908209375	0.217922432670627\\
0.090863165625	0.221338981124111\\
0.0926442375	0.224685619814896\\
0.094425309375	0.227959733976114\\
0.09620638125	0.231158708840897\\
0.097987453125	0.234279929642378\\
0.099768525	0.237320781613689\\
0.101549596875	0.240278649987961\\
0.10333066875	0.243150919998328\\
0.105111740625	0.245934976877921\\
0.1068928125	0.248628205859685\\
0.108673884375	0.251227797975044\\
0.11045495625	0.253730190406972\\
0.112236028125	0.256131637447091\\
0.1140171	0.258428393387023\\
0.115798171875	0.260616712518394\\
0.11757924375	0.262692849132825\\
0.119360315625	0.26465305752194\\
0.1211413875	0.266493591977361\\
0.122922459375	0.268210706790714\\
0.12470353125	0.269800656253619\\
0.126484603125	0.271259694657701\\
0.128265675	0.272584076294583\\
0.130046746875	0.273770055455888\\
0.13182781875	0.27481388643324\\
0.133608890625	0.27571182351826\\
0.1353899625	0.276460121002574\\
0.137171034375	0.277055033177803\\
0.13895210625	0.277492814335571\\
0.140733178125	0.277769718767501\\
0.14251425	0.277882000764981\\
0.144295321875	0.277825442748266\\
0.14607639375	0.277593984414415\\
0.147857465625	0.277181115705891\\
0.1496385375	0.276580326565154\\
0.151419609375	0.275785106934667\\
0.15320068125	0.274788946756889\\
0.154981753125	0.273585335974283\\
0.156762825	0.27216776452931\\
0.158543896875	0.270529722364431\\
0.16032496875	0.268664699422107\\
0.162106040625	0.266566185644801\\
0.1638871125	0.264227670974972\\
0.165668184375	0.261642645355083\\
0.16744925625	0.258804598727595\\
0.169230328125	0.255707021034969\\
0.1710114	0.252343402219666\\
0.172792471875	0.248707232224148\\
0.17457354375	0.244792000990876\\
0.176354615625	0.240591198462312\\
0.1781356875	0.236098314581365\\
0.179916759375	0.231308953237683\\
0.18169783125	0.226227023256999\\
0.183478903125	0.220858472656758\\
0.185259975	0.215209249454404\\
0.187041046875	0.209285301667378\\
0.18882211875	0.203092577313126\\
0.190603190625	0.19663702440909\\
0.1923842625	0.189924590972714\\
0.194165334375	0.182961225021442\\
0.19594640625	0.175752874572716\\
0.197727478125	0.168305487643981\\
0.19950855	0.16062501225268\\
0.201289621875	0.152717396416257\\
0.20307069375	0.144588588152154\\
0.204851765625	0.136244535477816\\
0.2066328375	0.127691186410685\\
0.208413909375	0.118934488968206\\
0.21019498125	0.109980391167822\\
0.211976053125	0.100834841026976\\
0.213757125	0.091503786562969\\
0.215538196875	0.0819909138978008\\
0.21731926875	0.0722909695030271\\
0.219100340625	0.0623964915991695\\
0.2208814125	0.0523000184067497\\
0.222662484375	0.0419940881462886\\
0.22444355625	0.0314712390383077\\
0.226224628125	0.0207240093033287\\
0.2280057	0.00974493716187316\\
0.229786771875	-0.00147343916553791\\
0.23156784375	-0.0129385814583829\\
0.233348915625	-0.0246579514961404\\
0.2351299875	-0.0366390110582889\\
0.236911059375	-0.0488892219243073\\
0.23869213125	-0.0614160458736737\\
0.240473203125	-0.0742269446858668\\
0.242254275	-0.087329380140365\\
0.244035346875	-0.100730814016647\\
0.24581641875	-0.114438708094192\\
0.247597490625	-0.128460524152478\\
0.2493785625	-0.142803723970983\\
0.251159634375	-0.157475769329187\\
0.25294070625	-0.172484122006568\\
0.254721778125	-0.187836243782603\\
0.25650285	-0.203539596436773\\
0.258283921875	-0.219601641748555\\
0.26006499375	-0.236029841497427\\
0.261846065625	-0.25283165746287\\
0.2636271375	-0.270014551424361\\
0.265408209375	-0.287585985161378\\
0.26718928125	-0.305553420453401\\
0.268970353125	-0.323924319079908\\
0.270751425	-0.342706142820377\\
0.272532496875	-0.361906353454287\\
0.27431356875	-0.381532412761116\\
0.276094640625	-0.401591782520344\\
0.2778757125	-0.422091924511448\\
0.279656784375	-0.443040300513908\\
0.28143785625	-0.464444372307202\\
0.283218928125	-0.486311601670808\\
0.285	-0.508649450384204\\
nan	nan\\
};
\addlegendentry{$Re=64$};

\addplot[solid,draw=green]
table[row sep=crcr] {%
x	y\\
2.85e-05	8.81022456444041e-05\\
0.001809571875	0.00559639838691593\\
0.00359064375	0.0111082811006288\\
0.005371715625	0.0166218918953865\\
0.0071527875	0.0221353722797924\\
0.00893385937500001	0.02764686376245\\
0.01071493125	0.0331545078519627\\
0.012496003125	0.0386564460569341\\
0.014277075	0.0441508198859674\\
0.016058146875	0.0496357708476662\\
0.01783921875	0.0551094404506339\\
0.019620290625	0.0605699702034739\\
0.0214013625	0.0660155016147898\\
0.023182434375	0.0714441761931849\\
0.02496350625	0.0768541354472627\\
0.026744578125	0.0822435208856266\\
0.02852565	0.0876104740168801\\
0.030306721875	0.0929531363496266\\
0.03208779375	0.0982696493924696\\
0.033868865625	0.103558154654012\\
0.0356499375	0.108816793642859\\
0.037431009375	0.114043707867612\\
0.03921208125	0.119237038836875\\
0.040993153125	0.124394928059252\\
0.042774225	0.129515517043346\\
0.044555296875	0.134596947297761\\
0.04633636875	0.139637360331099\\
0.048117440625	0.144634897651965\\
0.0498985125	0.149587700768963\\
0.051679584375	0.154493911190694\\
0.05346065625	0.159351670425763\\
0.055241728125	0.164159119982774\\
0.0570228	0.168914401370329\\
0.058803871875	0.173615656097032\\
0.06058494375	0.178261025671487\\
0.062366015625	0.182848651602297\\
0.0641470875	0.187376675398065\\
0.065928159375	0.191843238567395\\
0.06770923125	0.196246482618891\\
0.069490303125	0.200584549061155\\
0.071271375	0.204855579402544\\
0.073052446875	0.20905756635963\\
0.07483351875	0.213187928525284\\
0.076614590625	0.217243946037581\\
0.0783956625	0.221222899034591\\
0.080176734375	0.225122067654389\\
0.08195780625	0.228938732035047\\
0.083738878125	0.232670172314639\\
0.08551995	0.236313668631237\\
0.087301021875	0.239866501122915\\
0.08908209375	0.243325949927744\\
0.090863165625	0.246689295183799\\
0.0926442375	0.249953817029152\\
0.094425309375	0.253116795601876\\
0.09620638125	0.256175511040045\\
0.097987453125	0.25912724348173\\
0.099768525	0.261969273065005\\
0.101549596875	0.264698879927944\\
0.10333066875	0.267313344208618\\
0.105111740625	0.269809946045101\\
0.1068928125	0.272185965575462\\
0.108673884375	0.274438678179197\\
0.11045495625	0.276565340764059\\
0.112236028125	0.278563205756363\\
0.1140171	0.280429525582425\\
0.115798171875	0.282161552668562\\
0.11757924375	0.283756539441091\\
0.119360315625	0.285211738326327\\
0.1211413875	0.286524401750588\\
0.122922459375	0.287691782140188\\
0.12470353125	0.288711131921446\\
0.126484603125	0.289579703520677\\
0.128265675	0.290294749364197\\
0.130046746875	0.290853521878323\\
0.13182781875	0.291253273489371\\
0.133608890625	0.291491256623658\\
0.1353899625	0.2915647237075\\
0.137171034375	0.291470927167213\\
0.13895210625	0.291207119429113\\
0.140733178125	0.290770552919518\\
0.14251425	0.290158480064561\\
0.144295321875	0.289367789612975\\
0.14607639375	0.288393950102047\\
0.147857465625	0.287232083437168\\
0.1496385375	0.285877311523732\\
0.151419609375	0.28432475626713\\
0.15320068125	0.282569539572755\\
0.154981753125	0.280606783346\\
0.156762825	0.278431609492257\\
0.158543896875	0.276039139916919\\
0.16032496875	0.273424496525378\\
0.162106040625	0.270582801223026\\
0.1638871125	0.267509175915256\\
0.165668184375	0.264198742507461\\
0.16744925625	0.260646622905033\\
0.169230328125	0.256847939013365\\
0.1710114	0.252797812737849\\
0.172792471875	0.248491365983877\\
0.17457354375	0.243923720656842\\
0.176354615625	0.239089998662136\\
0.1781356875	0.233985321905376\\
0.179916759375	0.228605864183007\\
0.18169783125	0.222951931792052\\
0.183478903125	0.217024845722583\\
0.185259975	0.210825926964674\\
0.187041046875	0.204356496508394\\
0.18882211875	0.197617875343817\\
0.190603190625	0.190611384461015\\
0.1923842625	0.183338344850059\\
0.194165334375	0.175800077501021\\
0.19594640625	0.167997903403975\\
0.197727478125	0.159933143548991\\
0.19950855	0.151607118926141\\
0.201289621875	0.143021150525499\\
0.20307069375	0.134176559337134\\
0.204851765625	0.125074666351121\\
0.2066328375	0.115716792557531\\
0.208413909375	0.106104258946435\\
0.21019498125	0.0962383865079061\\
0.211976053125	0.0861204962320161\\
0.213757125	0.0757519091086688\\
0.215538196875	0.0651312844045024\\
0.21731926875	0.0542467615024251\\
0.219100340625	0.043083881188871\\
0.2208814125	0.0316281842502737\\
0.222662484375	0.0198652114730667\\
0.22444355625	0.0077805036436835\\
0.226224628125	-0.00464039845144175\\
0.2280057	-0.0174119540258753\\
0.229786771875	-0.030548622293184\\
0.23156784375	-0.0440648624669339\\
0.233348915625	-0.0579751337606914\\
0.2351299875	-0.0722938953880226\\
0.236911059375	-0.0870356065624943\\
0.23869213125	-0.102214726497672\\
0.240473203125	-0.117845714407123\\
0.242254275	-0.133943029504412\\
0.244035346875	-0.150521131003108\\
0.24581641875	-0.167594478116774\\
0.247597490625	-0.185177530058979\\
0.2493785625	-0.203284746043289\\
0.251159634375	-0.221930585283269\\
0.25294070625	-0.241129506992486\\
0.254721778125	-0.260895970384507\\
0.25650285	-0.281244434672896\\
0.258283921875	-0.302189359071222\\
0.26006499375	-0.32374520279305\\
0.261846065625	-0.345926425051947\\
0.2636271375	-0.368747485061479\\
0.265408209375	-0.392222842035211\\
0.26718928125	-0.416366955186713\\
0.268970353125	-0.441194283729547\\
0.270751425	-0.466719286877281\\
0.272532496875	-0.492956423843482\\
0.27431356875	-0.519920153841716\\
0.276094640625	-0.547624936085549\\
0.2778757125	-0.576085229788547\\
0.279656784375	-0.605315494164276\\
0.28143785625	-0.635330188426306\\
0.283218928125	-0.666143771788198\\
0.285	-0.697770703463521\\
nan	nan\\
};
\addlegendentry{$Re=32$};

\addplot[solid,draw=red]
table[row sep=crcr] {%
x	y\\
2.85e-05	9.26916990190841e-05\\
0.001809571875	0.00588671326637759\\
0.00359064375	0.011682016554514\\
0.005371715625	0.0174764965973446\\
0.0071527875	0.0232680484287854\\
0.00893385937500001	0.0290545670827526\\
0.01071493125	0.0348339475931624\\
0.012496003125	0.0406040849939308\\
0.014277075	0.0463628743189742\\
0.016058146875	0.0521082106022086\\
0.01783921875	0.0578379888775502\\
0.019620290625	0.0635501041789151\\
0.0214013625	0.0692424515402194\\
0.023182434375	0.0749129259953795\\
0.02496350625	0.0805594225783113\\
0.026744578125	0.0861798363229311\\
0.02852565	0.091772062263155\\
0.030306721875	0.0973339954328992\\
0.03208779375	0.10286353086608\\
0.033868865625	0.108358563596613\\
0.0356499375	0.113816988658415\\
0.037431009375	0.119236701085402\\
0.03921208125	0.12461559591149\\
0.040993153125	0.129951568170595\\
0.042774225	0.135242512896633\\
0.044555296875	0.140486325123521\\
0.04633636875	0.145680899885174\\
0.048117440625	0.15082413221551\\
0.0498985125	0.155913917148443\\
0.051679584375	0.160948149717891\\
0.05346065625	0.165924724957768\\
0.055241728125	0.170841537901992\\
0.0570228	0.175696483584479\\
0.058803871875	0.180487457039145\\
0.06058494375	0.185212353299905\\
0.062366015625	0.189869067400677\\
0.0641470875	0.194455494375375\\
0.065928159375	0.198969529257917\\
0.06770923125	0.203409067082219\\
0.069490303125	0.207772002882196\\
0.071271375	0.212056231691672\\
0.073052446875	0.216259592408845\\
0.07483351875	0.220379707328609\\
0.076614590625	0.224414146510135\\
0.0783956625	0.228360480012593\\
0.080176734375	0.232216277895153\\
0.08195780625	0.235979110216986\\
0.083738878125	0.239646547037262\\
0.08551995	0.24321615841515\\
0.087301021875	0.246685514409823\\
0.08908209375	0.250052185080449\\
0.090863165625	0.253313740486199\\
0.0926442375	0.256467750686243\\
0.094425309375	0.259511785739752\\
0.09620638125	0.262443415705896\\
0.097987453125	0.265260210643845\\
0.099768525	0.26795974061277\\
0.101549596875	0.270539575671841\\
0.10333066875	0.272997285880227\\
0.105111740625	0.275330441297101\\
0.1068928125	0.27753661198166\\
0.108673884375	0.279613397462111\\
0.11045495625	0.281558511659001\\
0.112236028125	0.283369696245631\\
0.1140171	0.285044692895301\\
0.115798171875	0.286581243281309\\
0.11757924375	0.287977089076956\\
0.119360315625	0.289229971955541\\
0.1211413875	0.290337633590364\\
0.122922459375	0.291297815654725\\
0.12470353125	0.292108259821924\\
0.126484603125	0.292766707765259\\
0.128265675	0.293270901158032\\
0.130046746875	0.293618581673541\\
0.13182781875	0.293807490985087\\
0.133608890625	0.293835370765969\\
0.1353899625	0.293699962689487\\
0.137171034375	0.29339900842894\\
0.13895210625	0.292930249657628\\
0.140733178125	0.292291428048852\\
0.14251425	0.291480285275695\\
0.144295321875	0.290494132600413\\
0.14607639375	0.289328600470385\\
0.147857465625	0.287978909095457\\
0.1496385375	0.286440278685472\\
0.151419609375	0.284707929450275\\
0.15320068125	0.282777081599712\\
0.154981753125	0.280642955343626\\
0.156762825	0.278300770891863\\
0.158543896875	0.275745748454268\\
0.16032496875	0.272973108240684\\
0.162106040625	0.269978070460958\\
0.1638871125	0.266755855324934\\
0.165668184375	0.263301683042455\\
0.16744925625	0.259610773823368\\
0.169230328125	0.255678347877517\\
0.1710114	0.251499625414747\\
0.172792471875	0.247069826644902\\
0.17457354375	0.242384171777827\\
0.176354615625	0.237437881023367\\
0.1781356875	0.232226174591525\\
0.179916759375	0.226745020009477\\
0.18169783125	0.220993320744662\\
0.183478903125	0.214970701154483\\
0.185259975	0.208676785596342\\
0.187041046875	0.202111198427639\\
0.18882211875	0.195273564005777\\
0.190603190625	0.188163506688156\\
0.1923842625	0.18078065083218\\
0.194165334375	0.173124620795248\\
0.19594640625	0.165195040934763\\
0.197727478125	0.156991535608126\\
0.19950855	0.148513729172739\\
0.201289621875	0.139761245986003\\
0.20307069375	0.130733710405321\\
0.204851765625	0.121430746788093\\
0.2066328375	0.111851979491721\\
0.208413909375	0.101997032873607\\
0.21019498125	0.0918655312911515\\
0.211976053125	0.0814570991017574\\
0.213757125	0.0707713606625551\\
0.215538196875	0.0598036576232576\\
0.21731926875	0.0485324051616688\\
0.219100340625	0.0369318373190559\\
0.2208814125	0.0249761881366853\\
0.222662484375	0.0126396916558234\\
0.22444355625	-0.000103418082263329\\
0.226224628125	-0.0132789070363077\\
0.2280057	-0.0269125411650431\\
0.229786771875	-0.0410300864272035\\
0.23156784375	-0.0556573087815218\\
0.233348915625	-0.0708199741867315\\
0.2351299875	-0.086543848601566\\
0.236911059375	-0.102854697984759\\
0.23869213125	-0.119778288295043\\
0.240473203125	-0.137340385491152\\
0.242254275	-0.155566755531819\\
0.244035346875	-0.174483164375778\\
0.24581641875	-0.194115377981762\\
0.247597490625	-0.214489162308504\\
0.2493785625	-0.235630283314738\\
0.251159634375	-0.257564506959198\\
0.25294070625	-0.280317599200616\\
0.254721778125	-0.303915325997725\\
0.25650285	-0.328383453309259\\
0.258283921875	-0.353747747093953\\
0.26006499375	-0.380033973310538\\
0.261846065625	-0.407267897917749\\
0.2636271375	-0.435475286874318\\
0.265408209375	-0.464681906138979\\
0.26718928125	-0.494913521670467\\
0.268970353125	-0.526195899427513\\
0.270751425	-0.55855480536885\\
0.272532496875	-0.592016005453213\\
0.27431356875	-0.626605265639336\\
0.276094640625	-0.662348351885951\\
0.2778757125	-0.699271030151791\\
0.279656784375	-0.73739906639559\\
0.28143785625	-0.776758226576083\\
0.283218928125	-0.817374276652001\\
0.285	-0.859272982582077\\
nan	nan\\
};
\addlegendentry{$Re=16$};

\addplot[solid,draw=blue]
table[row sep=crcr] {%
x	y\\
2.85e-05	9.43076113432056e-05\\
0.001809571875	0.00598876150581535\\
0.00359064375	0.011883359387044\\
0.005371715625	0.0177759204897093\\
0.0071527875	0.0236642640484915\\
0.00893385937500001	0.0295462092980709\\
0.01071493125	0.0354195754731277\\
0.012496003125	0.041282181808342\\
0.014277075	0.0471318475383942\\
0.016058146875	0.0529663918979645\\
0.01783921875	0.058783634121733\\
0.019620290625	0.06458139344438\\
0.0214013625	0.0703574891005858\\
0.023182434375	0.0761097403250305\\
0.02496350625	0.0818359663523944\\
0.026744578125	0.0875339864173577\\
0.02852565	0.0932016197546006\\
0.030306721875	0.0988366855988034\\
0.03208779375	0.104437003184646\\
0.033868865625	0.11000039174681\\
0.0356499375	0.115524670519973\\
0.037431009375	0.121007658738818\\
0.03921208125	0.126447175638023\\
0.040993153125	0.13184104045227\\
0.042774225	0.137187072416238\\
0.044555296875	0.142483090764608\\
0.04633636875	0.14772691473206\\
0.048117440625	0.152916363553274\\
0.0498985125	0.15804925646293\\
0.051679584375	0.163123412695708\\
0.05346065625	0.16813665148629\\
0.055241728125	0.173086792069355\\
0.0570228	0.177971653679582\\
0.058803871875	0.182789055551654\\
0.06058494375	0.187536816920249\\
0.062366015625	0.192212757020048\\
0.0641470875	0.196814695085731\\
0.065928159375	0.201340450351978\\
0.06770923125	0.20578784205347\\
0.069490303125	0.210154689424887\\
0.071271375	0.214438811700882\\
0.073052446875	0.218638011758288\\
0.07483351875	0.222750029356125\\
0.076614590625	0.226772589032017\\
0.0783956625	0.230703415323588\\
0.080176734375	0.234540232768462\\
0.08195780625	0.238280765904263\\
0.083738878125	0.241922739268616\\
0.08551995	0.245463877399144\\
0.087301021875	0.248901904833471\\
0.08908209375	0.252234546109221\\
0.090863165625	0.25545952576402\\
0.0926442375	0.258574568335489\\
0.094425309375	0.261577398361254\\
0.09620638125	0.264465740378939\\
0.097987453125	0.267237318926167\\
0.099768525	0.269889858540563\\
0.101549596875	0.27242108375975\\
0.10333066875	0.274828719121353\\
0.105111740625	0.277110489162997\\
0.1068928125	0.279264118422318\\
0.108673884375	0.281287346507804\\
0.11045495625	0.283177971529725\\
0.112236028125	0.284933805791481\\
0.1140171	0.286552661596474\\
0.115798171875	0.288032351248104\\
0.11757924375	0.289370687049773\\
0.119360315625	0.290565481304881\\
0.1211413875	0.291614546316829\\
0.122922459375	0.29251569438902\\
0.12470353125	0.293266737824852\\
0.126484603125	0.293865488927729\\
0.128265675	0.29430976000105\\
0.130046746875	0.294597363348217\\
0.13182781875	0.29472611127263\\
0.133608890625	0.294693816077691\\
0.1353899625	0.2944982900668\\
0.137171034375	0.294137345543359\\
0.13895210625	0.293608794810769\\
0.140733178125	0.292910450172431\\
0.14251425	0.292040123931524\\
0.144295321875	0.290995185582272\\
0.14607639375	0.28977127538759\\
0.147857465625	0.288363611555833\\
0.1496385375	0.286767412295352\\
0.151419609375	0.284977895814502\\
0.15320068125	0.282990280321634\\
0.154981753125	0.280799784025103\\
0.156762825	0.27840162513326\\
0.158543896875	0.27579102185446\\
0.16032496875	0.272963192397055\\
0.162106040625	0.269913354969397\\
0.1638871125	0.266636727779841\\
0.165668184375	0.263128529036739\\
0.16744925625	0.259383976948444\\
0.169230328125	0.255398289723309\\
0.1710114	0.251166685569688\\
0.172792471875	0.246684382695932\\
0.17457354375	0.241946599310396\\
0.176354615625	0.236948553621432\\
0.1781356875	0.231685463837557\\
0.179916759375	0.226153318480518\\
0.18169783125	0.22035113235557\\
0.183478903125	0.214278663340783\\
0.185259975	0.207935669314227\\
0.187041046875	0.201321908153973\\
0.18882211875	0.194437137738091\\
0.190603190625	0.18728111594465\\
0.1923842625	0.179853600651722\\
0.194165334375	0.172154349737377\\
0.19594640625	0.164183121079684\\
0.197727478125	0.155939672556714\\
0.19950855	0.147423762046537\\
0.201289621875	0.138635147427224\\
0.20307069375	0.129573586576844\\
0.204851765625	0.120238837373468\\
0.2066328375	0.110630657695166\\
0.208413909375	0.100748805420009\\
0.21019498125	0.0905930384260653\\
0.211976053125	0.0801631145914069\\
0.213757125	0.0694587917937528\\
0.215538196875	0.0584742828284942\\
0.21731926875	0.0471818847561558\\
0.219100340625	0.035548481064944\\
0.2208814125	0.0235409552430646\\
0.222662484375	0.0111261907787231\\
0.22444355625	-0.00172892883987479\\
0.226224628125	-0.0150575201245226\\
0.2280057	-0.0288926995870146\\
0.229786771875	-0.0432675837391453\\
0.23156784375	-0.0582152890927087\\
0.233348915625	-0.0737689321594989\\
0.2351299875	-0.08996162945131\\
0.236911059375	-0.106826497479937\\
0.23869213125	-0.124396652757172\\
0.240473203125	-0.142705211794811\\
0.242254275	-0.161785291104647\\
0.244035346875	-0.181670007198475\\
0.24581641875	-0.202392476588089\\
0.247597490625	-0.223985815785283\\
0.2493785625	-0.246483141301851\\
0.251159634375	-0.269917569649587\\
0.25294070625	-0.294322217340286\\
0.254721778125	-0.319730200885741\\
0.25650285	-0.346174636797747\\
0.258283921875	-0.373688641588097\\
0.26006499375	-0.402305331768587\\
0.261846065625	-0.43205782385101\\
0.2636271375	-0.46297923434716\\
0.265408209375	-0.495102679768831\\
0.26718928125	-0.528461276627819\\
0.268970353125	-0.563088141435916\\
0.270751425	-0.599016390704916\\
0.272532496875	-0.636279140946615\\
0.27431356875	-0.674909508672805\\
0.276094640625	-0.714940610395283\\
0.2778757125	-0.75640556262584\\
0.279656784375	-0.799337481876271\\
0.28143785625	-0.843769484658374\\
0.283218928125	-0.889734687483937\\
0.285	-0.937266206864757\\
nan	nan\\
};
\addlegendentry{$Re=8$};

\addplot[solid,draw=black]
table[row sep=crcr] {%
x	y\\
2.85e-05	9.45942251547604e-05\\
0.001809571875	0.00600717532820987\\
0.00359064375	0.0119202889650147\\
0.005371715625	0.0178317008970399\\
0.0071527875	0.0237391768857559\\
0.00893385937500001	0.0296404826926334\\
0.01071493125	0.0355333840791429\\
0.012496003125	0.041415646806755\\
0.014277075	0.0472850366369403\\
0.016058146875	0.0531393193311693\\
0.01783921875	0.0589762606509126\\
0.019620290625	0.0647936263576408\\
0.0214013625	0.0705891822128245\\
0.023182434375	0.0763606939779342\\
0.02496350625	0.0821059274144404\\
0.026744578125	0.0878226482838139\\
0.02852565	0.0935086223475251\\
0.030306721875	0.0991616153670446\\
0.03208779375	0.104779393103843\\
0.033868865625	0.110359721319391\\
0.0356499375	0.115900365775159\\
0.037431009375	0.121399092232617\\
0.03921208125	0.126853666453237\\
0.040993153125	0.132261854198488\\
0.042774225	0.137621421229842\\
0.044555296875	0.142930133308768\\
0.04633636875	0.148185756196738\\
0.048117440625	0.153386055655222\\
0.0498985125	0.158528797445691\\
0.051679584375	0.163611747329615\\
0.05346065625	0.168632671068464\\
0.055241728125	0.17358933442371\\
0.0570228	0.178479503156823\\
0.058803871875	0.183300943029273\\
0.06058494375	0.188051419802531\\
0.062366015625	0.192728699238068\\
0.0641470875	0.197330547097354\\
0.065928159375	0.201854729141859\\
0.06770923125	0.206299011133055\\
0.069490303125	0.210661158832412\\
0.071271375	0.21493893800144\\
0.073052446875	0.219130137800838\\
0.07483351875	0.223232637678759\\
0.076614590625	0.227244338856936\\
0.0783956625	0.231163142557104\\
0.080176734375	0.234986950000996\\
0.08195780625	0.238713662410347\\
0.083738878125	0.242341181006891\\
0.08551995	0.245867407012363\\
0.087301021875	0.249290241648496\\
0.08908209375	0.252607586137025\\
0.090863165625	0.255817341699684\\
0.0926442375	0.258917409558207\\
0.094425309375	0.261905690934329\\
0.09620638125	0.264780087049784\\
0.097987453125	0.267538499126305\\
0.099768525	0.270178828385628\\
0.101549596875	0.272698976049486\\
0.10333066875	0.275096843339613\\
0.105111740625	0.277370331477745\\
0.1068928125	0.279517341685559\\
0.108673884375	0.281535718355418\\
0.11045495625	0.28342308528051\\
0.112236028125	0.285177012734411\\
0.1140171	0.286795070990695\\
0.115798171875	0.288274830322937\\
0.11757924375	0.289613861004711\\
0.119360315625	0.290809733309591\\
0.1211413875	0.291860017511152\\
0.122922459375	0.292762283882969\\
0.12470353125	0.293514102698615\\
0.126484603125	0.294113044231665\\
0.128265675	0.294556678755695\\
0.130046746875	0.294842576544277\\
0.13182781875	0.294968307870987\\
0.133608890625	0.294931443009399\\
0.1353899625	0.294729552233087\\
0.137171034375	0.294360205815627\\
0.13895210625	0.293820974030592\\
0.140733178125	0.293109427151557\\
0.14251425	0.292223135451916\\
0.144295321875	0.291159309501829\\
0.14607639375	0.289913755179684\\
0.147857465625	0.288481935519876\\
0.1496385375	0.286859313556797\\
0.151419609375	0.28504135232484\\
0.15320068125	0.2830235148584\\
0.154981753125	0.280801264191868\\
0.156762825	0.27837006335964\\
0.158543896875	0.275725375396107\\
0.16032496875	0.272862663335664\\
0.162106040625	0.269777390212703\\
0.1638871125	0.266465019061619\\
0.165668184375	0.262921012916803\\
0.16744925625	0.25914083481265\\
0.169230328125	0.255119947783553\\
0.1710114	0.250853814863906\\
0.172792471875	0.2463378990881\\
0.17457354375	0.241567663490531\\
0.176354615625	0.236538571105591\\
0.1781356875	0.231246084967835\\
0.179916759375	0.225686429561529\\
0.18169783125	0.219858820832862\\
0.183478903125	0.213763209250758\\
0.185259975	0.207399545284141\\
0.187041046875	0.200767779401935\\
0.18882211875	0.193867862073064\\
0.190603190625	0.186699743766452\\
0.1923842625	0.179263374951022\\
0.194165334375	0.171558706095697\\
0.19594640625	0.163585687669402\\
0.197727478125	0.155344270141061\\
0.19950855	0.146834403979598\\
0.201289621875	0.138056039653935\\
0.20307069375	0.129009127632997\\
0.204851765625	0.119693618385708\\
0.2066328375	0.110109462380991\\
0.208413909375	0.100256610087771\\
0.21019498125	0.0901350119749701\\
0.211976053125	0.0797446185115133\\
0.213757125	0.0690853801659386\\
0.215538196875	0.0581511520559816\\
0.21731926875	0.0469116987477506\\
0.219100340625	0.0353308340170097\\
0.2208814125	0.0233723716395227\\
0.222662484375	0.0110001253910527\\
0.22444355625	-0.00182209095263638\\
0.226224628125	-0.0151304636157806\\
0.2280057	-0.0289611788226161\\
0.229786771875	-0.0433504227973798\\
0.23156784375	-0.0583343817643076\\
0.233348915625	-0.073949241947636\\
0.2351299875	-0.0902311895716011\\
0.236911059375	-0.10721641086044\\
0.23869213125	-0.124941092038387\\
0.240473203125	-0.143441419329681\\
0.242254275	-0.162753578958556\\
0.244035346875	-0.18291375714925\\
0.24581641875	-0.203958140125998\\
0.247597490625	-0.225922914113037\\
0.2493785625	-0.248844265334604\\
0.251159634375	-0.272758380014935\\
0.25294070625	-0.297701444378265\\
0.254721778125	-0.323709644648831\\
0.25650285	-0.350819167050869\\
0.258283921875	-0.379066197808616\\
0.26006499375	-0.408486923146309\\
0.261846065625	-0.439117529288183\\
0.2636271375	-0.470994202458474\\
0.265408209375	-0.504153128881419\\
0.26718928125	-0.538630494781256\\
0.268970353125	-0.574462486382219\\
0.270751425	-0.611685289908543\\
0.272532496875	-0.650335091584468\\
0.27431356875	-0.690448077634228\\
0.276094640625	-0.732060434282059\\
0.2778757125	-0.775208347752199\\
0.279656784375	-0.819928004268883\\
0.28143785625	-0.86625559005635\\
0.283218928125	-0.914227291338832\\
0.285	-0.963879294340567\\
nan	nan\\
};
\addlegendentry{Linear $Re$};

\end{axis}
\end{tikzpicture}%

%% file: figures/pressurefieldCa.tex
\begin{tikzpicture}[scale=1.08,]
\node[inner sep=0pt,anchor=north west] (fig1) at (.00\textwidth,.12\textwidth)   {\includegraphics[width=.2592\textwidth]{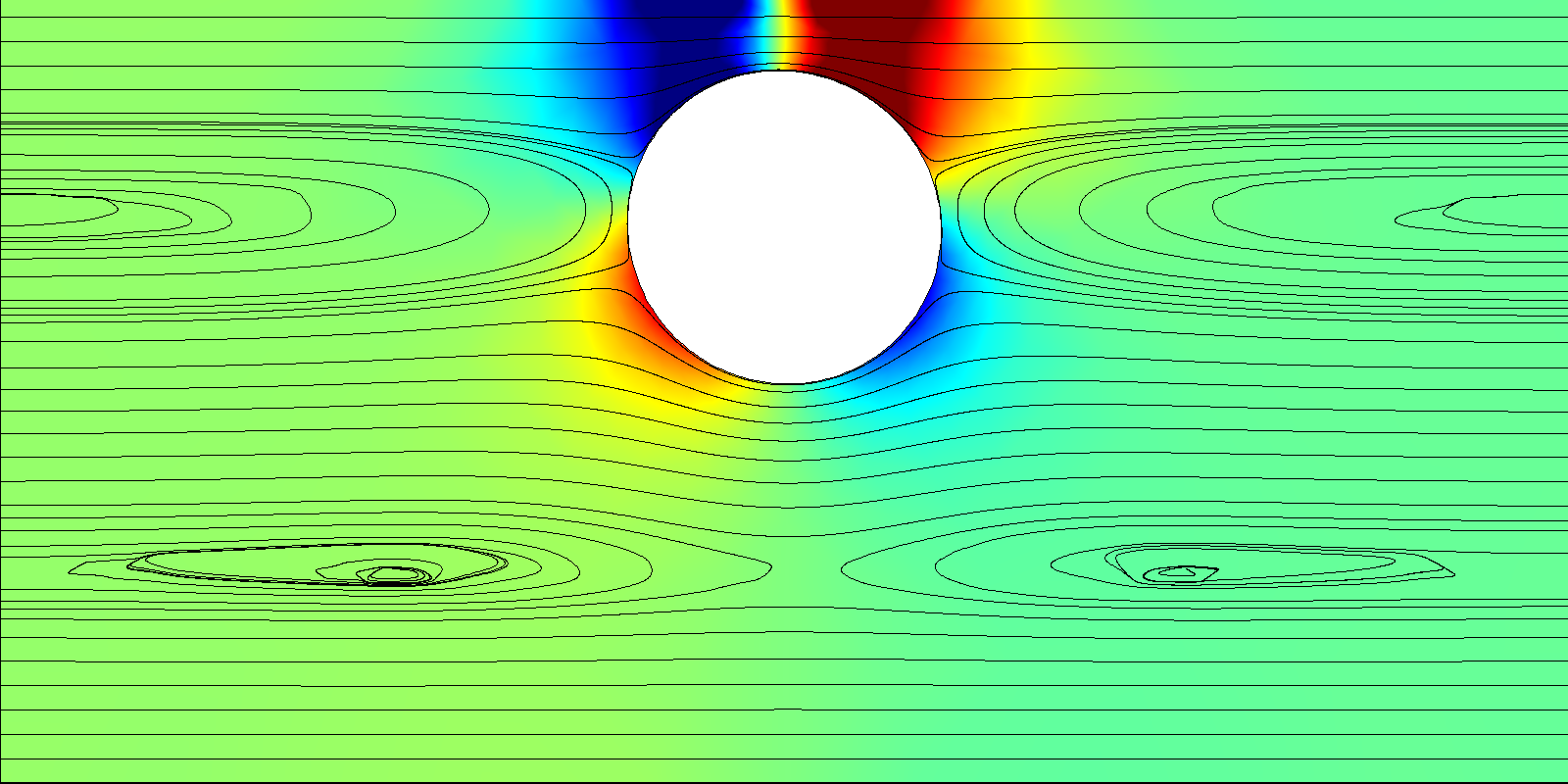}};
\node[inner sep=0pt,anchor=north west] (fig2) at (.25\textwidth,.12\textwidth)   {\includegraphics[width=.2592\textwidth]{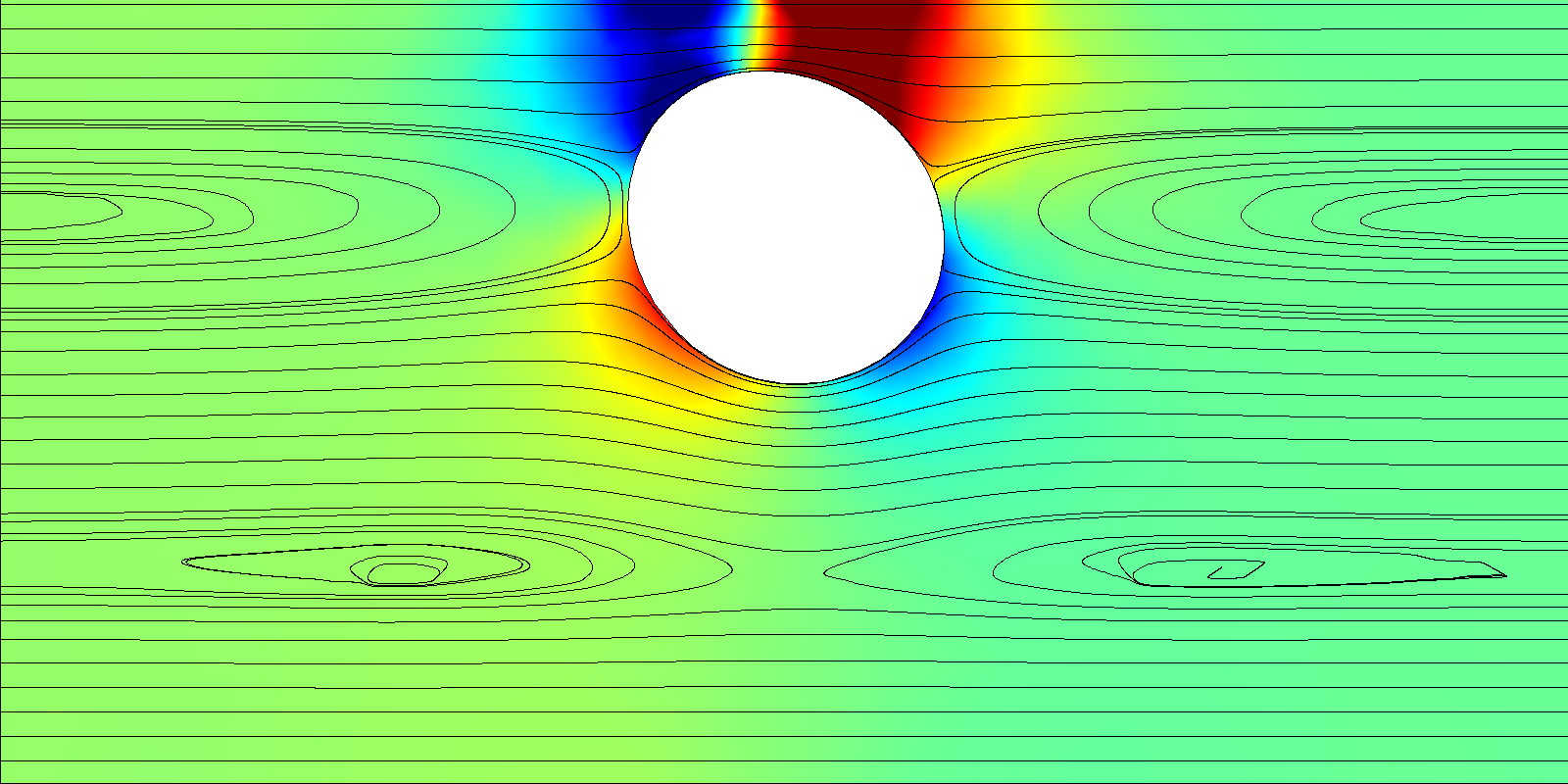}};
\node[inner sep=0pt,anchor=north west] (fig3) at (.50\textwidth,.12\textwidth)   {\includegraphics[width=.2592\textwidth]{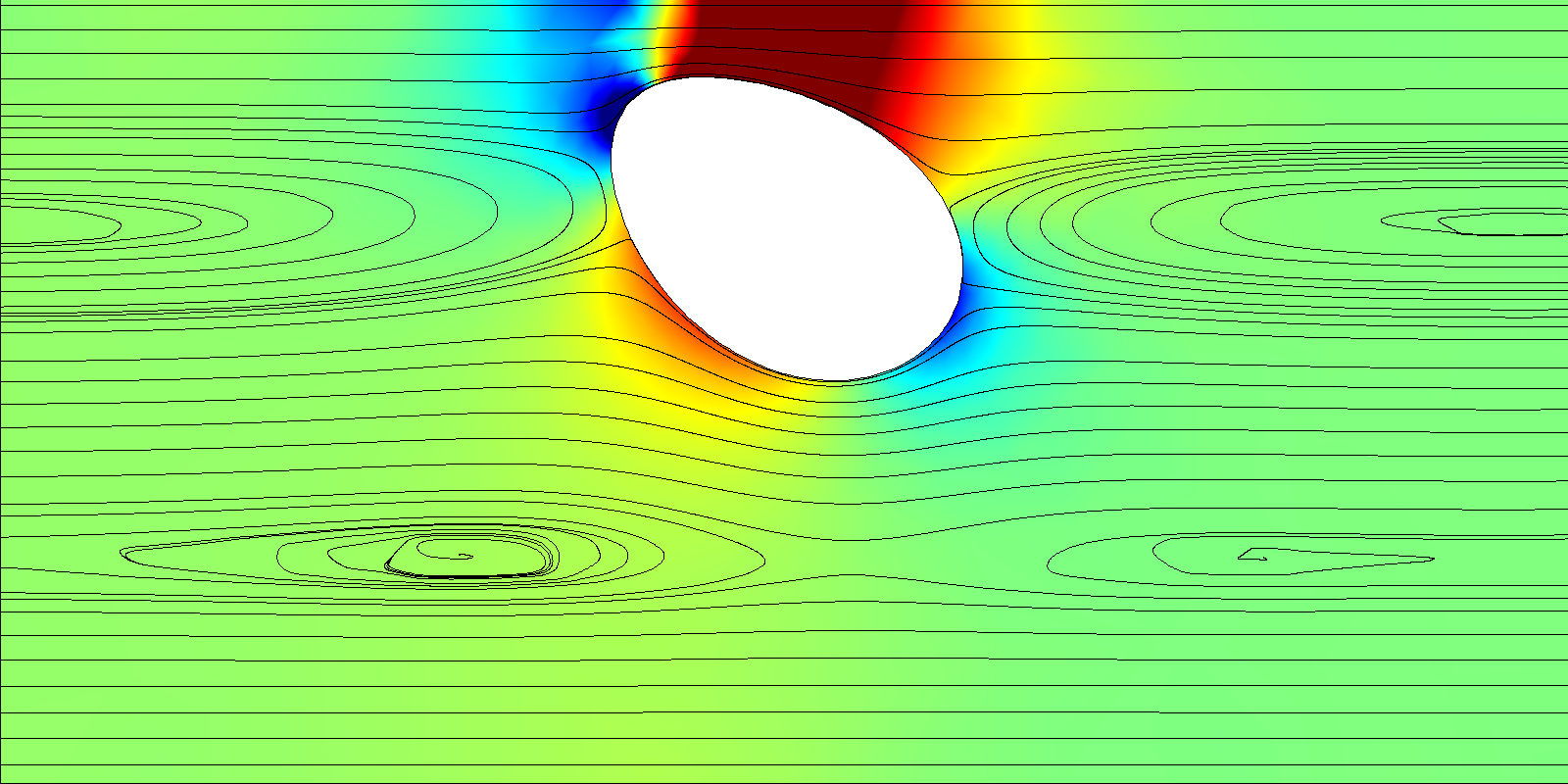}};

\draw[dashdot] (-.01\textwidth,.06\textwidth) -- (.75\textwidth,.06\textwidth);
\node[inner sep=0pt,anchor=north west] (fig1) at (.00\textwidth,.12\textwidth)   {(a)}; 
\node[inner sep=0pt,anchor=north west] (fig2) at (.25\textwidth,.12\textwidth)   {(b)}; 
\node[inner sep=0pt,anchor=north west] (fig3) at (.50\textwidth,.12\textwidth)   {(c)}; 

\draw[<-] (.05\textwidth,.12\textwidth)  -- (.07\textwidth,.12\textwidth);
\draw[<-] (.11\textwidth,.12\textwidth)  -- (.13\textwidth,.12\textwidth);
\draw[<-] (.17\textwidth,.12\textwidth)  -- (.19\textwidth,.12\textwidth);
\draw[<-] (.05\textwidth,.00\textwidth)  -- (.07\textwidth,.00\textwidth);
\draw[<-] (.11\textwidth,.00\textwidth)  -- (.13\textwidth,.00\textwidth);
\draw[<-] (.17\textwidth,.00\textwidth)  -- (.19\textwidth,.00\textwidth);

\draw[<-] (.30\textwidth,.12\textwidth)  -- (.32\textwidth,.12\textwidth);
\draw[<-] (.36\textwidth,.12\textwidth)  -- (.38\textwidth,.12\textwidth);
\draw[<-] (.42\textwidth,.12\textwidth)  -- (.44\textwidth,.12\textwidth);
\draw[<-] (.30\textwidth,.00\textwidth)  -- (.32\textwidth,.00\textwidth);
\draw[<-] (.36\textwidth,.00\textwidth)  -- (.38\textwidth,.00\textwidth);
\draw[<-] (.42\textwidth,.00\textwidth)  -- (.44\textwidth,.00\textwidth);

\draw[<-] (.55\textwidth,.12\textwidth)  -- (.57\textwidth,.12\textwidth);
\draw[<-] (.61\textwidth,.12\textwidth)  -- (.63\textwidth,.12\textwidth);
\draw[<-] (.67\textwidth,.12\textwidth)  -- (.69\textwidth,.12\textwidth);
\draw[<-] (.55\textwidth,.00\textwidth)  -- (.57\textwidth,.00\textwidth);
\draw[<-] (.61\textwidth,.00\textwidth)  -- (.63\textwidth,.00\textwidth);
\draw[<-] (.67\textwidth,.00\textwidth)  -- (.69\textwidth,.00\textwidth);
\end{tikzpicture}
\begin{tikzpicture}
\hspace{-.05\textwidth}
\begin{axis}[hide axis ,  scale only axis,    height=.100\textwidth,    width=.2cm,    colormap/jet,    colorbar,    point meta min=-1,    point meta max=1,
    colorbar style={  title=$\hat{p} - x \partial_x p_P$,        width=.2cm,        height=.100\textwidth,        ytick={-1,-.5,0,.5,1},        yticklabels={$\leq-5$,$-2.5$,$0$,$+2.5$,$\geq+5$},   yticklabel style={xshift=0.5ex} }
    ]
\end{axis}
\end{tikzpicture}

%% file: figures/Volume2.tex
\begin{tikzpicture}[x=.7\textwidth,y=.7\textwidth]

\definecolor{mycolor1}{rgb}{0,0,1}%
\definecolor{mycolor2}{rgb}{0,1,0}%

	\draw[line width=.1,fill=mycolor1!40,draw=mycolor1!40,opacity=.5] (-.5,-.3) rectangle (.5,.3)
	plot [domain = 0:360, samples = 80, variable = \i] ({.15*(1+.1*cos (45+2*\i))*cos \i}, {.1+.15*(1+.1*cos (45+2*\i))*sin \i}, 0) ;

	\draw[line width=.1,fill=mycolor2!40,draw=mycolor2!40,opacity=.5]  
	(-.5,-.3) rectangle (.5,.3)
	plot [domain = 0:360, samples = 80, variable = \i]  (.15*cos \i, .1+.15*sin \i, 0);

	\draw[line width=.2] plot [domain = 0:360, samples = 80, variable = \i] 
                   ({.15*(1+.1*cos (45+2*\i))*cos \i}, {.1+.15*(1+.1*cos (45+2*\i))*sin \i}, 0) ;
	
	\draw[line width=.2,dashed]  plot [domain = 0:360, samples = 80, variable = \i] 
                    (.15*cos \i, .1+.15*sin \i, 0);

	\draw[line width=.2,fill=mycolor1!40,draw=mycolor1!40,dashed,opacity=.5]  	(.51,-.02) rectangle (.65,.08);
	\node at (.58,.03) {$  \mathcal{V}$} ; 

	\draw[line width=.2,dashed] (.51,-.13) rectangle (.65,-.03);
	\node at (.58,-.08) {$  \Sigma_B$} ; 

	\draw[line width=.2,fill=mycolor2!40,draw=mycolor2!40,dashed,opacity=.5]  	(.51,.2) rectangle (.65,.3);
	\node at (.58,.25) {$  \mathcal{V}_0$} ; 

	\draw[line width=.2] (.51,.09) rectangle (.65,.19);
	\node at (.58,.14) {$  \Sigma_{B_0}$} ; 

\node[anchor=south east] at (-.5,.3) {(a)};
                    
\end{tikzpicture}

%% file: figures/SketchSurfPert.tex
\begin{tikzpicture}[x=.8\textwidth,y=.8\textwidth]

	\draw[line width=.1] plot [domain = 70:110, samples = 80, variable = \i] 
                    ({(1+.4*\i/100)*cos \i}, {(1+.4*\i/100)*sin \i}, 0) ;
	\draw[line width=.1,dashed]  plot [domain = 70:110, samples = 80, variable = \i] 
                    (1*cos \i, 1*sin \i, 0);

	\draw[blue,line width=1]  plot [domain = 80:100, samples = 40, variable = \i] 
                    ({(1+.4*\i/100)*cos \i}, {(1+.4*\i/100)*sin \i}, 0) ;
	\draw[blue,line width=1]  plot [domain = 80:100, samples = 40, variable = \i] 
                    (1*cos \i, 1*sin \i, 0) ;
         \draw[line width=1,->] ({cos 100},{sin 100})  -- ({1.4*cos 100},{1.4*sin 100}) node[above] {$ \dn \boldsymbol{n}_{0} $}  ;         
         \draw[line width=1,->] ({cos 80},{sin 80})  -- ({1.32*cos 80},{1.32*sin 80})   node[above] {$ \dn \boldsymbol{n}_{0} $}   ;     
          
          \draw[line width=.5,->] ({cos 90},{sin 90})  -- ({.9*cos 90},{.9*sin 90})   node[right] {$- \dd \Sigma_{B_0} \boldsymbol{n}_0 $}   ;   
          \draw[line width=.5,->] ({1.36*cos 90},{1.36*sin 90})  -- ({1.46*cos 89.5},{1.46*sin 89.5})   node[right] {$\dd \Sigma_B \boldsymbol{n} $}    ;  
          
          \draw[line width=.5,->] ({1.16*cos 80},{1.16*sin 80})  --++ ({.1*sin 80},{-.1*cos 80})   node[right] {$ \dn \boldsymbol{n}_{S0} $}   ;  
          \draw[line width=.5,->] ({1.2*cos 100},{1.2*sin 100})  --++ ({-.1*sin 100},{.1*cos 100})   node[left] {$ \dn \boldsymbol{n}_{S0} $}   ;

          \draw[line width=.5,->] ({1.*cos 80},{1.*sin 80})  --++ ({.1*sin 80},{-.1*cos 80})   node[above] {$  \boldsymbol{n}_{S0}$}   ;  
          \draw[line width=.5,->] ({1.*cos 100},{1.*sin 100})  --++ ({-.1*sin 100},{.1*cos 100})   node[above] {$ \boldsymbol{n}_{S0} $}   ;
                    
            
         \node at  (1.2*cos 90, 1.2*sin 90)   {$\dd \Sigma_{B_0} \delta $};   
            
           \node at  (0.9*cos 98, 0.9*sin 98)   {$\mathcal{V}_0$};

              \node[anchor=south east] at (-.45,1.35) {(b)};      
                    
\end{tikzpicture}

%% file: figures/PureNonLinearCa_d04_f.tex
%
%
\definecolor{mycolor1}{rgb}{1.00000,0.00000,1.00000}%
\begin{tikzpicture}[baseline]

\begin{axis}[%
width=.25\textwidth,
height=.25\textwidth,
scale only axis,
every outer x axis line/.append style={black},
every x tick label/.append style={font=\color{black}},
xmin=-.05,
xmax=1,
every outer y axis line/.append style={black},
every y tick label/.append style={font=\color{black}},
ymin=-500,
ymax=80,
axis background/.style={fill=white},
axis x line*=bottom,
axis y line*=left,
legend style={xshift=5pt,legend cell align=left,align=left,draw=black},
legend pos=south west,
axis lines = center,
xlabel=$\varepsilon/\varepsilon_*$,
ylabel=$f/Ca$,
minor xtick={ 0, 0.1, 0.2, 0.3, 0.4, 0.50, 0.6, 0.70, 0.8, .90,1},
xtick={ 0, 0.5, 1},
xticklabels={ $0$, $0.5$, $1$},
x tick label style={anchor=south,yshift=5pt},
x label style={anchor=north,yshift=-3pt},
axis line style=-,
]
\node[anchor=south east] at (rel axis cs:1,0) {(a)};
\addplot [color=mycolor1,solid]
  table[row sep=crcr]{%
0	0.000278123736784063\\
0.018	-1.00669536420998\\
0.036	-2.02144579296324\\
0.054	-3.04373843852411\\
0.072	-4.07132972752806\\
0.09	-5.10893184502355\\
0.108	-6.16242840523912\\
0.126	-7.22920699051957\\
0.144	-8.31549465240713\\
0.162	-9.42656524117554\\
0.18	-10.5605964854404\\
0.198	-11.72511064387\\
0.216	-12.9234442315238\\
0.234	-14.1561742169103\\
0.252	-15.4311245012237\\
0.27	-16.7497077726278\\
0.288	-18.1160449955549\\
0.306	-19.5381812553472\\
0.324	-21.0175201130506\\
0.342	-22.5612430901288\\
0.36	-24.1775274257167\\
0.378	-25.8696037521671\\
0.396	-27.6469896686549\\
0.414	-29.5183878651026\\
0.432	-31.4903599640856\\
0.45	-33.5731440481004\\
0.468	-35.7761685239342\\
0.486	-38.1106185904612\\
0.504	-40.5895564398356\\
0.522	-43.2274576373122\\
0.54	-46.0384461757772\\
0.558	-49.0369815273305\\
0.576	-52.2444122634932\\
0.594	-55.6789024475807\\
0.612	-59.3608248393231\\
0.63	-63.3209596454557\\
0.648	-67.583008686993\\
0.666	-72.182639301968\\
0.684	-77.1597253447572\\
0.702	-82.5338779215603\\
0.72	-88.3614059787851\\
0.738	-94.6869636786671\\
0.756	-101.535538942995\\
0.774	-108.99153716767\\
0.792	-117.108352338371\\
0.81	-125.941811302578\\
0.828	-135.618655102401\\
0.846	-146.178285353164\\
0.864	-157.741503573223\\
0.882	-170.458705900067\\
0.9	-184.202749972446\\
};
\addlegendentry{$Ca=1/4$};

\addplot [color=green,solid]
  table[row sep=crcr]{%
0	0.000276911076841414\\
0.018	-1.01708756613552\\
0.036	-2.04246360956007\\
0.054	-3.07560817593774\\
0.072	-4.11421829932161\\
0.09	-5.16313734253607\\
0.108	-6.22839661430188\\
0.126	-7.30727405619399\\
0.144	-8.40633558870418\\
0.162	-9.53110745629875\\
0.18	-10.6796803795111\\
0.198	-11.8600970027616\\
0.216	-13.075852307857\\
0.234	-14.3276544362684\\
0.252	-15.6241874326751\\
0.27	-16.9669111982388\\
0.288	-18.3605548607659\\
0.306	-19.8143006246743\\
0.324	-21.329507350307\\
0.342	-22.9149409520805\\
0.36	-24.5804506879095\\
0.378	-26.32959990867\\
0.396	-28.1747394170983\\
0.414	-30.1266190801984\\
0.432	-32.1933959346094\\
0.45	-34.390091007993\\
0.468	-36.7287074147851\\
0.486	-39.2249918302359\\
0.504	-41.8996881605484\\
0.522	-44.7714068045531\\
0.54	-47.8647198388688\\
0.558	-51.2065888829516\\
0.576	-54.82816262099\\
0.594	-58.7661731617621\\
0.612	-63.0591452074731\\
0.63	-67.7601247136544\\
0.648	-72.9306137891552\\
0.666	-78.6426865911606\\
0.684	-84.9872734326402\\
0.702	-92.049211512195\\
0.72	-99.9666326314428\\
0.738	-108.882408836488\\
0.756	-118.921132733826\\
0.774	-130.342413332817\\
0.792	-143.387153886763\\
0.81	-158.341045523791\\
0.828	-175.755102604175\\
0.846	-195.852365686127\\
0.864	-219.333048745709\\
0.882	-247.041678665306\\
0.9	-278.257150604585\\
};
\addlegendentry{$Ca=1/8$};

\addplot [color=red,solid]
  table[row sep=crcr]{%
0	0.000272210156019732\\
0.018	-1.01969755428455\\
0.036	-2.04773728286275\\
0.054	-3.08360204315385\\
0.072	-4.12497701627446\\
0.09	-5.1767324436907\\
0.108	-6.24492831521742\\
0.126	-7.32681626177425\\
0.144	-8.4290471112676\\
0.162	-9.55720922247863\\
0.18	-10.7093710897134\\
0.198	-11.8937059106378\\
0.216	-13.1137455237041\\
0.234	-14.3702325564362\\
0.252	-15.6720766065524\\
0.27	-17.0207428010617\\
0.288	-18.4211272591976\\
0.306	-19.8827275612077\\
0.324	-21.4069007439521\\
0.342	-23.0028074549377\\
0.36	-24.6807048378713\\
0.378	-26.4442428455662\\
0.396	-28.3066035769356\\
0.414	-30.2791385177244\\
0.432	-32.3704480587459\\
0.45	-34.5968962859409\\
0.468	-36.9711242687529\\
0.486	-39.5103607989232\\
0.504	-42.237920220291\\
0.522	-45.1736541134788\\
0.54	-48.3456593339979\\
0.558	-51.7850505646247\\
0.576	-55.525968455546\\
0.594	-59.6132288899935\\
0.612	-64.0925934675233\\
0.63	-69.0256766293306\\
0.648	-74.4934915122442\\
0.666	-80.583977441291\\
0.684	-87.4147882047122\\
0.702	-95.1124876217668\\
0.72	-103.845970089057\\
0.738	-113.848999393378\\
0.756	-125.362163575446\\
0.774	-138.771485137399\\
0.792	-154.567539894863\\
0.81	-173.305314001273\\
0.828	-196.016550794638\\
0.846	-223.734077664087\\
0.864	-258.973865982705\\
0.882	-304.278177033397\\
0.9	-357.342747632845\\
};
\addlegendentry{$Ca=1/16$};

\addplot [color=blue,solid]
  table[row sep=crcr]{%
0	0.000262659225016127\\
0.018	-1.02034174044331\\
0.036	-2.04903723209389\\
0.054	-3.08557851407983\\
0.072	-4.12765177366285\\
0.09	-5.1801265892841\\
0.108	-6.24905902081109\\
0.126	-7.33169374896935\\
0.144	-8.43471075509937\\
0.162	-9.56372111409944\\
0.18	-10.7167860351389\\
0.198	-11.90210422337\\
0.216	-13.1232123475291\\
0.234	-14.3808640804499\\
0.252	-15.6840272126779\\
0.27	-17.0341680827568\\
0.288	-18.4362244938259\\
0.306	-19.8997733575007\\
0.324	-21.4261685522036\\
0.342	-23.0246762493783\\
0.36	-24.7056643750942\\
0.378	-26.4728056500572\\
0.396	-28.3394855640522\\
0.414	-30.3171997430215\\
0.432	-32.4146663926158\\
0.45	-34.6486095799696\\
0.468	-37.0318415466754\\
0.486	-39.5819751582203\\
0.504	-42.3229928458177\\
0.522	-45.2750560497902\\
0.54	-48.4672343714829\\
0.558	-51.9317852013231\\
0.576	-55.70363887067\\
0.594	-59.8299180885835\\
0.612	-64.3583971938238\\
0.63	-69.3531125270275\\
0.648	-74.9010079866534\\
0.666	-81.0958680570493\\
0.684	-88.0625597981869\\
0.702	-95.9360712085313\\
0.72	-104.881844251108\\
0.738	-115.178512017114\\
0.756	-127.132378450498\\
0.774	-141.178464544773\\
0.792	-157.934564392868\\
0.81	-178.077085500921\\
0.828	-202.880262001348\\
0.846	-234.024958747165\\
0.864	-275.614927131381\\
0.882	-331.735399020249\\
0.9	-398.663706597578\\
};
\addlegendentry{$Ca=1/32$};

\addplot [color=black,solid]
  table[row sep=crcr]{%
0	0.000250635750855315\\
0.018	-1.02199274372501\\
0.036	-2.05191196667474\\
0.054	-3.08925830120177\\
0.072	-4.13151734621026\\
0.09	-5.18392632138909\\
0.108	-6.25308018268574\\
0.126	-7.33619029340455\\
0.144	-8.43970362317641\\
0.162	-9.5690788133624\\
0.18	-10.722430120343\\
0.198	-11.9079637072753\\
0.216	-13.1291992843035\\
0.234	-14.386892943657\\
0.252	-15.6900562997526\\
0.27	-17.0401380692405\\
0.288	-18.4420956378387\\
0.306	-19.9055657302512\\
0.324	-21.4318810781873\\
0.342	-23.0303497369979\\
0.36	-24.7113927657227\\
0.378	-26.4786742217185\\
0.396	-28.3456396080651\\
0.414	-30.3238226412561\\
0.432	-32.421967437867\\
0.45	-34.6569174521909\\
0.468	-37.0415416958661\\
0.486	-39.5935492872229\\
0.504	-42.3370999206959\\
0.522	-45.2924501674949\\
0.54	-48.4889192063839\\
0.558	-51.9588109000703\\
0.576	-55.7365451343141\\
0.594	-59.8724685579967\\
0.612	-64.4188967303683\\
0.63	-69.441365581271\\
0.648	-75.0209736614774\\
0.666	-81.2452424047208\\
0.684	-88.2446078577005\\
0.702	-96.1678101193748\\
0.72	-105.194815392175\\
0.738	-115.600696992554\\
0.756	-127.678767049786\\
0.774	-141.850252319497\\
0.792	-158.800476680489\\
0.81	-179.265954958252\\
0.828	-204.578263883906\\
0.846	-236.702035936797\\
0.864	-280.465571064205\\
0.882	-340.664063314983\\
0.9	-412.924172322603\\
};
\addlegendentry{Linear $Ca$};

\end{axis}
\end{tikzpicture}%

%% file: figures/PureNonLinearCa_d04_V.tex
%
%
\definecolor{mycolor1}{rgb}{1.00000,0.00000,1.00000}%
\begin{tikzpicture}[baseline]

\begin{axis}[%
width=.25\textwidth,
height=.25\textwidth,
scale only axis,
every outer x axis line/.append style={black},
every x tick label/.append style={font=\color{black}},
xmin=-.05,
xmax=1,
every outer y axis line/.append style={black},
every y tick label/.append style={font=\color{black}},
ymin=1.0,
ymax=2.3,
axis background/.style={fill=white},
axis x line*=bottom,
axis y line*=left,
axis lines = center,
xlabel=$\varepsilon/\varepsilon_*$,
ylabel=$V$,
minor xtick={ 0, 0.1, 0.2, 0.3, 0.4, 0.50, 0.6, 0.70, 0.8, .90,1},
xtick={ 0, 0.5, 1},
xticklabels={ $0$, $0.5$, $1$},
axis line style=-,
]
\node[anchor=north east] at (rel axis cs:1,1) {(b)};
\addplot [color=mycolor1,solid,forget plot]
  table[row sep=crcr]{%
0	1.98422141864979\\
0.018	1.98392006147505\\
0.036	1.98318028436571\\
0.054	1.98222337245012\\
0.072	1.9806409075321\\
0.09	1.97848187557981\\
0.108	1.97603806355878\\
0.126	1.97303401745897\\
0.144	1.96956280540029\\
0.162	1.96572167790533\\
0.18	1.96133273233069\\
0.198	1.95651126997102\\
0.216	1.95125565465149\\
0.234	1.94547387127994\\
0.252	1.93927532752017\\
0.27	1.9325994780863\\
0.288	1.92542619350744\\
0.306	1.91783033844707\\
0.324	1.90972811243724\\
0.342	1.90114300112737\\
0.36	1.89211782157063\\
0.378	1.88257245491777\\
0.396	1.87255326284221\\
0.414	1.86207247145434\\
0.432	1.85107037653399\\
0.45	1.83959731156305\\
0.468	1.82762865233281\\
0.486	1.81513873271213\\
0.504	1.80217699446465\\
0.522	1.78869620193082\\
0.54	1.77469986273024\\
0.558	1.7602229526363\\
0.576	1.74521364202614\\
0.594	1.72968538634936\\
0.612	1.71364355508897\\
0.63	1.69704123722116\\
0.648	1.67992033375776\\
0.666	1.66227854325911\\
0.684	1.6440814723145\\
0.702	1.62536694745397\\
0.72	1.60610601843875\\
0.738	1.58629093803626\\
0.756	1.56595202981935\\
0.774	1.54504623520789\\
0.792	1.52358937572805\\
0.81	1.50160455839467\\
0.828	1.47904715246546\\
0.846	1.45594962959927\\
0.864	1.43232582643734\\
0.882	1.40813933777441\\
0.9	1.38340684524299\\
};
\addplot [color=green,solid,forget plot]
  table[row sep=crcr]{%
0	1.98151930668772\\
0.018	1.98120241136856\\
0.036	1.98042449132881\\
0.054	1.97941824741699\\
0.072	1.97775429031966\\
0.09	1.9754839668656\\
0.108	1.97291385544317\\
0.126	1.96975410789746\\
0.144	1.9661024062211\\
0.162	1.96206076980054\\
0.18	1.95744173937898\\
0.198	1.9523663270807\\
0.216	1.94683244958983\\
0.234	1.94074281247105\\
0.252	1.93421207450796\\
0.27	1.92717604144382\\
0.288	1.91961276731449\\
0.306	1.91160020737527\\
0.324	1.90304956137576\\
0.342	1.89398413851905\\
0.36	1.88444761515487\\
0.378	1.87435477330702\\
0.396	1.86375202314982\\
0.414	1.85265050430987\\
0.432	1.84098577109623\\
0.45	1.82880660945789\\
0.468	1.81608485209099\\
0.486	1.80279055111746\\
0.504	1.78897064459102\\
0.522	1.77457428655865\\
0.54	1.7595977970934\\
0.558	1.74406893791882\\
0.576	1.72793134279394\\
0.594	1.71118798286671\\
0.612	1.69383621382953\\
0.63	1.6758230104638\\
0.648	1.65716630845342\\
0.666	1.63784855077739\\
0.684	1.61782602719172\\
0.702	1.59711317515059\\
0.72	1.575670013523\\
0.738	1.55346805202041\\
0.756	1.53050732066145\\
0.774	1.50672699259828\\
0.792	1.48212505363686\\
0.81	1.45669536146599\\
0.828	1.43035918662364\\
0.846	1.40319548822137\\
0.864	1.37523911938125\\
0.882	1.34641423772966\\
0.9	1.31675362371424\\
};
\addplot [color=red,solid,forget plot]
  table[row sep=crcr]{%
0	1.98083840330547\\
0.018	1.98051750727667\\
0.036	1.97972976684626\\
0.054	1.97871082313022\\
0.072	1.97702589250921\\
0.09	1.97472692941171\\
0.108	1.97212431861781\\
0.126	1.96892451073344\\
0.144	1.96522637437157\\
0.162	1.96113315614957\\
0.18	1.95645494015512\\
0.198	1.95131420504976\\
0.216	1.94570875849345\\
0.234	1.93953994273772\\
0.252	1.9329237385722\\
0.27	1.92579501634114\\
0.288	1.918131337084\\
0.306	1.91001140422421\\
0.324	1.90134514551849\\
0.342	1.89215572960422\\
0.36	1.88248696738059\\
0.378	1.8722523401679\\
0.396	1.86149801685236\\
0.414	1.85023472414202\\
0.432	1.83839673120917\\
0.45	1.82603192937836\\
0.468	1.8131112514165\\
0.486	1.79960317241747\\
0.504	1.78555281026473\\
0.522	1.7709086313278\\
0.54	1.75566358926129\\
0.558	1.73984170074418\\
0.576	1.7233860903023\\
0.594	1.70629256609042\\
0.612	1.68855353230455\\
0.63	1.67011453543749\\
0.648	1.65097658836744\\
0.666	1.63111718159408\\
0.684	1.6104842415102\\
0.702	1.58905789046376\\
0.72	1.56680075114241\\
0.738	1.54365016865568\\
0.756	1.51954720611915\\
0.774	1.4944517558985\\
0.792	1.46824937887342\\
0.81	1.44082088931583\\
0.828	1.41209864509352\\
0.846	1.38195553365836\\
0.864	1.35021457537848\\
0.882	1.31678231280289\\
0.9	1.28177109801844\\
};
\addplot [color=blue,solid,forget plot]
  table[row sep=crcr]{%
0	1.98066785196015\\
0.018	1.98034594832713\\
0.036	1.97955573475625\\
0.054	1.97853359327086\\
0.072	1.97684338399965\\
0.09	1.97453721340693\\
0.108	1.97192642263993\\
0.126	1.96871653024766\\
0.144	1.96500670429882\\
0.162	1.9609005005075\\
0.18	1.95620738412195\\
0.198	1.95105020513083\\
0.216	1.94542674658136\\
0.234	1.9392380064547\\
0.252	1.93260029369508\\
0.27	1.9254482421003\\
0.288	1.91775928879639\\
0.306	1.90961232467885\\
0.324	1.90091695904994\\
0.342	1.89169631806954\\
0.36	1.88199424000354\\
0.378	1.87172387996748\\
0.396	1.86093133080966\\
0.414	1.84962720765096\\
0.432	1.83774545203926\\
0.45	1.8253336976686\\
0.468	1.81236265774199\\
0.486	1.79880037195259\\
0.504	1.784691404048\\
0.522	1.76998407083463\\
0.54	1.75467035497784\\
0.558	1.73877318084319\\
0.576	1.72223561786339\\
0.594	1.7050512387897\\
0.612	1.68721097477479\\
0.63	1.66866004593175\\
0.648	1.64939385798562\\
0.666	1.62938874977472\\
0.684	1.60858963814956\\
0.702	1.58696367837677\\
0.72	1.56447660144248\\
0.738	1.54105012687295\\
0.756	1.51659917190113\\
0.774	1.49110037426605\\
0.792	1.46437164218307\\
0.81	1.43624409077087\\
0.828	1.40668114689457\\
0.846	1.37528970182378\\
0.864	1.34163077737244\\
0.882	1.30560189816395\\
0.9	1.26740639315442\\
};
\addplot [color=black,solid,forget plot]
  table[row sep=crcr]{%
0	1.98062699333249\\
0.018	1.98030467916614\\
0.036	1.9795134863057\\
0.054	1.97849015242322\\
0.072	1.97679827655136\\
0.09	1.97448999983284\\
0.108	1.97187693115272\\
0.126	1.96866448836931\\
0.144	1.96495178268687\\
0.162	1.96084231172644\\
0.18	1.956145438718\\
0.198	1.95098411711484\\
0.216	1.94535612585198\\
0.234	1.93916237649229\\
0.252	1.93251925906791\\
0.27	1.92536134813269\\
0.288	1.91766605103776\\
0.306	1.90951230688249\\
0.324	1.90080964694067\\
0.342	1.89158118517404\\
0.36	1.88187076511274\\
0.378	1.87159146166234\\
0.396	1.8607893474962\\
0.414	1.84947500962142\\
0.432	1.83758230817652\\
0.45	1.82515880916901\\
0.468	1.81217517326773\\
0.486	1.79859932720396\\
0.504	1.78447568349191\\
0.522	1.76975252071634\\
0.54	1.75442157480359\\
0.558	1.738505458158\\
0.576	1.72194712981958\\
0.594	1.70473993344135\\
0.612	1.68687491678724\\
0.63	1.66829720626751\\
0.648	1.64899975801203\\
0.666	1.62895790455193\\
0.684	1.60811615570819\\
0.702	1.5864388006408\\
0.72	1.56389309223918\\
0.738	1.54039552146869\\
0.756	1.51585258517868\\
0.774	1.49024661673059\\
0.792	1.46337362474492\\
0.81	1.43505195271921\\
0.828	1.4052592616058\\
0.846	1.37348905110597\\
0.864	1.33918891105038\\
0.882	1.30224941635879\\
0.9	1.26291505850382\\
};
\end{axis}
\end{tikzpicture}%

%% file: figures/PureNonLinearCa_d04_beta.tex
%
%
\definecolor{mycolor1}{rgb}{1.00000,0.00000,1.00000}%
\begin{tikzpicture}[baseline]

\begin{axis}[%
width=.25\textwidth,
height=.25\textwidth,
scale only axis,
every outer x axis line/.append style={black},
every x tick label/.append style={font=\color{black}},
xmin=-.05,
xmax=1,
every outer y axis line/.append style={black},
every y tick label/.append style={font=\color{black}},
ymin=-0.4,
ymax=1,
axis background/.style={fill=white},
axis x line*=bottom,
axis y line*=left,
axis lines = center,
xlabel=$\varepsilon/\varepsilon_*$,
ylabel=$\beta$,
minor xtick={ 0, 0.1, 0.2, 0.3, 0.4, 0.50, 0.6, 0.70, 0.8, .90,1},
xtick={ 0, 0.5, 1},
xticklabels={ $0$, $0.5$, $1$},
y filter/.code={\pgfmathparse{#1*3/64/.4^3}\pgfmathresult},
axis line style=-,
]
\node[anchor=south east] at (rel axis cs:1,0) {(c)};
\addplot [color=mycolor1,solid,forget plot]
  table[row sep=crcr]{%
0	-0.221020572150202\\
0.018	-0.220284934661573\\
0.036	-0.219075580665882\\
0.054	-0.217572854687683\\
0.072	-0.215455760070305\\
0.09	-0.212744654359297\\
0.108	-0.209640327157213\\
0.126	-0.205861899009488\\
0.144	-0.201484891877923\\
0.162	-0.196544348486898\\
0.18	-0.19080613385702\\
0.198	-0.184523398776821\\
0.216	-0.17780367894651\\
0.234	-0.170466240802405\\
0.252	-0.162471025024115\\
0.27	-0.153570284068969\\
0.288	-0.143905718099905\\
0.306	-0.133845926094474\\
0.324	-0.123306004915312\\
0.342	-0.112250402926371\\
0.36	-0.100728401185204\\
0.378	-0.0886946403965228\\
0.396	-0.0760416579471778\\
0.414	-0.0625043511034402\\
0.432	-0.0481116824172149\\
0.45	-0.0336580730093061\\
0.468	-0.0196350592588856\\
0.486	-0.00556402511010667\\
0.504	0.00918611521178493\\
0.522	0.0247869269321002\\
0.54	0.0409882848721412\\
0.558	0.0576416549334106\\
0.576	0.0749520598122074\\
0.594	0.0924431764634345\\
0.612	0.109698741992591\\
0.63	0.126909948584699\\
0.648	0.143863328363949\\
0.666	0.160474183786221\\
0.684	0.17679830622649\\
0.702	0.192674986332774\\
0.72	0.208317754151839\\
0.738	0.223416203252754\\
0.756	0.237321020784107\\
0.774	0.250238123790327\\
0.792	0.262228721077798\\
0.81	0.273564575658243\\
0.828	0.284872545550466\\
0.846	0.293497924585518\\
0.864	0.298493774411502\\
0.882	0.302337308983505\\
0.9	0.303902784858991\\
};
\addplot [color=green,solid,forget plot]
  table[row sep=crcr]{%
0	-0.215300433359161\\
0.018	-0.214507355178696\\
0.036	-0.213140113302557\\
0.054	-0.211428332029344\\
0.072	-0.208958770168086\\
0.09	-0.205765415827328\\
0.108	-0.202118450487756\\
0.126	-0.197677887031183\\
0.144	-0.192535137067811\\
0.162	-0.186741820875811\\
0.18	-0.180024313588835\\
0.198	-0.172664301841627\\
0.216	-0.164764598814396\\
0.234	-0.15612772476715\\
0.252	-0.14676133735201\\
0.27	-0.136463571817264\\
0.288	-0.125296545614837\\
0.306	-0.113505410274666\\
0.324	-0.100950308083869\\
0.342	-0.087658647800878\\
0.36	-0.0737395776209751\\
0.378	-0.0591069605084033\\
0.396	-0.0436116486271056\\
0.414	-0.0270500235137341\\
0.432	-0.0093804601191005\\
0.45	0.00910398422078629\\
0.468	0.028327093496157\\
0.486	0.0484521927318005\\
0.504	0.0695348588129331\\
0.522	0.0916076008874616\\
0.54	0.114822762622987\\
0.558	0.139430323544569\\
0.576	0.165732339399432\\
0.594	0.193120999063869\\
0.612	0.220942576290129\\
0.63	0.249430505718524\\
0.648	0.279301441527681\\
0.666	0.311079694632869\\
0.684	0.34462219974976\\
0.702	0.379699682952388\\
0.72	0.41641554847464\\
0.738	0.454866367502683\\
0.756	0.495005356378413\\
0.774	0.537131610829251\\
0.792	0.580954212418558\\
0.81	0.626253589009259\\
0.828	0.673444207847754\\
0.846	0.721399276109637\\
0.864	0.769277374912938\\
0.882	0.817510696493511\\
0.9	0.866122759152333\\
};
\addplot [color=red,solid,forget plot]
  table[row sep=crcr]{%
0	-0.21385551478344\\
0.018	-0.213047754043504\\
0.036	-0.211639973494247\\
0.054	-0.209874653320971\\
0.072	-0.207314577522835\\
0.09	-0.203997130674282\\
0.108	-0.200210530674097\\
0.126	-0.195599874744775\\
0.144	-0.190258929866362\\
0.162	-0.184242818143267\\
0.18	-0.177267131875238\\
0.198	-0.169613387198848\\
0.216	-0.16137637568459\\
0.234	-0.152352985687121\\
0.252	-0.142563213645327\\
0.27	-0.131801041933712\\
0.288	-0.12012113810205\\
0.306	-0.107767948267981\\
0.324	-0.0945903675137675\\
0.342	-0.0806184128924237\\
0.36	-0.0659701449207302\\
0.378	-0.0505550033490204\\
0.396	-0.0341856279885175\\
0.414	-0.0166127922346017\\
0.432	0.00220849538626647\\
0.45	0.0219357850131843\\
0.468	0.0424583288654379\\
0.486	0.0639932046710002\\
0.504	0.0866766576259563\\
0.522	0.110572936030872\\
0.54	0.135840476477864\\
0.558	0.162739406660134\\
0.576	0.191590866630149\\
0.594	0.221892642554294\\
0.612	0.253077884333631\\
0.63	0.285395789150892\\
0.648	0.319667084633978\\
0.666	0.356466726080855\\
0.684	0.395785115239732\\
0.702	0.437692494035463\\
0.72	0.48236939565616\\
0.738	0.530234965109137\\
0.756	0.581727563603792\\
0.774	0.637224870742044\\
0.792	0.697319711065964\\
0.81	0.762727131467539\\
0.828	0.834247915899307\\
0.846	0.912156120244821\\
0.864	0.997325521225693\\
0.882	1.09081236496671\\
0.9	1.19171541589827\\
};
\addplot [color=blue,solid,forget plot]
  table[row sep=crcr]{%
0	-0.213493284321245\\
0.018	-0.212681880926077\\
0.036	-0.211263900277255\\
0.054	-0.209485099782548\\
0.072	-0.206902124375576\\
0.09	-0.203553273257898\\
0.108	-0.199731457683912\\
0.126	-0.195078092519561\\
0.144	-0.189687406638899\\
0.162	-0.18361527574514\\
0.18	-0.176574600402081\\
0.198	-0.168846910427059\\
0.216	-0.160524956131204\\
0.234	-0.151404209972447\\
0.252	-0.141507934979261\\
0.27	-0.130629261170603\\
0.288	-0.118820783633687\\
0.306	-0.106326670317781\\
0.324	-0.0929944907852927\\
0.342	-0.0788509934278313\\
0.36	-0.064010754936917\\
0.378	-0.0483806523491824\\
0.396	-0.0317721522517645\\
0.414	-0.0139307763313777\\
0.432	0.00519186899032514\\
0.45	0.0252416795543411\\
0.468	0.0460953044759621\\
0.486	0.0679877658794477\\
0.504	0.0910887701333255\\
0.522	0.115478480574241\\
0.54	0.141306925146325\\
0.558	0.168823984279026\\
0.576	0.198354933449003\\
0.594	0.22943730301789\\
0.612	0.26153880612977\\
0.63	0.294913935028048\\
0.648	0.330404951845247\\
0.666	0.368590399351965\\
0.684	0.409511738344948\\
0.702	0.453358017939138\\
0.72	0.500343348351033\\
0.738	0.55100806455934\\
0.756	0.605962233324022\\
0.774	0.665568047260675\\
0.792	0.730969719809183\\
0.81	0.803369989814434\\
0.828	0.883748453123377\\
0.846	0.974160624996731\\
0.864	1.07788344633208\\
0.882	1.19709559982074\\
0.9	1.32945565344227\\
};
\addplot [color=black,solid,forget plot]
  table[row sep=crcr]{%
0	-0.213447146834255\\
0.018	-0.212629148536214\\
0.036	-0.211201701053751\\
0.054	-0.209411325704144\\
0.072	-0.206812271364892\\
0.09	-0.203444768771183\\
0.108	-0.199605977093287\\
0.126	-0.194939573476183\\
0.144	-0.189535997663515\\
0.162	-0.183447617535517\\
0.18	-0.176388139325604\\
0.198	-0.168639441886878\\
0.216	-0.160293855449368\\
0.234	-0.151146532789918\\
0.252	-0.141221617153146\\
0.27	-0.130312184439136\\
0.288	-0.11847031014257\\
0.306	-0.105940000823269\\
0.324	-0.0925686353764591\\
0.342	-0.0783820966773721\\
0.36	-0.0634940165418269\\
0.378	-0.0478107596510061\\
0.396	-0.0311433848245899\\
0.414	-0.0132361024638513\\
0.432	0.00596024355282112\\
0.45	0.0260889611851667\\
0.468	0.0470236952469775\\
0.486	0.0690038229779103\\
0.504	0.0922073336585613\\
0.522	0.116718254392285\\
0.54	0.142684799620799\\
0.558	0.17035697663168\\
0.576	0.200068162018169\\
0.594	0.23134093348511\\
0.612	0.263617040615334\\
0.63	0.297152887372482\\
0.648	0.332861454767201\\
0.666	0.371373693248315\\
0.684	0.412720913526794\\
0.702	0.457079671855787\\
0.72	0.504657037006957\\
0.738	0.556048137912974\\
0.756	0.611929107725009\\
0.774	0.672663920756574\\
0.792	0.739556604984294\\
0.81	0.813928728591218\\
0.828	0.896805577491118\\
0.846	0.990996492722865\\
0.864	1.10085293512034\\
0.882	1.22911731382875\\
0.9	1.37276821082861\\
};

\end{axis}
\end{tikzpicture}%

%% file: figures/Forma_p2_Ca_Ca_1_16_d04.tex
%
%
\begin{tikzpicture}[baseline]

\begin{axis}[%
width=.35\textwidth,
height=.35\textwidth,
scale only axis,
every outer x axis line/.append style={black},
every x tick label/.append style={font=\color{black}},
xmin=-0.05,
xmax=1.2,
every outer y axis line/.append style={black},
every y tick label/.append style={font=\color{black}},
ymin=-3600,
ymax=800,
ytick={ -500,0,500,-1500,-2500,-3500},
yticklabels={ $-500$,$0$,$500$,$-1500$,$-2500$,$-3500$},
axis background/.style={fill=white},
axis x line*=bottom,
axis y line*=left,
axis on top,
axis lines = center,
xlabel=$\varepsilon/\varepsilon_*$,
ylabel=$f/Ca$,
minor xtick={ 0, 0.1, 0.2, 0.3, 0.4, 0.50, 0.6, 0.70, 0.8, .90,1,1.1,1.2},
xtick={ 0, 0.5, 1},
xticklabels={ $0$, $0.5$, $1$},
y filter/.code={\pgfmathparse{#1*-1}\pgfmathresult},
]
\node[anchor=north east,yshift=4pt] at (rel axis cs:1,1) {(b)};
\addplot [color=blue,only marks,mark=x,mark options={solid},forget plot]
  table[row sep=crcr]{%
0	-0.000278022157248644\\
0.15	8.80526828752248\\
0.3	19.3876426151637\\
0.45	34.5879474120929\\
0.6	61.039603202238\\
0.75	121.312373738282\\
0.9	356.938127002606\\
0.95	613.996377055052\\
1	1111.24102252526\\
1.05	1926.28929463237\\
1.1	3143.29973578041\\
};
\label{discretos};
\addplot [color=blue,solid,forget plot]
  table[row sep=crcr]{%
0	-0.000278022157248644\\
0.03	1.6212904291062\\
0.06	3.31653743185252\\
0.09	5.0815548399351\\
0.12	6.91243450720729\\
0.15	8.80526828752248\\
0.18	10.736951414887\\
0.21	12.7166028969762\\
0.24	14.7886502481474\\
0.27	16.9975209827575\\
0.3	19.3876426151637\\
0.33	21.9478277830466\\
0.36	24.6815428073817\\
0.39	27.6551957274683\\
0.42	30.9351945826056\\
0.45	34.5879474120929\\
0.48	38.6335738725857\\
0.51	43.1493740294626\\
0.54	48.2875265351076\\
0.57	54.2002100419047\\
0.6	61.039603202238\\
0.63	68.9424786415636\\
0.66	78.3349712961812\\
0.69	89.8033032847404\\
0.72	103.933696725891\\
0.75	121.312373738282\\
0.7824	150.236290308738\\
0.8172	194.47128217515\\
0.8508	247.806891082767\\
0.8796	304.032658776836\\
0.9	356.938127002606\\
0.9132	404.299133642995\\
0.9236	450.293278037829\\
0.9324	498.312744117713\\
0.9408	551.749715813252\\
0.95	613.996377055052\\
0.96	689.325185221279\\
0.97	778.533906582381\\
0.98	879.802688407392\\
0.99	991.311677965342\\
1	1111.24102252526\\
1.01	1243.82635847822\\
1.02	1393.35947588772\\
1.03	1558.10475253144\\
1.04	1736.3265661871\\
1.05	1926.28929463237\\
1.06	2132.44562455394\\
1.07	2359.13312792381\\
1.08	2604.44351044771\\
1.09	2866.46847783134\\
1.1	3143.29973578041\\
};
\end{axis}
\end{tikzpicture}%

%% file: figures/LandauLevich1.tex
%
\begin{tikzpicture}[%
baseline
]

\begin{axis}[%
width=.3\textwidth,
height=.3\textwidth,
scale only axis,
unbounded coords=jump,
every outer x axis line/.append style={black},
every x tick label/.append style={font=\color{black}},
xmode=log,
xmin=0.001,
xmax=1,
xminorticks=true,
xlabel={$Ca$},
every outer y axis line/.append style={black},
every y tick label/.append style={font=\color{black}},
ymode=log,
ymin=0.008,
ymax=.2,
yminorticks=true,
ylabel={$h_{min}/d$},
axis background/.style={fill=white},
axis x line*=bottom,
axis y line*=left,
legend style={legend cell align=left,align=left,draw=black},
legend pos=south east,
name=one,
]
\node[anchor=north west] at (rel axis cs:0,1) {(a)};
\addplot [color=black,solid]
  table[row sep=crcr,x=x,y expr=\thisrow{y}/0.5]{%
    x y \\
0.0078125	0.00537052887913086\\
0.00929068413810788	0.00634154018144156\\
0.0110485395487997	0.00737377851206561\\
0.0131389987820148	0.00830332664845501\\
0.015625	0.00916432670830852\\
0.0185813510128694	0.0102417871940783\\
0.0220971035116717	0.0115823952029033\\
0.0262780320907328	0.0131614817615101\\
0.03125	0.0146742740225888\\
0.0371627020257389	0.016513825873369\\
0.0441942070233434	0.0186867594588193\\
0.0525560641814656	0.0208655955092307\\
0.0625	0.0232531740711604\\
0.0743256802657886	0.0260822391278563\\
0.0883884140466868	0.028884090549062\\
0.105111575937858	0.0321047073626945\\
0.125	0.0355332766878915\\
0.14865025567844	0.0392294160076399\\
0.176775265604837	0.0433177770098552\\
0.210225361587622	0.0474688889782055\\
0.25	0.0518643216741277\\
};
\addlegendentry{$d=0.5$};

\addplot [color=blue,solid]
  table[row sep=crcr,x=x,y expr=\thisrow{y}/0.6]{%
    x y \\
0.0078125	0.00777116728240107\\
0.00929068413810788	0.00901139183522831\\
0.0110485395487997	0.0103275468205971\\
0.0131389987820148	0.0115188009098361\\
0.015625	0.0129244735834636\\
0.0185813510128694	0.0145714328277336\\
0.0220971035116717	0.0164874557435045\\
0.0262780320907328	0.0182768174000126\\
0.03125	0.0203684323752968\\
0.0371627020257389	0.0228532390164881\\
0.0441942070233434	0.0255520916088033\\
0.0525560641814656	0.0282936118959524\\
0.0625	0.0315358021211868\\
0.0743256802657886	0.0350557112581485\\
0.0883884140466868	0.0386675902763632\\
0.105111575937858	0.0428156730361008\\
0.125	0.0468252179219579\\
0.14865025567844	0.0514284501343355\\
0.176775265604837	0.0562085044597567\\
0.210225361587622	0.0612646915548647\\
0.25	0.0664781150874341\\
0.29730051135688	inf\\
0.353556781219064	inf\\
0.420450723175244	inf\\
0.5	inf\\
0.594601022713759	inf\\
0.707113562438128	inf\\
0.840901446350488	inf\\
1	inf\\
};
\addlegendentry{$d=0.6$};

\addplot [color=red,solid]
  table[row sep=crcr,x=x,y expr=\thisrow{y}/0.7]{%
  x y \\
0.0078125	0.0109221153908832\\
0.00929068413810788	0.0124625227641469\\
0.0110485395487997	0.013787552156334\\
0.0131389987820148	0.0153235421348433\\
0.015625	0.017130955270782\\
0.0185813510128694	0.0192357042175897\\
0.0220971035116717	0.0214927727462669\\
0.0262780320907328	0.0237422555723654\\
0.03125	0.0264113073135988\\
0.0371627020257389	0.0295157978438919\\
0.0441942070233434	0.0326589183440956\\
0.0525560641814656	0.0360483039818021\\
0.0625	0.0399648140254921\\
0.0743256802657886	0.0439732710612688\\
0.0883884140466868	0.0482676042012889\\
0.105111575937858	0.0530552649934715\\
0.125	0.0577478106200199\\
0.14865025567844	0.0630174119807046\\
0.176775265604837	0.0685057571607799\\
0.210225361587622	0.0741247032875629\\
0.25	0.0799654559143707\\
0.29730051135688	inf\\
0.353556781219064	inf\\
0.420450723175244	inf\\
0.5	inf\\
0.594601022713759	inf\\
0.707113562438128	inf\\
0.840901446350488	inf\\
1	inf\\
};
\addlegendentry{$d=0.7$};

\draw (0.01,0.07)--(0.05,0.07) node[midway,above] {$3$};
\draw (0.01,0.07)--(0.01,{0.07/5^(2/3)}) node[midway,left] {$2$};
\draw (0.05,0.07)--(0.01,{0.07/5^(2/3)});
\end{axis}

%% file: figures/LandauLevich2.tex
%

\begin{axis}[%
width=.3\textwidth,
height=.3\textwidth,
scale only axis,
at=(one.south west), anchor=south west, xshift=.5\textwidth, 
unbounded coords=jump,
every outer x axis line/.append style={black},
every x tick label/.append style={font=\color{black}},
xmode=log,
xmin=0.001,
xmax=1,
xminorticks=true,
xlabel={$Ca$},
every outer y axis line/.append style={black},
every y tick label/.append style={font=\color{black}},
ylabel={$f$},
axis background/.style={fill=white},
axis x line*=bottom,
axis y line*=left,
legend style={legend cell align=left,align=left,draw=black},
]
\node[anchor=north west] at (rel axis cs:0,1) {(b)};
\addplot [color=black,solid]
  table[row sep=crcr,x=x,y expr=\thisrow{y}/0.0654]{%
x y \\
0.0078125	nan\\
0.00929068413810788	nan\\
0.0110485395487997	nan\\
0.0131389987820148	nan\\
0.015625	nan\\
0.0185813510128694	nan\\
0.0220971035116717	nan\\
0.0262780320907328	nan\\
0.03125	nan\\
0.0371627020257389	nan\\
0.0441942070233434	nan\\
0.0525560641814656	nan\\
0.0625	nan\\
0.0743256802657886	nan\\
0.0883884140466868	nan\\
0.105111575937858	nan\\
0.125	nan\\
0.14865025567844	nan\\
0.176775265604837	nan\\
0.210225361587622	nan\\
0.25	nan\\
0.29730051135688	nan\\
0.353556781219064	nan\\
0.420450723175244	nan\\
0.5	nan\\
0.594601022713759	nan\\
0.707113562438128	nan\\
0.840901446350488	nan\\
1	nan\\
0.0078125	58.0257789104181\\
0.00929068413810788	58.373122165669\\
0.0110485395487997	58.3637170014713\\
0.0131389987820148	58.2235417977769\\
0.015625	58.1067509466614\\
0.0185813510128694	58.0871770357891\\
0.0220971035116717	58.172109518714\\
0.0262780320907328	58.3328468975874\\
0.03125	58.5395510835594\\
0.0371627020257389	58.7731787232013\\
0.0441942070233434	59.0282278891371\\
0.0525560641814656	59.2965186518079\\
0.0625	59.5662843157636\\
0.0743256802657886	59.819709324465\\
0.0883884140466868	60.0352170466553\\
0.105111575937858	60.1904404035161\\
0.125	60.2655623999759\\
0.14865025567844	60.248415113187\\
0.176775265604837	60.1402099643934\\
0.210225361587622	59.9580920902787\\
0.25	59.7384646341339\\
};

\addplot [color=blue,solid]
  table[row sep=crcr,x=x,y expr=\thisrow{y}/0.1131]{%
x y \\
0.0078125	53.4182059178489\\
0.00929068413810788	53.3540469717501\\
0.0110485395487997	53.2338335100618\\
0.0131389987820148	53.0945013319959\\
0.015625	52.9660408577417\\
0.0185813510128694	52.8691067239477\\
0.0220971035116717	52.8138318926119\\
0.0262780320907328	52.7979872060526\\
0.03125	52.8190258198476\\
0.0371627020257389	52.855334888318\\
0.0441942070233434	52.8891852258024\\
0.0525560641814656	52.9017662510249\\
0.0625	52.8695025760756\\
0.0743256802657886	52.7694061159759\\
0.0883884140466868	52.5798984265311\\
0.105111575937858	52.2851258851572\\
0.125	51.8804059095787\\
0.14865025567844	51.3765130706846\\
0.176775265604837	50.8011430884338\\
0.210225361587622	50.1978891858529\\
0.25	49.6260382004294\\
0.29730051135688	nan\\
0.353556781219064	nan\\
0.420450723175244	nan\\
0.5	nan\\
0.594601022713759	nan\\
0.707113562438128	nan\\
0.840901446350488	nan\\
1	nan\\
};

\addplot [color=red,solid]
   table[row sep=crcr,x=x,y expr=\thisrow{y}/0.1796]{%
x y \\
0.0078125	50.916479042114\\
0.00929068413810788	50.5747525399469\\
0.0110485395487997	50.2516717316594\\
0.0131389987820148	49.9594733667827\\
0.015625	49.6999453713414\\
0.0185813510128694	49.4699930816471\\
0.0220971035116717	49.2622536131461\\
0.0262780320907328	49.0649236354549\\
0.03125	48.8615821043292\\
0.0371627020257389	48.631470134562\\
0.0441942070233434	48.3496193657066\\
0.0525560641814656	47.9909248101863\\
0.0625	47.5278415746013\\
0.0743256802657886	46.938003093063\\
0.0883884140466868	46.2088067135581\\
0.105111575937858	45.3429323848414\\
0.125	44.3616563980145\\
0.14865025567844	43.3043659698279\\
0.176775265604837	42.2236605113282\\
0.210225361587622	41.178787686936\\
0.25	40.2310244131771\\
0.29730051135688	nan\\
0.353556781219064	nan\\
0.420450723175244	nan\\
0.5	nan\\
0.594601022713759	nan\\
0.707113562438128	nan\\
0.840901446350488	nan\\
1	nan\\
};

\end{axis}

%% file: figures/LandauLevich3.tex
%

\begin{axis}[%
width=.3\textwidth,
height=.3\textwidth,
scale only axis,
name=three,
at=(one.south west), anchor=south west, yshift=-.4\textwidth, 
unbounded coords=jump,
every outer x axis line/.append style={black},
every x tick label/.append style={font=\color{black}},
xmode=log,
xmin=0.001,
xmax=1,
xminorticks=true,
xlabel={$Ca$},
every outer y axis line/.append style={black},
every y tick label/.append style={font=\color{black}},
ymin=1,
ymax=1.9,
ylabel={$V$},
axis background/.style={fill=white},
axis x line*=bottom,
axis y line*=left,
legend style={legend cell align=left,align=left,draw=black},
]
\node[anchor=north west] at (rel axis cs:0,1) {(c)};
\addplot [color=black,solid]
  table[row sep=crcr]{%
0.0078125	nan\\
0.00929068413810788	nan\\
0.0110485395487997	nan\\
0.0131389987820148	nan\\
0.015625	nan\\
0.0185813510128694	nan\\
0.0220971035116717	nan\\
0.0262780320907328	nan\\
0.03125	nan\\
0.0371627020257389	nan\\
0.0441942070233434	nan\\
0.0525560641814656	nan\\
0.0625	nan\\
0.0743256802657886	nan\\
0.0883884140466868	nan\\
0.105111575937858	nan\\
0.125	nan\\
0.14865025567844	nan\\
0.176775265604837	nan\\
0.210225361587622	nan\\
0.25	nan\\
0.29730051135688	nan\\
0.353556781219064	nan\\
0.420450723175244	nan\\
0.5	nan\\
0.594601022713759	nan\\
0.707113562438128	nan\\
0.840901446350488	nan\\
1	nan\\
0.0078125	1.07394589940539\\
0.00929068413810788	1.08941762832376\\
0.0110485395487997	1.10460251492751\\
0.0131389987820148	1.12001179526116\\
0.015625	1.1360245485216\\
0.0185813510128694	1.15279653073792\\
0.0220971035116717	1.17019705743534\\
0.0262780320907328	1.18809835442233\\
0.03125	1.2063729767461\\
0.0371627020257389	1.2250353424245\\
0.0441942070233434	1.24418066329513\\
0.0525560641814656	1.26395704419781\\
0.0625	1.28450875577477\\
0.0743256802657886	1.30598405413869\\
0.0883884140466868	1.32855052842835\\
0.105111575937858	1.35240859598186\\
0.125	1.3777957788306\\
0.14865025567844	1.40498927942948\\
0.176775265604837	1.43432122919931\\
0.210225361587622	1.46616269807925\\
0.25	1.50089312211186\\
};

\addplot [color=blue,solid]
  table[row sep=crcr]{%
0.0078125	1.24050464744945\\
0.00929068413810788	1.25580037034019\\
0.0110485395487997	1.27118146737119\\
0.0131389987820148	1.28675134986154\\
0.015625	1.30262848454014\\
0.0185813510128694	1.31893058781356\\
0.0220971035116717	1.33576705864195\\
0.0262780320907328	1.35323650341093\\
0.03125	1.37142722524114\\
0.0371627020257389	1.39044815456796\\
0.0441942070233434	1.41043470412962\\
0.0525560641814656	1.43157039433855\\
0.0625	1.4540773398958\\
0.0743256802657886	1.47820774785902\\
0.0883884140466868	1.50422832972051\\
0.105111575937858	1.5324158953268\\
0.125	1.56305137288881\\
0.14865025567844	1.59640196337977\\
0.176775265604837	1.63271173595313\\
0.210225361587622	1.67215468216777\\
0.25	1.71474663651242\\
0.29730051135688	nan\\
0.353556781219064	nan\\
0.420450723175244	nan\\
0.5	nan\\
0.594601022713759	nan\\
0.707113562438128	nan\\
0.840901446350488	nan\\
1	nan\\
};

\addplot [color=red,solid]
  table[row sep=crcr]{%
0.0078125	1.34918883552215\\
0.00929068413810788	1.36145095937229\\
0.0110485395487997	1.37408220136879\\
0.0131389987820148	1.38717682729276\\
0.015625	1.4008282781618\\
0.0185813510128694	1.41514846733366\\
0.0220971035116717	1.4302750779668\\
0.0262780320907328	1.4463724964317\\
0.03125	1.46363421436514\\
0.0371627020257389	1.48228512311289\\
0.0441942070233434	1.50257932989563\\
0.0525560641814656	1.52480446143095\\
0.0625	1.54925133037015\\
0.0743256802657886	1.57620861196514\\
0.0883884140466868	1.60593816188294\\
0.105111575937858	1.63866479424683\\
0.125	1.67456660146352\\
0.14865025567844	1.71375049415735\\
0.176775265604837	1.75622379295511\\
0.210225361587622	1.80181368213713\\
0.25	1.85005720413261\\
0.29730051135688	nan\\
0.353556781219064	nan\\
0.420450723175244	nan\\
0.5	nan\\
0.594601022713759	nan\\
0.707113562438128	nan\\
0.840901446350488	nan\\
1	nan\\
};

\end{axis}

%% file: figures/LandauLevich4a.tex
%

\begin{axis}[%
width=.3\textwidth,
height=.3\textwidth,
scale only axis,
at=(three.south west), anchor=south west, xshift=.5\textwidth, 
unbounded coords=jump,
every outer x axis line/.append style={black},
every x tick label/.append style={font=\color{black}},
xmode=log,
xmin=0.001,
xmax=1,
xminorticks=true,
xlabel={$Ca$},
every outer y axis line/.append style={black},
every y tick label/.append style={font=\color{black}},
ymin=-1,
ymax=3,
ylabel={$\beta$},
axis background/.style={fill=white},
axis x line*=center,
axis y line*=left,
legend style={legend cell align=left,align=left,draw=black},
]
\node[anchor=north west] at (rel axis cs:0,1) {(d)};
\addplot [color=black,solid]
  table[row sep=crcr,x=x,y expr=\thisrow{y}*3/64/.5^3]{%
  x y \\
0.0078125	nan\\
0.00929068413810788	nan\\
0.0110485395487997	nan\\
0.0131389987820148	nan\\
0.015625	nan\\
0.0185813510128694	nan\\
0.0220971035116717	nan\\
0.0262780320907328	nan\\
0.03125	nan\\
0.0371627020257389	nan\\
0.0441942070233434	nan\\
0.0525560641814656	nan\\
0.0625	nan\\
0.0743256802657886	nan\\
0.0883884140466868	nan\\
0.105111575937858	nan\\
0.125	nan\\
0.14865025567844	nan\\
0.176775265604837	nan\\
0.210225361587622	nan\\
0.25	nan\\
0.29730051135688	nan\\
0.353556781219064	nan\\
0.420450723175244	nan\\
0.5	nan\\
0.594601022713759	nan\\
0.707113562438128	nan\\
0.840901446350488	nan\\
1	nan\\
0.0078125	5.5788266664612\\
0.00929068413810788	5.32078561546334\\
0.0110485395487997	5.06834923073826\\
0.0131389987820148	4.81324430437549\\
0.015625	4.54938670709059\\
0.0185813510128694	4.27435244828403\\
0.0220971035116717	3.991275291161\\
0.0262780320907328	3.70225657244119\\
0.03125	3.41017804502566\\
0.0371627020257389	3.11543388314506\\
0.0441942070233434	2.81760157266538\\
0.0525560641814656	2.51528921394937\\
0.0625	2.20769772543298\\
0.0743256802657886	1.89430326759724\\
0.0883884140466868	1.57464413749852\\
0.105111575937858	1.24839869810259\\
0.125	0.915194731978559\\
0.14865025567844	0.57462862083193\\
0.176775265604837	0.225964116713783\\
0.210225361587622	-0.131391904436697\\
0.25	-0.498033330257982\\
};

\addplot [color=blue,solid]
  table[row sep=crcr,x=x,y expr=\thisrow{y}*3/64/.6^3]{%
  x y \\
0.0078125	7.7604451263576\\
0.00929068413810788	7.29512159715071\\
0.0110485395487997	6.8326723669373\\
0.0131389987820148	6.37115504460849\\
0.015625	5.90822585437748\\
0.0185813510128694	5.44187464155043\\
0.0220971035116717	4.97072155418368\\
0.0262780320907328	4.49397263006581\\
0.03125	4.01279983576642\\
0.0371627020257389	3.52665304664598\\
0.0441942070233434	3.03605351277282\\
0.0525560641814656	2.54089486631509\\
0.0625	2.04009900788162\\
0.0743256802657886	1.53288548791365\\
0.0883884140466868	1.01871197455051\\
0.105111575937858	0.497009734567186\\
0.125	-0.0324769066197348\\
0.14865025567844	-0.569284159072481\\
0.176775265604837	-1.11318808802768\\
0.210225361587622	-1.66377619975404\\
0.25	-2.21993771300799\\
0.29730051135688	nan\\
0.353556781219064	nan\\
0.420450723175244	nan\\
0.5	nan\\
0.594601022713759	nan\\
0.707113562438128	nan\\
0.840901446350488	nan\\
1	nan\\
};

\addplot [color=red,solid]
  table[row sep=crcr,x=x,y expr=\thisrow{y}*3/64/.7^3]{%
  x y \\
0.0078125	9.49111120268572\\
0.00929068413810788	8.78579279513913\\
0.0110485395487997	8.08016207530002\\
0.0131389987820148	7.37314224189046\\
0.015625	6.6644660200962\\
0.0185813510128694	5.95365292976155\\
0.0220971035116717	5.24041375687152\\
0.0262780320907328	4.52365307846239\\
0.03125	3.80238610331425\\
0.0371627020257389	3.07557927796519\\
0.0441942070233434	2.34199855562777\\
0.0525560641814656	1.60027198671271\\
0.0625	0.849650199970171\\
0.0743256802657886	0.0898906062915756\\
0.0883884140466868	-0.677938269233394\\
0.105111575937858	-1.45227876167401\\
0.125	-2.23134078820441\\
0.14865025567844	-3.01283856339087\\
0.176775265604837	-3.79520906607596\\
0.210225361587622	-4.57588693922638\\
0.25	-5.35147317466517\\
0.29730051135688	nan\\
0.353556781219064	nan\\
0.420450723175244	nan\\
0.5	nan\\
0.594601022713759	nan\\
0.707113562438128	nan\\
0.840901446350488	nan\\
1	nan\\
};

\end{axis}
\end{tikzpicture}%

%% file: figures/Centred_Stability_Oh.tex
%
%
\begin{tikzpicture}[baseline]

\begin{axis}[%
width=.3\textwidth,
height=.3\textwidth,
scale only axis,
separate axis lines,
every outer x axis line/.append style={black},
every x tick label/.append style={font=\color{black}},
every outer y axis line/.append style={black},
every y tick label/.append style={font=\color{black}},
ymode=log,
ymin=0.002,
ymax=.2,
yminorticks=true,
ylabel={$Oh$},
axis background/.style={fill=white},
legend style={legend cell align=left,align=left,draw=black},
xmin=0,
xmax=1,
minor xtick={ 0, 0.1, 0.2, 0.3, 0.4, 0.5, 0.6, 0.7, 0.8, 0.9, 1},
xtick={ 0, 0.5, 1},
xticklabels={ $0$, $0.5$, $1$},
xlabel={$d$},
axis x line*=bottom,
axis y line*=left,
colormap/jet,
point meta=y,
point meta rel=axis wide,
extra x ticks ={0.85}, extra x tick labels={{\color{blue}$0.85$}},
]
\node[anchor=north east] at (rel axis cs:1,1) {(b)};   

\node[anchor=north west] at (rel axis cs:.7,.85) {Stable};
\node[anchor=north west] at (rel axis cs:.3,.5) {Unstable};

\addplot[solid,mesh, draw=mapped color, forget plot]
table[row sep=crcr] {%
x	y\\
0.850611954158382	0.001\\
0.850569518442762	0.00115139539932645\\
0.850513334224882	0.00132571136559011\\
0.850438978686889	0.00152641796717523\\
0.850340629893861	0.00175751062485479\\
0.850251256281407	0.00194819885705138\\
0.850208792869906	0.00202358964772516\\
0.85002933423446	0.00232995181051537\\
0.849792664187667	0.00268269579527972\\
0.849481067993644	0.00308884359647748\\
0.849071731914657	0.00355648030622313\\
0.848535557673915	0.00409491506238043\\
0.847835909647728	0.00471486636345739\\
0.846927464891721	0.00542867543932386\\
0.845755480868214	0.00625055192527398\\
0.84572864321608	0.00626962116086952\\
0.844185981109829	0.00719685673001152\\
0.842196304390248	0.00828642772854684\\
0.841206030150754	0.00881869170686216\\
0.839635423050632	0.00954095476349994\\
0.836683417085427	0.01087909292045\\
0.83641125272617	0.0109854114198756\\
0.832339074539281	0.012648552168553\\
0.832160804020101	0.0127250375787888\\
0.827638190954774	0.0144417839632473\\
0.82727446868107	0.0145634847750124\\
0.823115577889447	0.0160910574232604\\
0.821043868562352	0.0167683293681101\\
0.818592964824121	0.0176585649803626\\
0.814070351758794	0.0191464143159798\\
0.813537669294377	0.0193069772888325\\
0.809547738693467	0.0206689350678607\\
0.805025125628141	0.0220854776193031\\
0.804527894171468	0.0222299648252619\\
0.800502512562814	0.0235687506835614\\
0.795979899497487	0.0249727479651315\\
0.793834571275294	0.0255954792269954\\
0.791457286432161	0.0263894932748363\\
0.786934673366834	0.0278180485901336\\
0.782412060301508	0.029170303985145\\
0.781348903612196	0.0294705170255181\\
0.777889447236181	0.0306038391198957\\
0.773366834170854	0.0320205773494436\\
0.768844221105528	0.0333771409083905\\
0.766901242105035	0.0339322177189533\\
0.764321608040201	0.0347905570920674\\
0.759798994974874	0.0362374633358462\\
0.755276381909548	0.0376356669468514\\
0.750753768844221	0.0389855960173518\\
0.750459345722781	0.0390693993705462\\
0.746231155778894	0.0404789286403272\\
0.741708542713568	0.0419450349839397\\
0.737185929648241	0.043370747951766\\
0.732663316582915	0.0447556885192901\\
0.731884026391489	0.0449843266896945\\
0.728140703517588	0.0462732814639298\\
0.723618090452261	0.0477912351543587\\
0.719095477386935	0.0492739808215995\\
0.714572864321608	0.0507204790688508\\
0.711111265899565	0.0517947467923121\\
0.710050251256282	0.0521809129707959\\
0.705527638190955	0.0537693567799315\\
0.701005025125628	0.0553253365054302\\
0.696482412060301	0.0568471366453236\\
0.691959798994975	0.0583320854080258\\
0.687871327013968	0.0596362331659464\\
0.687437185929648	0.059798602275431\\
0.682914572864322	0.0614204423519584\\
0.678391959798995	0.0630013550945189\\
0.673869346733668	0.0645378944102597\\
0.669346733668342	0.0660269820169661\\
0.664824120603015	0.0674659566001393\\
0.660904978839052	0.06866488450043\\
0.660301507537688	0.0688811355277411\\
0.655778894472362	0.0704213743278378\\
0.651256281407035	0.0719064762254626\\
0.646733668341709	0.0733348317981551\\
0.642211055276382	0.0747056280841337\\
0.637688442211055	0.0760189406296897\\
0.633165829145729	0.077275816295146\\
0.628643216080402	0.0784783630995831\\
0.626340526945739	0.079060432109077\\
0.624120603015075	0.0797167772874764\\
0.619597989949749	0.0809945281585418\\
0.615075376884422	0.0822305306918811\\
0.610552763819096	0.0834331627745436\\
0.606030150753769	0.0846130733817727\\
0.601507537688442	0.085784296847805\\
0.596984924623116	0.0869621091169921\\
0.592462311557789	0.0881517511765585\\
0.587939698492462	0.0893478616709222\\
0.583417085427136	0.0905450044962518\\
0.581559077850865	0.0910298177991522\\
0.578894472361809	0.0918451563765249\\
0.574371859296482	0.0932107940611197\\
0.569849246231156	0.0945619781389544\\
0.565326633165829	0.0958922375463929\\
0.560804020100503	0.0971959322016158\\
0.556281407035176	0.0984684803213417\\
0.551758793969849	0.0997064366689971\\
0.547236180904523	0.100907477645665\\
0.542713567839196	0.102070305384193\\
0.538190954773869	0.103194505804678\\
0.533668341708543	0.104280396474897\\
0.531358752018436	0.104811313415469\\
0.529145728643216	0.105407679042638\\
0.524623115577889	0.106576822500654\\
0.520100502512563	0.107709503883369\\
0.515577889447236	0.10880760987456\\
0.51105527638191	0.109873258607757\\
0.506532663316583	0.110908733695159\\
0.502010050251256	0.111916346404472\\
0.49748743718593	0.112898578064354\\
0.492964824120603	0.113857846461065\\
0.488442211055276	0.114796149101387\\
0.48391959798995	0.115715377560239\\
0.479396984924623	0.116617043058974\\
0.474874371859296	0.11750250929304\\
0.47035175879397	0.118372943537105\\
0.465829145728643	0.119229392551436\\
0.461306532663317	0.120072775192106\\
0.457991489063992	0.120679264063933\\
0.45678391959799	0.1209379828458\\
0.452261306532663	0.121883189134214\\
0.447738693467337	0.122817773811462\\
0.44321608040201	0.123742368272919\\
0.438693467336683	0.124657517847663\\
0.434170854271357	0.125563748866956\\
0.42964824120603	0.126461517388385\\
0.425125628140704	0.127351251431629\\
0.420603015075377	0.128233332266724\\
0.41608040201005	0.129108096782132\\
0.411557788944724	0.129975868694125\\
0.407035175879397	0.130836915191503\\
0.3 0.1480\\
0.2 0.1620\\
0.1 0.172\\
0 0.17212\\
};

\end{axis}
\end{tikzpicture}%

%% file: figures/StabReCad.tex
%
%
\definecolor{mycolor1}{rgb}{1.00000,0.00000,1.00000}%
\definecolor{mycolor2}{rgb}{0.00000,1.00000,1.00000}%
\begin{tikzpicture}[baseline]

\begin{axis}[%
width=.4\textwidth,
height=.3\textwidth,
scale only axis,
every outer x axis line/.append style={black},
every x tick label/.append style={font=\color{black}},
xmode=log,
xmin=1,
xmax=256,
xlabel={$Re$},
every outer y axis line/.append style={black},
every y tick label/.append style={font=\color{black}},
ymode=log,
ymin=0.015625,
ymax=1,
ylabel={$Ca$},
axis background/.style={fill=white},
axis x line*=bottom,
axis y line*=left,
legend style={legend cell align=left,align=left,draw=black},
legend pos=south east,
xtick={1,2,4,8,16,32,64,128,256},
xticklabels={$1$,$2$,$4$,$8$,$16$,$32$,$64$,$128$,$256$},
ytick={  1,.5,.25,   .125,  0.0625,  0.03125, 0.015625},
yticklabels={  $1$,  $1/2$,  $1/4$,   $1/8$,  $1/16$,  $1/32$, $1/64$},
]
\node[anchor=north west] at (rel axis cs:.25,.9) {Stable};
\node[anchor=north west] at (rel axis cs:.3,.2) {Unstable};

\addplot [color=black,dashed,forget plot,thin]
  table[row sep=crcr]{%
1	0.0324\\
2	0.0648\\
4	0.1296\\
8	0.2592\\
16	0.5184\\
32	1.0368\\
64	2.0736\\
128	4.1472\\
256	8.2944\\
};
\label{OhLinLimit},

\addplot [color=blue,dashed,forget plot,thin]
  table[row sep=crcr]{%
1	0.030625\\
2	0.06125\\
4	0.1225\\
8	0.245\\
16	0.49\\
32	0.98\\
64	1.96\\
128	3.92\\
256	7.84\\
};
\addplot [color=red,dashed,forget plot,thin]
  table[row sep=crcr]{%
1	0.024336\\
2	0.048672\\
4	0.097344\\
8	0.194688\\
16	0.389376\\
32	0.778752\\
64	1.557504\\
128	3.115008\\
256	6.230016\\
};
\addplot [color=green,dashed,forget plot,thin]
  table[row sep=crcr]{%
1	0.019044\\
2	0.038088\\
4	0.076176\\
8	0.152352\\
16	0.304704\\
32	0.609408\\
64	1.218816\\
128	2.437632\\
256	4.875264\\
};
\addplot [color=mycolor1,dashed,forget plot,thin]
  table[row sep=crcr]{%
1	0.013225\\
2	0.02645\\
4	0.0529\\
8	0.1058\\
16	0.2116\\
32	0.4232\\
64	0.8464\\
128	1.6928\\
256	3.3856\\
};
\addplot [color=mycolor2,dashed,forget plot,thin]
  table[row sep=crcr]{%
1	0.007744\\
2	0.015488\\
4	0.030976\\
8	0.061952\\
16	0.123904\\
32	0.247808\\
64	0.495616\\
128	0.991232\\
256	1.982464\\
};

\addplot[solid,draw=black]
table[row sep=crcr] {%
x	y\\
1	0.0323076898539557\\
1.05462575847443	0.0340169270833333\\
1.07259114583333	0.03457820924044\\
1.15478967052182	0.037109375\\
1.1640625	0.0373946912290789\\
1.26672843561109	0.04052734375\\
1.2724609375	0.0407021924102136\\
1.39006419076346	0.0442708333333333\\
1.39583333333333	0.0444458654185013\\
1.52459483858437	0.04833984375\\
1.5322265625	0.0485706212819891\\
1.67018744724734	0.052734375\\
1.6796875	0.0530210751595773\\
1.82673763317917	0.0574544270833333\\
1.83626302083333	0.0577415783945682\\
1.99415084457012	0.0625\\
2	0.0626762544973696\\
2.17708333333333	0.0680150661114159\\
2.17770697657255	0.0680338541666667\\
2.375	0.0739764761211475\\
2.38306435496094	0.07421875\\
2.59375	0.0805470364916685\\
2.61071900921006	0.0810546875\\
2.83333333333333	0.0877130186913297\\
2.86118086382827	0.0885416666666667\\
3.09375	0.0954603606445544\\
3.13498614813661	0.0966796875\\
3.375	0.103774615691798\\
3.43270435126711	0.10546875\\
3.67708333333333	0.11264090445461\\
3.75494467402559	0.114908854166667\\
4	0.122043869762428\\
4.10216259074741	0.125\\
4.35416666666667	0.132288175877696\\
4.48572356058019	0.136067708333333\\
4.75	0.143655460184397\\
4.91811144005739	0.1484375\\
5.1875	0.156094776253211\\
5.40152110870615	0.162109375\\
5.66666666666667	0.169554292914788\\
5.93834589574532	0.177083333333333\\
6.1875	0.183981287039587\\
6.53121334028952	0.193359375\\
6.75	0.199322152614344\\
7.18302259545298	0.2109375\\
7.35416666666667	0.215522417126729\\
7.89698584331186	0.229817708333333\\
8	0.232526765346859\\
8.67942193624412	0.25\\
8.70833333333333	0.250742283757258\\
9.5	0.270560286194346\\
9.56444287406566	0.272135416666667\\
10.375	0.291900591683374\\
10.5838145774897	0.296875\\
11.3333333333333	0.314682686480041\\
11.7440362215476	0.32421875\\
12.375	0.338825973955962\\
13.0529454352194	0.354166666666667\\
13.5	0.36425034195712\\
14.5190938426135	0.38671875\\
14.7083333333333	0.39087663950984\\
16	0.418626716441383\\
16.1566465994163	0.421875\\
17.4166666666667	0.447897130947039\\
18.0105133583278	0.459635416666667\\
19	0.47910343286119\\
20.1029029935923	0.5\\
20.75	0.512195279419102\\
22.5406642690166	0.544840494791667\\
22.6666666666667	0.547124192615248\\
24.75	0.58384128205246\\
25.4406152294883	0.595703125\\
27	0.622314111465963\\
28.7875615436321	0.652099609375\\
29.4166666666667	0.662512771263233\\
32	0.704411342128351\\
32.5803166987364	0.713541666666667\\
34.8333333333333	0.748738517035248\\
36.8879242137111	0.779541015625\\
38	0.796088260984705\\
41.5	0.846238753972738\\
41.7431158481964	0.849609375\\
45.3333333333333	0.898972090003574\\
47.1607691720714	0.923258463541667\\
49.5	0.954078647473919\\
53.1015808153922	1\\
nan	nan\\
};
\addlegendentry{$d=0.1$};
\label{OhNLinLimit},

\addplot[solid,draw=blue]
table[row sep=crcr] {%
x	y\\
1	0.0300417523805359\\
1.04208246654781	0.03125\\
1.07259114583333	0.0321255056256908\\
1.13909335099475	0.0340169270833333\\
1.1640625	0.0347266743487038\\
1.24832620624898	0.037109375\\
1.2724609375	0.037791570331513\\
1.36956712544208	0.04052734375\\
1.39583333333333	0.0412671360761712\\
1.50270845602051	0.0442708333333333\\
1.5322265625	0.0451002302077353\\
1.64768304029267	0.04833984375\\
1.6796875	0.049237644655363\\
1.80443870896626	0.052734375\\
1.83626302083333	0.0536261657062242\\
1.9729254009288	0.0574544270833333\\
2	0.0582126343567478\\
2.15305685508937	0.0625\\
2.17708333333333	0.063172766316454\\
2.35079774537079	0.0680338541666667\\
2.375	0.0687108021066089\\
2.57233054198541	0.07421875\\
2.59375	0.0748163052458695\\
2.81808486871387	0.0810546875\\
2.83333333333333	0.0814784664312494\\
3.08853553097819	0.0885416666666667\\
3.09375	0.0886858854410537\\
3.375	0.0964265097558894\\
3.38424947415925	0.0966796875\\
3.67708333333333	0.104687599367497\\
3.70582356309059	0.10546875\\
4	0.113455680564529\\
4.05376036709774	0.114908854166667\\
4.35416666666667	0.123018449113085\\
4.4279251107389	0.125\\
4.75	0.133637975756375\\
4.84131067214253	0.136067708333333\\
5.1875	0.145260776720086\\
5.30837355334926	0.1484375\\
5.66666666666667	0.157832211568745\\
5.83183579140521	0.162109375\\
6.1875	0.171295462473547\\
6.41483471304144	0.177083333333333\\
6.75	0.185591287142102\\
7.06098934240264	0.193359375\\
7.35416666666667	0.200658097524678\\
7.77448517571193	0.2109375\\
8	0.216432071855036\\
8.56103902499946	0.229817708333333\\
8.70833333333333	0.233316426973201\\
9.42776857390772	0.25\\
9.5	0.251665364682651\\
10.375	0.271331172818334\\
10.4117214625616	0.272135416666667\\
11.3333333333333	0.292166462149251\\
11.5561826856592	0.296875\\
12.375	0.314035178360657\\
12.8758491762534	0.32421875\\
13.5	0.336798793806331\\
14.3900171737662	0.354166666666667\\
14.7083333333333	0.360320232336987\\
16	0.384464089818329\\
16.1263101562231	0.38671875\\
17.4166666666667	0.409495038679826\\
18.1602458505973	0.421875\\
19	0.435678682752928\\
20.538595484167	0.459635416666667\\
20.75	0.462880647613864\\
22.6666666666667	0.490968548376034\\
23.311177417787	0.5\\
24.75	0.519802600256458\\
26.6550789457827	0.544840494791667\\
27	0.549284530939142\\
29.4166666666667	0.579272907789582\\
30.7966799295374	0.595703125\\
32	0.609734605045438\\
34.8333333333333	0.640764684177137\\
35.9483973800971	0.652099609375\\
38	0.672484392086759\\
41.5	0.70486678287868\\
42.489728969083	0.713541666666667\\
45.3333333333333	0.737854241533619\\
49.5	0.771431558419309\\
50.5484586438555	0.779541015625\\
54	0.805542639680866\\
58.8333333333333	0.840174882089845\\
60.1940288702091	0.849609375\\
64	0.875283631681178\\
69.6666666666667	0.911361849268816\\
71.644009671387	0.923258463541667\\
76	0.948720213960658\\
83	0.987078400011096\\
85.4840029446811	1\\
nan	nan\\
};
\addlegendentry{$d=0.2$};

\addplot[solid,draw=red]
table[row sep=crcr] {%
x	y\\
1	0.0242104723957664\\
1.07259114583333	0.0259651641648975\\
1.08424647523913	0.0262451171875\\
1.1640625	0.0281617553675064\\
1.18621314004607	0.0286916097005208\\
1.2724609375	0.0307540700955535\\
1.29325271138814	0.03125\\
1.39583333333333	0.0336955515150194\\
1.40933734635847	0.0340169270833333\\
1.5322265625	0.0369397882113919\\
1.53936543021183	0.037109375\\
1.6796875	0.0404409665326791\\
1.68332864839632	0.04052734375\\
1.83626302083333	0.0441533900592047\\
1.84121898927602	0.0442708333333333\\
2	0.0480314850625089\\
2.01302327996074	0.04833984375\\
2.17708333333333	0.0522220211529733\\
2.19875855390987	0.052734375\\
2.375	0.0568972297451469\\
2.39863851903662	0.0574544270833333\\
2.59375	0.0620493600098084\\
2.6129399446863	0.0625\\
2.83333333333333	0.067669100481318\\
2.84894074641833	0.0680338541666667\\
3.09375	0.0737466563170266\\
3.11406651673144	0.07421875\\
3.375	0.0802723603211043\\
3.40888677812135	0.0810546875\\
3.67708333333333	0.0872354561810508\\
3.73407778431202	0.0885416666666667\\
4	0.0946239916096\\
4.09033288103896	0.0966796875\\
4.35416666666667	0.102669597392206\\
4.47819059429055	0.10546875\\
4.75	0.111585799323275\\
4.89886483878849	0.114908854166667\\
5.1875	0.121329948917875\\
5.35413327299466	0.125\\
5.66666666666667	0.131851029682604\\
5.86131393489218	0.136067708333333\\
6.1875	0.143094387472137\\
6.43895787169376	0.1484375\\
6.75	0.155008157417337\\
7.09145234894045	0.162109375\\
7.35416666666667	0.167538795346259\\
7.82421068732275	0.177083333333333\\
8	0.180628311724978\\
8.64560246405101	0.193359375\\
8.70833333333333	0.194586532943815\\
9.5	0.209675676215042\\
9.56794831876841	0.2109375\\
10.375	0.225759728222903\\
10.6023815169938	0.229817708333333\\
11.3333333333333	0.242699488241391\\
11.760528766154	0.25\\
12.375	0.26032063370241\\
13.1025181649278	0.272135416666667\\
13.5	0.27846305833582\\
14.7009093477327	0.296875\\
14.7083333333333	0.296986656823704\\
16	0.315761631623778\\
16.6148738790779	0.32421875\\
17.4166666666667	0.335014410925407\\
18.9411833638223	0.354166666666667\\
19	0.35488871717179\\
20.75	0.375194143112217\\
21.8104252151867	0.38671875\\
22.6666666666667	0.395775452104502\\
24.75	0.416477956529628\\
25.3236649652357	0.421875\\
27	0.437175382688154\\
29.4166666666667	0.457757469653461\\
29.6477751018794	0.459635416666667\\
32	0.478134283381013\\
34.8333333333333	0.498381022099888\\
35.0791932027206	0.5\\
38	0.518461687752831\\
41.5	0.538314385583043\\
42.7464506249845	0.544840494791667\\
45.3333333333333	0.557735606091706\\
49.5	0.576665834467528\\
54	0.595078159994067\\
54.1619895926002	0.595703125\\
58.8333333333333	0.612792467765595\\
64	0.629867541019055\\
69.6666666666667	0.646594329260587\\
71.7098725529178	0.652099609375\\
76	0.663032382851399\\
83	0.678935561534778\\
90.6666666666667	0.694073853091398\\
99	0.70821511122776\\
102.572659945377	0.713541666666667\\
108	0.72117381064858\\
117.666666666667	0.732765367571193\\
128	0.742839707222049\\
139.479166666667	0.751554735527021\\
152.5	0.759177144098031\\
166.9375	0.765788473568424\\
182.666666666667	0.771482555156335\\
199.5625	0.776356762599451\\
213.094479619131	0.779541015625\\
217.5	0.78048982304808\\
236.354166666667	0.783995566251533\\
256	0.787035601567512\\
nan	nan\\
};
\addlegendentry{$d=0.3$};

\addplot[solid,draw=green]
table[row sep=crcr] {%
x	y\\
1	0.0187407062886588\\
1.06028018600307	0.0198822021484375\\
1.07259114583333	0.0201152869977541\\
1.16247919992557	0.0218098958333333\\
1.1640625	0.0218397363561558\\
1.2724609375	0.0238771179186948\\
1.27586831694961	0.0239410400390625\\
1.39583333333333	0.026190430619791\\
1.39875409498059	0.0262451171875\\
1.52949429968711	0.0286916097005208\\
1.5322265625	0.0287427088716839\\
1.66644908864124	0.03125\\
1.6796875	0.0314970889577884\\
1.81482104445909	0.0340169270833333\\
1.83626302083333	0.0344163745269318\\
1.98093160761652	0.037109375\\
2	0.0374640249641557\\
2.16485969415332	0.04052734375\\
2.17708333333333	0.0407542753915808\\
2.36681855924307	0.0442708333333333\\
2.375	0.0444223140610388\\
2.58706772966612	0.04833984375\\
2.59375	0.0484631361900031\\
2.82590419074266	0.052734375\\
2.83333333333333	0.0528708609951383\\
3.08367043912084	0.0574544270833333\\
3.09375	0.0576386621210059\\
3.36076204209285	0.0625\\
3.375	0.0627585166868172\\
3.66686624697761	0.0680338541666667\\
3.67708333333333	0.0682179177145784\\
4	0.0740052334640449\\
4.01198010547313	0.07421875\\
4.35416666666667	0.0802949799640358\\
4.39724648922296	0.0810546875\\
4.75	0.0872500888394461\\
4.82414840707397	0.0885416666666667\\
5.1875	0.0948413201883936\\
5.29455608739335	0.0966796875\\
5.66666666666667	0.103035092924303\\
5.8107364055672	0.10546875\\
6.1875	0.111793293436643\\
6.37543902032953	0.114908854166667\\
6.75	0.121073235058306\\
6.99200127653531	0.125\\
7.35416666666667	0.130810036139408\\
7.68728157030511	0.136067708333333\\
8	0.140937232585082\\
8.49283469547457	0.1484375\\
8.70833333333333	0.151674535332545\\
9.42326030070558	0.162109375\\
9.5	0.163214044513186\\
10.375	0.175450463934053\\
10.4952010605335	0.177083333333333\\
11.3333333333333	0.188273428936668\\
11.7273175173102	0.193359375\\
12.375	0.201563537992072\\
13.1434672980396	0.2109375\\
13.5	0.215197178475799\\
14.7083333333333	0.229050738586507\\
14.777672596984	0.229817708333333\\
16	0.243005323312454\\
16.6874702764623	0.25\\
17.4166666666667	0.257090135165605\\
19	0.271371939368366\\
19.089831735033	0.272135416666667\\
20.75	0.285437000762748\\
22.3138207862699	0.296875\\
22.6666666666667	0.29933282952404\\
24.75	0.312812676202501\\
26.7113520799946	0.32421875\\
27	0.325824001623701\\
29.4166666666667	0.338245674655834\\
32	0.349955287413914\\
33.0261303629432	0.354166666666667\\
34.8333333333333	0.361273379701308\\
38	0.372337521554072\\
41.5	0.382856016782053\\
42.975677457942	0.38671875\\
45.3333333333333	0.392819292700677\\
49.5	0.40185910368673\\
54	0.409398595437625\\
58.8333333333333	0.415102072351188\\
64	0.418808548465717\\
69.6666666666667	0.421241454847525\\
71.6520181531081	0.421875\\
76	0.423258598386735\\
83	0.424874020520083\\
90.6666666666667	0.426052452885352\\
99	0.426830281965461\\
108	0.427249663762795\\
117.666666666667	0.427349395212499\\
128	0.427160439346111\\
139.479166666667	0.424945586107186\\
146.923858120676	0.421875\\
152.5	0.419777669190011\\
166.9375	0.41293629788352\\
182.666666666667	0.40548875931421\\
199.5625	0.398162092992245\\
217.5	0.391296172515599\\
230.813570176427	0.38671875\\
236.354166666667	0.385216036495384\\
256	0.380264800477807\\
nan	nan\\
};
\addlegendentry{$d=0.4$};

\addplot[solid,draw=mycolor1,forget plot]
table[row sep=crcr] {%
x	y\\
256	0.203210259021783\\
236.354166666667	0.208146844255981\\
226.423128093371	0.2109375\\
217.5	0.214004006048081\\
199.5625	0.220800292430806\\
182.666666666667	0.228095689408635\\
178.75953324605	0.229817708333333\\
166.9375	0.235244784114196\\
152.5	0.241546757370981\\
139.479166666667	0.246646079279761\\
128.099093652986	0.25\\
128	0.250034260364607\\
117.666666666667	0.252638308339465\\
108	0.255626520507637\\
99	0.259046106852604\\
90.6666666666667	0.262876742076229\\
83	0.266950161097696\\
76	0.270842045095163\\
73.1225729954785	0.272135416666667\\
69.6666666666667	0.273957074527289\\
64	0.275223349660223\\
58.8333333333333	0.274703790046642\\
54	0.273190813392167\\
52.0476953808991	0.272135416666667\\
49.5	0.270773933661787\\
45.3333333333333	0.267375880999677\\
41.5	0.26276022566764\\
38	0.256827406385254\\
34.9412201774859	0.25\\
34.8333333333333	0.249742300466952\\
32	0.242652667745407\\
29.4166666666667	0.235333714386674\\
27.7746493406586	0.229817708333333\\
27	0.22712114443644\\
24.75	0.218269702666795\\
23.1195607007856	0.2109375\\
22.6666666666667	0.208835589249601\\
20.75	0.19912415639335\\
19.7168837666046	0.193359375\\
19	0.189237243274266\\
17.4166666666667	0.179413810039751\\
17.0638263022528	0.177083333333333\\
16	0.16985864877305\\
14.928278315241	0.162109375\\
14.7083333333333	0.160469076801785\\
13.5	0.151086961918234\\
13.172992828234	0.1484375\\
12.375	0.141752902191976\\
11.7223102571743	0.136067708333333\\
11.3333333333333	0.132536406096413\\
10.5289190721879	0.125\\
10.375	0.123508275666616\\
9.51121783955155	0.114908854166667\\
9.5	0.114795294573681\\
8.70833333333333	0.106594684042141\\
8.60254518043545	0.10546875\\
8	0.0989809711829817\\
7.79085016416977	0.0966796875\\
7.35416666666667	0.0918271333481192\\
7.06267972004675	0.0885416666666667\\
6.75	0.0849854467360249\\
6.40853485152416	0.0810546875\\
6.1875	0.0784884450598669\\
5.82350342262581	0.07421875\\
5.66666666666667	0.0723631609812237\\
5.30396597979281	0.0680338541666667\\
5.1875	0.0666305531943703\\
4.84733412624124	0.0625\\
4.75	0.0613090002977612\\
4.43724830468161	0.0574544270833333\\
4.35416666666667	0.0564261707703878\\
4.05801131195317	0.052734375\\
4	0.0520087619488918\\
3.70838366049536	0.04833984375\\
3.67708333333333	0.0479449187551715\\
3.38700106049088	0.0442708333333333\\
3.375	0.0441184442497414\\
3.09375	0.0405354516844666\\
3.09311551525284	0.04052734375\\
2.83333333333333	0.0372010220589643\\
2.82619497371998	0.037109375\\
2.59375	0.03411925889286\\
2.5858128441862	0.0340169270833333\\
2.375	0.0312932561884085\\
2.37165830845749	0.03125\\
2.17708333333333	0.0287269803518179\\
2.17436020431889	0.0286916097005208\\
2	0.0264245895676898\\
1.98621453453319	0.0262451171875\\
1.83626302083333	0.0242911821442147\\
1.8094181882621	0.0239410400390625\\
1.6796875	0.0222476068361426\\
1.64618745571799	0.0218098958333333\\
1.5322265625	0.0203199531323685\\
1.49877460291241	0.0198822021484375\\
1.39583333333333	0.0185346068140224\\
1.36941460317707	0.0181884765625\\
1.2724609375	0.0169182128302617\\
1.260336162566	0.0167592366536458\\
1.17377589932709	0.015625\\
nan	nan\\
};
\addlegendentry{$d=0.5$};

\addplot[solid,draw=mycolor2]
table[row sep=crcr] {%
x	y\\
256	0.114695484111993\\
253.127566544891	0.114908854166667\\
236.354166666667	0.116428771985698\\
217.5	0.118651022581081\\
199.5625	0.121546924701958\\
183.982828795406	0.125\\
182.666666666667	0.125367866443056\\
166.9375	0.130278728506899\\
152.5	0.135950839035378\\
152.178907627371	0.136067708333333\\
139.479166666667	0.141904080423057\\
128	0.146231863033429\\
120.155845598507	0.1484375\\
117.666666666667	0.149310563021313\\
108	0.153278378372017\\
99	0.158124393536754\\
92.9836422829156	0.162109375\\
90.6666666666667	0.164047065957073\\
83	0.171092084539872\\
77.0556061181513	0.177083333333333\\
76	0.178395340566268\\
69.6666666666667	0.184618598575759\\
64	0.187123912641315\\
58.8333333333333	0.186759157909885\\
54	0.185683403812459\\
49.5	0.183848093004319\\
45.3333333333333	0.181120525211509\\
41.5	0.177302800689018\\
41.3291050240353	0.177083333333333\\
38	0.17238550055668\\
34.8333333333333	0.166195310059368\\
33.073676091572	0.162109375\\
32	0.159491338960352\\
29.4166666666667	0.152590378237792\\
28.113463393192	0.1484375\\
27	0.14492173706198\\
24.75	0.136798830732045\\
24.5655298113893	0.136067708333333\\
22.6666666666667	0.128778842641008\\
21.7672068187818	0.125\\
20.75	0.120785750376009\\
19.4276010943034	0.114908854166667\\
19	0.112987515376817\\
17.4174818766113	0.10546875\\
17.4166666666667	0.105464856044311\\
16	0.0984498135106704\\
15.6569662571712	0.0966796875\\
14.7083333333333	0.0918077797010437\\
14.094113692852	0.0885416666666667\\
13.5	0.0854158673763162\\
12.695664012232	0.0810546875\\
12.375	0.0793457554284189\\
11.4381436305163	0.07421875\\
11.3333333333333	0.0736593025463655\\
10.375	0.0684285957837021\\
10.3043410662267	0.0680338541666667\\
9.5	0.0637023340144943\\
9.2822611210935	0.0625\\
8.70833333333333	0.0593789569324869\\
8.36352307569421	0.0574544270833333\\
8	0.0553974899389562\\
7.53904785366323	0.052734375\\
7.35416666666667	0.0516515332194051\\
6.79569994749236	0.04833984375\\
6.75	0.0480648606067984\\
6.1875	0.0446487136996435\\
6.12588329838877	0.0442708333333333\\
5.66666666666667	0.0414126898892321\\
5.52570928609414	0.04052734375\\
5.1875	0.0383664590900649\\
4.99232310852341	0.037109375\\
4.75	0.0355161288682371\\
4.52366842679314	0.0340169270833333\\
4.35416666666667	0.0328646109111785\\
4.11842363399851	0.03125\\
4	0.0304235705358183\\
3.75322113010655	0.0286916097005208\\
3.67708333333333	0.0281526017820936\\
3.40846184234502	0.0262451171875\\
3.375	0.0260043620637086\\
3.09375	0.0239752483836225\\
3.08902077595388	0.0239410400390625\\
2.83333333333333	0.0220572278582588\\
2.79985062235872	0.0218098958333333\\
2.59375	0.0202468860105904\\
2.54576717213185	0.0198822021484375\\
2.375	0.0185329284930322\\
2.33147835326242	0.0181884765625\\
2.17708333333333	0.0168891670372271\\
2.16166031192174	0.0167592366536458\\
2.0410834917047	0.015625\\
nan	nan\\
};
\addlegendentry{$d=0.6$};


\addplot [color=black,dotted,forget plot,thin,smooth]
  table[row sep=crcr]{%
236.354166666667	0.783995566251533\\
102	0.457160439346111\\
58	0.273190813392167\\
58.8333333333333	0.186759157909885\\
};
\node[anchor=west] at (58.83,0.27) {$Ca_c (d)$};

\end{axis}
\end{tikzpicture}%

%% file: Tables/matrizLCircularLibre.tex
\begin{tabular}{|c|r|r|r|r|}
$\varphi=f_1$ & $i=1$ & $i=3$& $i=5$ &$i=7$\\
$j=1$&3.55&-7.42&-0.10&-0.51\\
$j=2$&-0.27&16.33&5.85&1.27\\
$j=3$&-4.67&7.65&-7.98&-38.37\\
$j=4$&0.45&-43.31&-14.05&100.91\\
$j=5$&-0.79&28.12&19.56&-68.44\\
\end{tabular}


%% file: Tables/matrizV_VPCircularLibre.tex
\begin{tabular}{|c|r|r|r|r|}
$ \varphi=V_0/V_P$&$i=0$&$i=2$&$i=4$&$i=6$\\
$j=0$&1.00&-0.01&0.04&-0.07\\
$j=2$&0.02&-0.17&0.26&-1.56\\
$j=4$&-0.32&-0.02&-2.70&7.93\\
$j=6$&-1.05&0.32&6.87&-15.38\\
$j=8$&0.94&0.07&-5.67&11.28\\
\end{tabular}

%% file: Tables/matrizbeta100CircularLibre.tex
\begin{tabular}{|c|r|r|r|r|}
$\varphi=100 \beta_0$&$i=0$&$i=2$&$i=4$&$i=6$\\
$j=0$&-0.03&0.31&-1.11&0.98\\
$j=2$&0.95&-5.67&31.40&-27.27\\
$j=4$&-8.76&73.27&-218.95&205.73\\
$j=6$&-6.43&-181.60&511.37&-393.88\\
$j=8$&18.21&163.32&-396.80&262.54\\
\end{tabular}

%% file: Tables/matrizLCircularRigido.tex
\begin{tabular}{|c|r|r|r|r|}
$\varphi=f_1$&$i=1$&$i=3$&$i=5$&$i=7$\\
$j=1$&13.28&-33.15&2.20&-0.63\\
$j=2$&-43.14&115.27&19.72&-28.76\\
$j=3$&48.60&-104.32&-137.39&122.41\\
$j=4$&-20.20&-36.05&255.14&-177.07\\
$j=5$&0.57&65.47&-150.87&89.05\\
\end{tabular}


%% file: Tables/matrizV_VPCircularRigido.tex
\begin{tabular}{|c|r|r|r|r|}
$ \varphi=V_0/V_P$  &$i=0$&$i=2$&$i=4$&$i=6$\\
$j=0$&1.00&-0.02&0.10&-0.20\\
$j=2$&-0.67&-0.63&-1.07&-0.47\\
$j=4$&0.05&2.42&4.19&5.87\\
$j=6$&0.31&-3.47&-7.35&-14.90\\
$i=8$&-0.22&1.95&4.88&12.49\\
\end{tabular}

%% file: Tables/matrizbeta100CircularRigido.tex
\begin{tabular}{|c|r|r|r|r|}
$\varphi=100 \beta_0$  &$i=0$&$i=2$&$i=4$&$i=6$\\
$j=0$&-0.01&0.14&-0.68&0.45\\
$j=2$&0.23&3.65&16.84&-9.38\\
$j=4$&3.44&60.06&-95.98&116.73\\
$j=6$&18.31&-130.40&171.85&-146.87\\
$j=8$&12.72&105.94&-100.60&50.42\\
\end{tabular}

%% file: Tables/matrizOmCircularRigido.tex
\begin{tabular}{|c|r|r|r|r|}
$\varphi=\Omega_0/8 \varepsilon$ &$i=0$&$i=2$&$i=4$&$i=6$\\
$j=0$&1.00&-0.02&0.12&-0.20\\
$j=2$&-0.14&-0.42&-0.93&-0.98\\
$j=4$&-0.77&1.30&1.00&10.63\\
$j=6$&0.94&-2.40&6.04&-31.08\\
$j=8$&-0.62&2.43&-10.59&29.65\\
\end{tabular}


%% file: Tables/matrizfdatos_Ca0.tex
\begin{tabular}{|c|c|c|c|c|}
$\varphi=f_1$& $i=1$&$i=3$&$i=5$&$i=7$\\
$j=1$&-104.96&-238.40&1115.32&-1865.91\\
$j=3$&9.20&-608.76&-2701.91&1727.04\\
$j=5$&-1287.13&-26449.45&162691.63&-246331.61\\
$j=7$&3666.66&94964.56&-547128.86&839403.47\\
$j=9$&-4931.48&-140975.28&775574.28&-1203791.84\\
\end{tabular}


%% file: figures/ReversibilityExample.tex
%
\begin{tikzpicture}

\hspace{-.2\textwidth}

\begin{axis}[hide axis, xmin=0, ymin=0, xmax=1, ymax=1,  scale only axis,    height=.25\textwidth,    width=.25\textwidth,    colormap/jet,    colorbar,    point meta min=-1,    point meta max=1,
    colorbar style={xshift=-.0\textwidth, title=$\hat{p}_0- x \partial_x p_P$,        width=.2cm,        height=.250\textwidth,        ytick={-1,-.5,0,.5,1},        yticklabels={$\leq-2$,$-1$,$0$,$+1$, $\geq + 2$},   yticklabel style={xshift=0.5ex} }
    ]
\node[inner sep=0pt,anchor=south west] (fig2) at (0,0)   {\includegraphics[width=.25\textwidth]{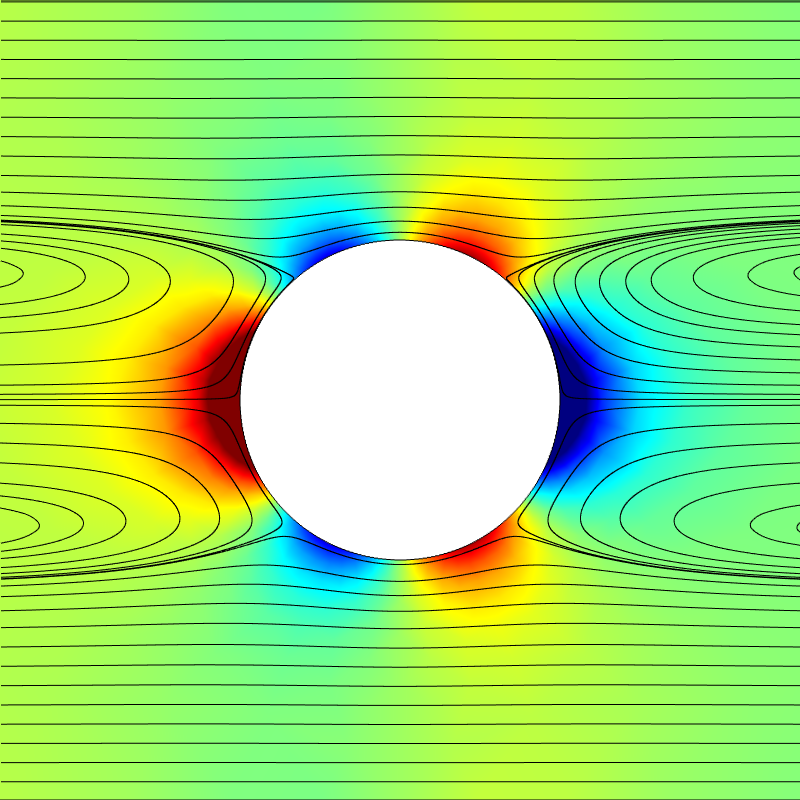}};  
\node[inner sep=0pt,anchor=north west] (fig2) at (0,1)   {(a)};   
\end{axis}

\hspace{.4\textwidth}
\begin{axis}[hide axis, xmin=0, ymin=0, xmax=1, ymax=1,  scale only axis,    height=.25\textwidth,    width=.25\textwidth,    colormap/jet,    colorbar,    point meta min=-1,    point meta max=2,
    colorbar style={xshift=-.0\textwidth, title=$\hat{p}_1$,        width=.2cm,        height=.250\textwidth,        ytick={-1,0,1,2},        yticklabels={$\leq-1 \cdot 10^{-2}$,$0$,$ +1 \cdot 10^{-2}$, $\geq + 2  \cdot 10^{-2}$},   yticklabel style={xshift=0.5ex} }
    ]
\node[inner sep=0pt,anchor=south west] (fig2) at (0,0)   {\includegraphics[width=.25\textwidth]{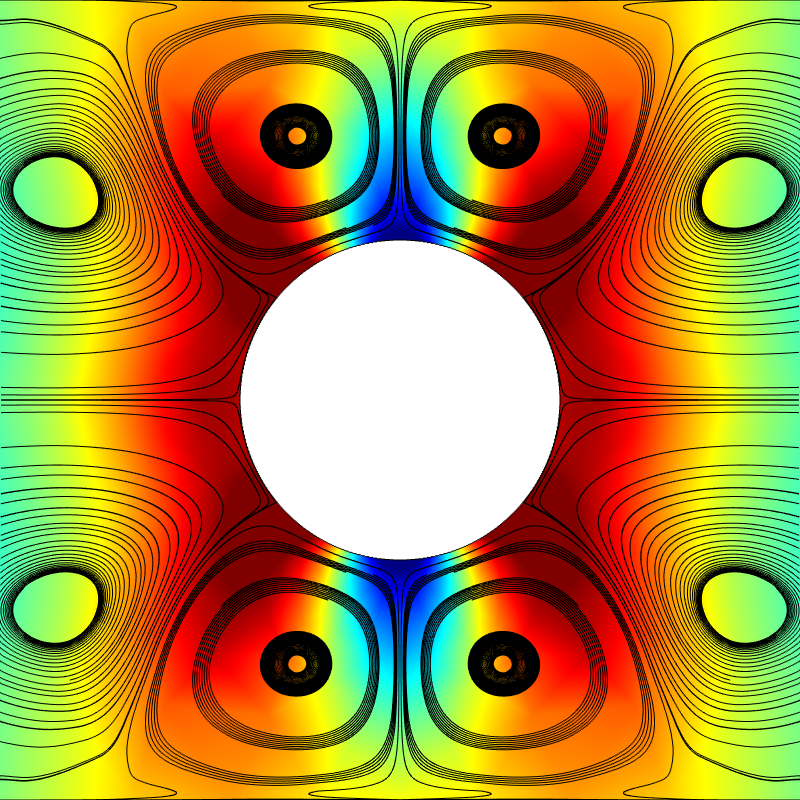}};  
\node[inner sep=0pt,anchor=north west] (fig2) at (0,1)   {(b)};   
\end{axis}

\end{tikzpicture}